\documentclass[sigconf, nonacm]{acmart}

\newcommand\vldbdoi{XX.XX/XXX.XX}
\newcommand\vldbpages{XXX-XXX}
\newcommand\vldbvolume{14}
\newcommand\vldbissue{1}
\newcommand\vldbyear{2020}
\newcommand\vldbauthors{\authors}
\newcommand\vldbtitle{\shorttitle} 

\newif\iftr     
\newif\ifall    
\newif\ifconf   
\newif\ifsq     
\newif\ifnonb   
\newif\iftodos  

\newif\ifsqCAP
\newif\ifsqVS
\newif\ifsqEN
\newif\ifsqTIT

\newcommand{\tr}[1]{\iftr #1 \fi}
\newcommand{\all}[1]{\ifall #1 \fi}
\newcommand{\cnf}[1]{\ifconf #1 \fi}

\sqfalse
\sqCAPfalse
\sqENfalse
\sqVSfalse
\sqTITfalse

\iftr
\renewcommand{\marginpar}{}
\fi

\usepackage{etex}
\usepackage{balance}
\usepackage{epstopdf}
\usepackage{placeins}

\usepackage{graphicx}
\usepackage{float}
\usepackage{dblfloatfix}
\usepackage{multirow}
\usepackage{rotating}
\usepackage[switch]{lineno} 
\usepackage{makecell}
\usepackage{tabulary}
\usepackage{parcolumns}
\usepackage{tikz}
\usetikzlibrary{tikzmark}

\usepackage{xpatch}
\expandafter\xpatchcmd
\csname pgfk@/tikz/every picture/.@cmd\endcsname
{\thepage}{\arabic{page}}{}{}

\tikzstyle{comment} = [draw, fill=blue!70, text=white, text width=3cm, minimum height=1cm, rounded corners, align=left, font=\scriptsize]
\tikzstyle{background_alg} = [draw, fill=blue!20, opacity=0.4, inner sep=4pt, rounded corners=2pt]

\usetikzlibrary{shapes}
\usetikzlibrary{plotmarks}
\usetikzlibrary{calc, fit}

\usepackage{enumitem}

\usepackage{amsmath,amsfonts}
\usepackage{mathtools,mathrsfs}

\newtheorem{theorem}{Theorem}[section]

\newtheorem{lemma}[theorem]{Lemma}

\usepackage{mleftright}
\newcommand{\bigTheta}[1]{\Theta\mleft( #1 \mright)}
\newcommand{\bigO}[1]{\mathcal{O}\mleft( #1 \mright)}

\usepackage{soul}
\usepackage{fontawesome}
\usepackage{pifont}
\usepackage{textcomp}
\usepackage{booktabs}
\usepackage{url}
\usepackage{pbox}
\usepackage[normalem]{ulem}
\usepackage[10pt]{moresize}

\usepackage{cleveref}
\usepackage[utf8]{inputenc}
\crefname{section}{§}{§§}
\Crefname{section}{§}{§§}

\newcommand{\macb}[1]{\textbf{{#1}}}

\definecolor{aablack}{rgb}{0.4 0.4 0.4}

\ifsqCAP
\usepackage[font={normalfont, scriptsize}]{caption}
\usepackage[font={normalfont, scriptsize}]{subfig}
\else
\usepackage[font={normalfont, footnotesize}]{caption}
\usepackage[font={normalfont, footnotesize}]{subfig}
\fi

\newcommand{\vspaceSQ}[1]{\ifsqVS\vspace{#1}\fi}
\newcommand{\enlargeSQ}[1]{\ifsqEN\enlargethispage{\baselineskip}\fi}

\ifsqTIT
\usepackage[compact]{titlesec}
\titlespacing*{\section}{0pt}{6pt}{2pt}
\titlespacing*{\subsection}{0pt}{5pt}{1pt}
\titlespacing*{\subsubsection}{0pt}{5pt}{1pt}
\fi

\usepackage{xcolor}
\definecolor{darkgrey}{RGB}{70,70,70}
\definecolor{lightgrey}{RGB}{200,200,200}
\definecolor{lyellow}{RGB}{255,255,100}
\definecolor{llyellow}{RGB}{250,250,180}
\definecolor{lgreen}{RGB}{144,238,144}

\usepackage[customcolors]{hf-tikz}
\hfsetbordercolor{white}
\hfsetfillcolor{vlgray}

\definecolor{vlgray}{rgb}{0.77 0.77 0.77}
\definecolor{ablack}{rgb}{0.2 0.2 0.2}
\definecolor{vllgray}{rgb}{0.9 0.9 0.9}
\definecolor{bblue}{rgb}{0.7 0.7 0.99}

\usepackage{colortbl}

\usepackage{listings}

\ifsq
\else
\fi

\newcommand{\maciej}[1]{\textcolor{blue}{[Maciej: #1]}}
\newcommand{\m}[1]{\textcolor{blue}{[Maciej: #1]}}

\newcommand{\zur}[1]{\textcolor{blue}{[Zur: #1]}}

\definecolor{hlL}{rgb}{0.8 0.8 0.99}

\newcounter{highlight}

\newcounter{hlLIR} 
\newcommand{\hlLIR}[3]{%
\stepcounter{hlLIR}\hfsetfillcolor{hlL}\tikzmarkin[set fill color=hlL, set border color=white, above offset=0.225, right offset=#1, below offset=-0.125]{\thehighlight}#2\tikzmarkend{\thehighlight}\hspace{-0.3em}\includegraphics[scale=0.13,trim=0 4 0 0]{#3}}

\newcounter{hlLIIR} 
\newcommand{\hlLIIR}[4]{%
\stepcounter{hlLIIR}\hfsetfillcolor{hlL}\tikzmarkin[set fill color=hlL, set border color=white, above offset=0.225, right offset=#1, below offset=-0.125]{\thehighlight}#2\tikzmarkend{\thehighlight}\hspace{-0.4em}\includegraphics[scale=0.13,trim=0 4 0 0]{#3}\includegraphics[scale=0.13,trim=0 4 0 0]{#4}}

\newcounter{Ahighlight}

\usepackage{scalerel,stackengine}
\stackMath
\newcommand\rwh[1]{%
\savestack{\tmpbox}{\stretchto{%
  \scaleto{%
        \scalerel*[\widthof{\ensuremath{#1}}]{\kern-.6pt\bigwedge\kern-.6pt}%
                  {\rule[-\textheight/2]{1ex}{\textheight}}
                              }{\textheight}%
}{0.5ex}}%
\stackon[1pt]{#1}{\tmpbox}%
}

\usepackage[hang,flushmargin]{footmisc}

\renewcommand{\epsilon}{\ensuremath\varepsilon}

\renewcommand{\phi}{\ensuremath{\varphi}}

\conffalse
\allfalse
\trtrue
\todosfalse

\newcommand{\VfaBatteryFull}{{\tiny\faBatteryFull}}
\newcommand{\VfaBatteryHalf}{{\tiny\faBatteryHalf}}

\newcolumntype{y}{>{\columncolor{yellow}}l}

\renewcommand{\marginpar}[1]{}
\renewcommand{\hl}[1]{#1}

\begin{document}




\ifconf
\title{\vspace{-1.5em}\hspace{-0.7em}\includegraphics[scale=0.35,trim=0 16 18 0]{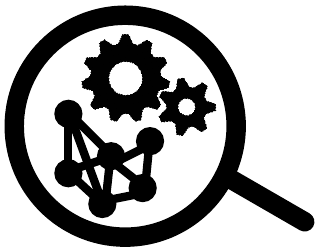}\hspace{-0.2em}~GraphMineSuite: Enabling High-Performance {\Huge\bf\emph{and}} {\hspace{-0.5em}Programmable Graph Mining Algorithms with Set Algebra [Benchmark \& Analysis]}\vspace{-0.5em}}
\else
\title{\vspace{-1.5em}\hspace{-0.7em}\includegraphics[scale=0.35,trim=0 16 18 0]{logo.pdf}\hspace{-0.2em}~GraphMineSuite: Enabling High-Performance {\Huge\bf\emph{and}} Programmable Graph Mining Algorithms with Set Algebra}
\fi


%

\iftr
\author{Maciej Besta$^1$$^*$, Zur Vonarburg-Shmaria$^1$, Yannick Schaffner$^1$, Leonardo Schwarz$^1$, \break Grzegorz Kwasniewski$^1$, Lukas Gianinazzi$^1$, Jakub Beranek$^2$, Kacper Janda$^3$, Tobias Holenstein$^1$,\break Sebastian Leisinger$^1$,
Peter Tatkowski$^1$, Esref Ozdemir$^1$, Adrian Balla$^1$, Marcin Copik$^1$,\break Philipp Lindenberger$^1$, Pavel Kalvoda$^1$, Marek Konieczny$^3$, Onur Mutlu$^1$, Torsten Hoefler$^1$$^*$}
       \affiliation{\vspace{0.3em}$^1$Department of Computer Science, ETH Zurich;
       {$^2$}Faculty of Electrical Engineering and Computer Science, VSB;\\
       {$^3$}Department of Computer Science, AGH-UST Krakow;
       {$^*$}Corresponding authors\\
}
\else
\author{Maciej Besta$^1$$^*$, Zur Vonarburg-Shmaria$^1$, Yannick Schaffner$^1$, Leonardo Schwarz$^1$, \break Grzegorz Kwasniewski$^1$, Lukas Gianinazzi$^1$, Jakub Beranek$^2$, Kacper Janda$^3$, Tobias Holenstein$^1$,\break Sebastian Leisinger$^1$,
Peter Tatkowski$^1$, Esref Ozdemir$^1$, Adrian Balla$^1$, Marcin Copik$^1$,\break Philipp Lindenberger$^1$, Marek Konieczny$^3$, Onur Mutlu$^1$, Torsten Hoefler$^1$$^*$}
       \affiliation{\vspace{0.3em}$^1$Department of Computer Science, ETH Zurich;
       {$^2$}Faculty of Electrical Engineering and Computer Science, VSB;\\
       {$^3$}Department of Computer Science, AGH-UST Krakow;
       {$^*$}Corresponding authors\\
}
\fi

\renewcommand{\shortauthors}{M. Besta et al.}
\renewcommand{\vldbauthors}{M. Besta et al.}

\ifconf
\renewcommand{\vldbtitle}{GraphMineSuite: Enabling High-Performance and Programmable Graph Mining Algorithms with Set Algebra [Ben.]}
\renewcommand{\shorttitle}{GraphMineSuite: Enabling High-Performance and Programmable Graph Mining Algorithms with Set Algebra [Ben.]}
\else
\renewcommand{\vldbtitle}{GraphMineSuite: Enabling High-Performance and Programmable Graph Mining Algorithms with Set Algebra}
\renewcommand{\shorttitle}{GraphMineSuite: Enabling High-Performance and Programmable Graph Mining Algorithms with Set Algebra}
\fi

\begin{abstract}
\all{Challenges in today's graph mining include fast-growing datasets inflating
compute times, massively parallel architectures that are hard to program, lack
of established comparison baselines, and many others. To address this,}
We propose GraphMineSuite (GMS): the first benchmarking suite for graph mining
that facilitates evaluating and constructing high-performance
graph mining algorithms. First, GMS comes with a benchmark specification based
on extensive literature review, prescribing representative problems,
algorithms, and datasets.
Second, GMS offers a carefully designed software platform for seamless testing
of different fine-grained elements of graph mining algorithms, such as graph
representations or algorithm subroutines. The platform includes parallel
implementations of more than 40 considered baselines, and it facilitates
developing complex and fast mining algorithms. High modularity is possible
by harnessing set algebra operations such as set intersection and
difference, which enables breaking complex graph mining algorithms into simple
building blocks that can be separately experimented with. GMS is supported with a broad
concurrency analysis for portability in performance insights, and a novel
performance metric to assess the throughput of graph mining algorithms,
enabling more insightful evaluation.
\all{illustrating theoretical tradeoffs between performance, storage, and
accuracy, and it can be used to quickly assess potential scalability of a new
algorithmic idea.}
As use cases, we harness GMS to rapidly redesign and accelerate
state-of-the-art baselines of core graph mining problems: degeneracy reordering
(by up to $>$2$\times$), maximal clique listing (by up to $>$9$\times$), $k$-clique listing
(by 1.1$\times$), and subgraph isomorphism (by up to 2.5$\times$), also
obtaining better theoretical performance bounds.
\all{GMS may become an established community benchmark and design platform,
propelling research into high-performance graph mining.}
\end{abstract}

\settopmatter{printfolios=true}

\maketitle
\pagestyle{plain}

\vspace{-1em}
%
{\normalsize\noindent\macb{Website:} {\small\url{http://spcl.inf.ethz.ch/Research/Parallel_Programming/GMS}}}
%
\vspace{0.5em}

\cnf{
%
\begingroup\small\noindent\raggedright\textbf{PVLDB Reference Format:}\\
\vldbauthors. \vldbtitle. PVLDB, \vldbvolume(\vldbissue): \vldbpages, \vldbyear.
\href{https://doi.org/\vldbdoi}{doi:\vldbdoi}
\endgroup
\begingroup
%
%
%
}



\vspaceSQ{-0.3em}
\section{INTRODUCTION AND MOTIVATION}
\label{sec:intro}
\vspaceSQ{-0.2em}

\marginpar{\vspace{-20em}\colorbox{yellow}{\textbf{R-2}}}


\iftr
Graph mining is used in many compute-related domains, such as social sciences
(e.g., studying human interactions), bioinformatics (e.g., analyzing protein
structures), chemistry (e.g., designing chemical compounds), medicine (e.g.,
drug discovery), cybersecurity (e.g., identifying intruder machines),
healthcare (e.g., exposing groups of people who submit fraudulent claims), web
graph analysis (e.g., providing accurate search services), entertainment
services (e.g., predicting movie popularity), and many
others~\cite{cook2006mining, jiang2013survey, horvath2004cyclic,
chakrabarti2006graph}.
\else
Graph mining is used in social sciences, bioinformatics, chemistry, medicine,
cybersecurity, web graph analysis, and many
others~\cite{cook2006mining, jiang2013survey, horvath2004cyclic,
chakrabarti2006graph}.
\fi
Yet, graphs can reach one trillion edges (the Facebook
graph (2015)~\cite{ching2015one}) or even 12 trillion edges (the Sogou
webgraph (2018)~\cite{lin2018shentu}), requiring unprecedented amounts of compute
power to solve even simple graph problems such as
BFS~\cite{lin2018shentu}.
\tr{For example, running PageRank on the Sogou webgraph using 38,656 compute
nodes (10,050,560 cores) on the Sunway TaihuLight
supercomputer~\cite{fu2016sunway} (nearly the full scale of TaihuLight) takes 8
minutes~\cite{lin2018shentu}.}
\all{Harder problems, such as mining $k$-cliques (time complexity is a high-degree
polynomial) or maximal cliques (NP-hard in the worst case), face even larger
challenges.}
{Harder problems, such as mining cliques, face even larger challenges.}

\tr{At the same time, massive parallelism has become prevalent in modern
compute devices, from smartphones to high-end
servers~\cite{bassini2018parallel}, bringing a promise of high-performance
parallel graph mining algorithms.}
\cnf{At the same time, massive parallelism has become prevalent in modern
compute devices~\cite{bassini2018parallel}, bringing a promise of fast 
parallel graph mining algorithms.}
\all{Parallel computing can thus be used to accelerate graph mining.}
Yet, several issues hinder achieving this.
First, a large number of graph mining algorithms and their
variants make it hard to identify the most relevant baselines as either
promising candidates for further improvement, or as appropriate comparison
targets. Similarly, a plethora of available networks hinder 
selecting relevant input datasets for evaluation.
Second, even when experimenting with a single specific algorithm, one often
faces numerous design choices, for example which graph representation to use,
whether to apply graph compression, how to represent auxiliary data
structures, etc.. Such choices may significantly impact performance, often in a
non-obvious way, and they may require a large coding effort when trying
different options~\cite{danisch2018listing}.
\tr{This is further aggravated by the fact that developing efficient parallel
algorithms is usually challenging~\cite{asanovic2009view} because one must
tackle issues such as deadlocks, data conflicts, and many
others~\cite{asanovic2009view}.}
\all{An example of this is a recent work by Danisch et
al.~\cite{danisch2018listing}, with a new way to parallelize an algorithm for
listing $k$-cliques. It takes \emph{a whole research paper} to design a
parallelization strategy for \emph{a single} graph pattern matching problem.}
\all{One such example is a recent work by Danisch et
al.~\cite{danisch2018listing}, where it takes \emph{a whole research paper} to
design a parallelization strategy for \emph{a single} problem of listing
$k$-cliques.}

\begin{table*}[t]
\vspace{-1.25em}
\setlength{\tabcolsep}{1pt}
\ifsq\renewcommand{\arraystretch}{0.5}\fi
\centering
\scriptsize
\sf
\begin{tabular}{llllllllllllll}
\toprule
\multirow{2}{*}{\makecell[c]{\textbf{Reference /}\\\textbf{Infrastructure}}} & \multirow{2}{*}{\makecell[c]{\textbf{Focus on}\\\textbf{what problems?}}} & \multicolumn{5}{c}{\textbf{Pattern Matching}} & \multicolumn{4}{c}{\textbf{Learning}} & \multirow{2}{*}{\makecell[l]{\textbf{Opt}}} & \multirow{2}{*}{\makecell[c]{\textbf{Vr}}} & \multirow{2}{*}{\makecell[c]{\textbf{Remarks}}} \\
\cmidrule(lr){3-7} \cmidrule(lr){8-11} 
 & &  
\textbf{mC?} & \textbf{kC?} & \textbf{dS?} & \textbf{sI?} & \textbf{fS?} & 
\textbf{vS?} & \textbf{lP?} & \textbf{cl?} & \textbf{cD?} & 
 & & \\
\midrule
\mbox{[B]} Cyclone~\cite{tang2016benchmarking} & \makecell[l]{Graph database queries} & \faTimes &  \faTimes & \faTimes &  \faTimes &  \faTimes &  \faTimes &  \faTimes &   \faTimes &  \faTimes  &  \faBatteryHalf$^{*}$ & \faBatteryHalf$^{**}$ & \makecell[l]{$^*$Only shortest paths. $^{**}$Only degree centrality.} \vspaceSQ{-0.15em} \\
\mbox{[B]} \makecell[l]{GBBS~\cite{dhulipala2020graph} + Ligra~\cite{shun2013ligra}} & \makecell[l]{More than 10 ``low-complexity'' algorithms} & \faTimes & \faBatteryFull & \faBatteryFull & \faTimes & \faTimes & \faTimes & \faTimes & \faTimes & \faTimes & \faBatteryFull $^\text{\faStar}$ & \faBatteryHalf$^*$ & \makecell[l]{$^*$Support for degeneracy, but no explicit rank derivation.\\ $^\text{\faStar}$GBBS offers a large number of optimization problems} \vspaceSQ{-0.15em} \\
\mbox{[B]} GraphBIG~\cite{nai2015graphbig} & \makecell[l]{Mostly vertex-centric schemes} & \faTimes &  \faBatteryHalf$^*$ &  \faTimes &  \faTimes &  \faTimes &  \faTimes &  \faTimes &   \faTimes &  \faTimes  &  \faBatteryHalf$^{**}$ & \faBatteryHalf & \makecell[l]{$^*$Only $k=3$. $^{**}$Only shortest paths and one coloring scheme.} \vspaceSQ{-0.15em} \\
\mbox{[B]} GAPBS~\cite{beamer2015gap} & \makecell[l]{Seven ``low-complexity'' algorithms} & \faTimes & \faBatteryHalf$^*$ & \faTimes & \faTimes & \faTimes & \faTimes & \faTimes & \faTimes & \faTimes & \faBatteryHalf$^{**}$ & \faTimes & \makecell[l]{$^*$Only $k=3$. $^{**}$Only shortest paths.} \vspaceSQ{-0.15em} \\
\mbox{[B]} LDBC~\cite{boncz2013ldbc} & \makecell[l]{Graph database queries} & \faTimes &  \faTimes &  \faTimes &  \faTimes &  \faTimes &  \faTimes &  \faTimes &   \faBatteryHalf$^*$ &  \faTimes  &  \faBatteryHalf$^{**}$ &  \faTimes &  \makecell[l]{$^*$Only one clustering coefficient. $^{**}$Only shortest paths.} \vspaceSQ{-0.15em} \\
\mbox{[B]} WGB~\cite{ammar2013wgb} & \makecell[l]{Mostly online queries} & \faTimes &  \faTimes &  \faTimes &  \faTimes &  \faTimes &  \faTimes &  \faTimes &   \faBatteryHalf$^*$ & \faTimes  &  \faBatteryHalf$^{**}$ &  \faTimes &  \makecell[l]{$^*$Only one clustering scheme. $^{**}$Only shortest paths.} \vspaceSQ{-0.15em} \\
\mbox{[B]} PBBS~\cite{blelloch2011problem} & \makecell[l]{General parallel problems}  & \faTimes & \faTimes & \faTimes & \faTimes & \faTimes & \faTimes & \faTimes & \faBatteryHalf & \faTimes & \faBatteryFull & \faTimes & Only graph optimization problems are considered \vspaceSQ{-0.15em} \\
\mbox{[B]} Graph500~\cite{murphy2010introducing} & \makecell[l]{Graph traversals} & \faTimes & \faTimes & \faTimes & \faTimes & \faTimes & \faTimes & \faTimes & \faTimes & \faTimes & \faBatteryHalf$^*$ & \faTimes & \makecell[l]{$^{*}$Support for shortest paths only.} \vspaceSQ{-0.15em} \\
\iftr
\mbox{[B]} HPCS~\cite{bader2005design} & \makecell[l]{Two ``low-complexity'' algorithms} & \faTimes &  \faTimes &  \faTimes &  \faTimes &  \faTimes &  \faTimes &  \faTimes &   \faBatteryHalf$^*$ &  \faTimes  &  \faTimes &  \faTimes &  \makecell[l]{$^*$Just one clustering scheme is considered} \vspaceSQ{-0.15em} \\
\mbox{[B]} Han at al. ~\cite{han2014experimental} & \makecell[l]{Evaluation of various graph processing systems} & \faTimes &  \faTimes &  \faTimes &  \faTimes &  \faTimes &  \faTimes &  \faTimes &   \faTimes &  \faTimes  &  \faBatteryHalf$^*$ &  \faTimes &  \makecell[l]{$^*$Support for Shortest Paths and Minimum ST} \vspaceSQ{-0.15em} \\
\fi
\mbox{[B]} CRONO ~\cite{ahmad2015crono} & \makecell[l]{Focus on futuristic multicores} & \faTimes &  \faTimes &  \faTimes &  \faTimes &  \faTimes &  \faTimes &  \faTimes &   \faTimes &  \faBatteryFull  &  \faBatteryHalf$^*$ &  \faBatteryHalf$^{**}$ &  \makecell[l]{$^*$Only shortest paths. $^{**}$Only triangle counting.} \vspaceSQ{-0.15em} \\
\mbox{[B]} GARDENIA ~\cite{xu2019gardenia} & \makecell[l]{Focus on future accelerators} & \faTimes &  \faTimes &  \faTimes &  \faTimes &  \faTimes &  \faTimes &  \faTimes &   \faTimes &  \faTimes  &  \faBatteryHalf$^*$ &  \faBatteryHalf$^{**}$ &  \makecell[l]{$^*$Only shortest paths. $^{**}$Triangle counting and vertex coloring.} \vspaceSQ{-0.15em} \\
\midrule
\multicolumn{2}{l}{\mbox{[F]} A framework, e.g., Peregrine~\cite{jamshidi2020peregrine} or Fractal~\cite{dias2019fractal} (more at the end of Section~\ref{sec:intro})} & \faBatteryHalf$^*$ & \faBatteryHalf$^*$ & \faBatteryHalf$^*$ & \faBatteryHalf$^*$ & \faBatteryHalf$^*$ & \faTimes & \faTimes & \faTimes & \faTimes & \faTimes & \faTimes & \makecell[l]{$^*$No good performance bounds (focus on \emph{expressiveness}),\\ not competitive to specific parallel mining algorithms} \\
%
%
%
\all{\midrule
\multicolumn{2}{l}{\mbox{[A]} A specific algorithm and problem, e.g., $k$-clique listing~\cite{danisch2018listing}} & \faTimes & \faBatteryFull & \faTimes & \faTimes & \faTimes & \faTimes & \faTimes & \faTimes & \faTimes & \faTimes & \faTimes & \makecell[l]{Each such work usually focuses on one specific problem.} \\
%
%
}
\midrule
\mbox{[B]} \textbf{GMS [This paper]} & \textbf{General graph mining} & \faBatteryFull & \faBatteryFull & \faBatteryFull & \faBatteryFull & \faBatteryFull & \faBatteryFull & \faBatteryFull & \faBatteryFull & \faBatteryFull & \faBatteryFull & \faBatteryFull & \makecell[l]{Details in Table~\ref{tab:problems} and Section~\ref{sec:bench_spec}} \\ 
\bottomrule
\end{tabular}
%
%
\caption{
\textmd{\textbf{\ul{Related work analysis, part~1}: a comparison of GMS
to selected existing graph-related \emph{benchmarks} (``[B]'') and graph mining
\emph{frameworks}
(``[F]'')}, focusing on 
\ul{supported graph mining problems}.
%
%
We exclude benchmarks only partially related to graph
processing, with no focus on mining algorithms
(Lonestar~\cite{burtscher2012quantitative}, Rodinia~\cite{che2009rodinia},
\cnf{HPCS~\cite{bader2005design}, work by Han et al~\cite{han2014experimental},}
Parboil~\cite{stratton2012parboil}, BigDataBench~\cite{wang2014bigdatabench},
BDGS~\cite{ming2013bdgs}, LinkBench~\cite{armstrong2013linkbench},
and SeBS~\cite{copik2020sebs}).
\textbf{mC}: maximal clique listing,
\textbf{kC}: $k$-clique listing,
\textbf{dS}: densest subgraph,
\textbf{sI}: subgraph isomorphism,
\textbf{fS}: frequent subgraph mining,
\textbf{vS}: vertex similarity,
\textbf{lP}: link prediction,
\textbf{cl}: clustering,
\textbf{cD}: community detection,
\textbf{Opt}: optimization,
\textbf{Vr}: vertex rankings,
\faBatteryFull: Supported. \faBatteryHalf: Partial support. 
\faTimes: no support.
}}
\vspace{-2.5em}
%
\label{tab:comparison_problems}
\end{table*}

\begin{figure}[b]
\vspaceSQ{-1.5em}
\centering
\includegraphics[width=1.0\columnwidth]{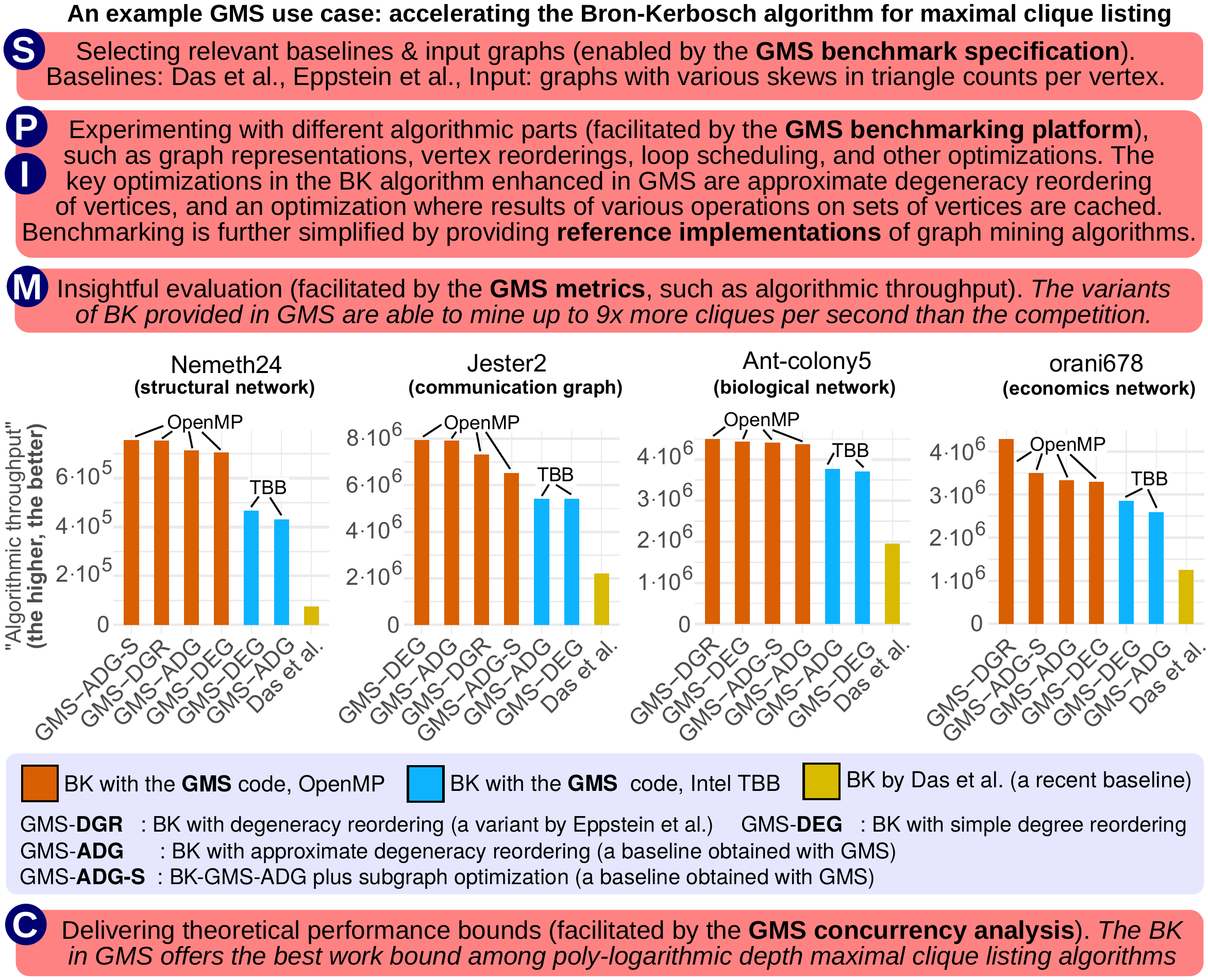}
\vspace{-2.25em}
\caption{\textmd{\textbf{Performance advantages of the parallel Bron-Kerbosch
(BK) algorithm implemented in GMS} over a state-of-the-art implementation by
Das et al.~\protect\cite{das2018shared} and a recent algorithm by Eppstein et
al.~\protect\cite{DBLP:conf/isaac/EppsteinLS10} (GMS-DGR) using a novel
performance metric ``algorithmic throughput'' that shows a number of maximal
cliques found per second. Details of experimental setup:
Section~\ref{sec:eval}.}}
\label{fig:posterchild}
\vspaceSQ{1em}
\end{figure}

\marginpar{\vspace{+10em}\colorbox{yellow}{\textbf{R-2}}}

To address these issues, we introduce \textbf{GraphMineSuite (GMS)}, \emph{a
benchmarking suite} for \emph{{high-performance}} graph mining
\emph{{algorithms}}.
GMS provides an exhaustive benchmark
specification~\includegraphics[scale=0.2,trim=0 16 0 0]{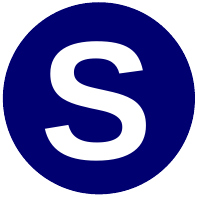}.
Moreover, GMS offers a novel performance
metric~\includegraphics[scale=0.2,trim=0 16 0 0]{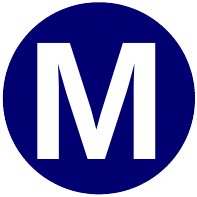} and a broad theoretical
concurrency analysis~\includegraphics[scale=0.2,trim=0 16 0 0]{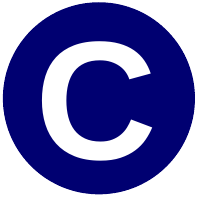} for
deeper performance insights beyond simple empirical run-times.
\hl{To maximize GMS' usability, we arm it with an} accompanying software
platform~\includegraphics[scale=0.2,trim=0 16 0 0]{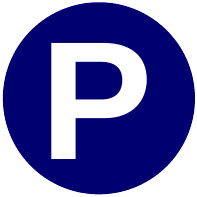} with reference
implementations of algorithms~\includegraphics[scale=0.2,trim=0 16 0 0]{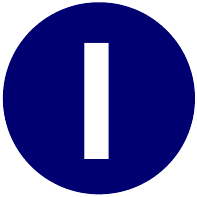}.
\hl{We motivate the GMS platform} in Figure~\ref{fig:posterchild}, which
illustrates example performance advantages (even more than 9$\times$) of the
GMS code over a state-of-the-art variant of the Bron-Kerbosch (BK) algorithm.
\hl{This shows the key benefit of the platform: it facilitates developing,
redesigning, and enhancing algorithms considered in the benchmark, and thus it
enabled us to rapidly obtain large speedups over fast existing BK baselines.}
\tr{GMS aims to propel research into different aspects of {high-performance graph
mining algorithms}, including {design}, {implementation}, {analysis}, and
{evaluation}.} 

\all{As use cases, we harness GMS to enhance existing state-of-the-art algorithms
for core graph mining problems (maximal clique listing, $k$-clique listing,
subgraph isomorphism, degeneracy). Here,
the GMS benchmark specification facilitated selecting the most promising
baselines for each considered problem, while the GMS benchmarking
platform enabled rapidly identifying the most effective optimizations and
appropriate design choices.}

\enlargeSQ

\ifall
\maciej{ADD these 4 new papers from RW: motifs and others}
\fi


To construct GMS, we first identify representative graph mining
\emph{problems}, \emph{algorithms}, and \emph{datasets}. We conduct an
extensive literature review~\cite{chakrabarti2006graph, washio2003state,
lee2010survey, rehman2012graph, gallagher2006matching, ramraj2015frequent,
jiang2013survey, aggarwal2010managing, tang2010graph, leicht2006vertex,
liben2007link, ribeiro2019survey, lu2011link, al2011survey}, and obtain a
\emph{{benchmark specification}}~\includegraphics[scale=0.2,trim=0 16 0
0]{S.pdf} that can be used as a reference point
when selecting relevant comparison targets.

Second, GMS comes with a \emph{{benchmarking platform}}~\includegraphics[scale=0.2,trim=0 16 0 0]{P.pdf}: a highly
modular infrastructure for easy experimenting with different design choices in
a given graph mining algorithm. A key idea for high modularity is exploiting
\emph{{set algebra}}. Here, we observe that data structures and subroutines in
many mining algorithms are ``set-centric'': they can be expressed with sets and
set operations, and the user can seamlessly use different implementations of
the same specific ``set-centric'' part. 
This enables the user to seamlessly use new graph representations, data
layouts, architectural features such as vectorization, and even use numerous
graph compression schemes.
We deliver ready-to-go parallel implementations of the above-mentioned elements,
including parallel \emph{{reference
implementations}}~\includegraphics[scale=0.2,trim=0 16 0 0]{I.pdf} of graph
mining algorithms, as well as representations, data layouts, and compression schemes.
\all{Moreover, the GMS platform facilitates \emph{interpretability} and
\emph{reproducibility}~\cite{hoefler2015scientific}, two methodological
properties relevant for statistically meaningful results.}
\tr{Our code is \emph{public} and can be reused by anyone willing to use it as
a basis for trying new algorithmic ideas, or simply as comparison
baselines.}  
%

For more insightful performance analyses, we propose a novel
\emph{{performance metric}}~\includegraphics[scale=0.2,trim=0 16 0
0]{M.pdf} that assesses ``algorithmic efficiency'', i.e., ``how efficiently a
given algorithm mines selected graph motifs''.

\all{Intuitively, the former tells ``how well a
given machine is utilized by a given algorithm'' while the latter}

To ensure performance insights that are portable across different machines and
independent of various implementation details, GMS also provides
\emph{{the first extensive concurrency
analysis}}~\includegraphics[scale=0.2,trim=0 16 0 0]{C.pdf} of a wide selection
of graph mining algorithms.
We use \emph{work-depth}, an established theoretical framework from parallel
computing~\cite{Bilardi2011, blelloch2010parallel}, to show which algorithms
come with more potential for high performance on today's massively parallel
systems. 
\tr{Our analysis enables developers to \emph{reduce time}
spent on implementation: instead of spending days or weeks to implement an
algorithm that would turn out not scalable, one can use our theoretical
insights and guidelines for deciding against mounting an implementation effort.}

To show the potential of GMS, we
\emph{{enhance state-of-the-art algorithms}} that target some of the
most researched graph mining problems. This includes maximal clique
listing~\cite{das2018shared}, $k$-clique listing~\cite{danisch2018listing},
degeneracy reordering (core decomposition)~\cite{matula1983smallest}, and
subgraph isomorphism~\cite{carletti2018vf3, carletti2019parallel}. By being
able to rapidly experiment with different design choices, we get \emph{speedups
of $>$9$\times$, up to 1.1$\times$, $>$2$\times$, and 2.5$\times$},
respectively. We also \emph{improve theoretical bounds}: for example, for
maximal clique listing, we obtain $O( d m 3^{{(2+\epsilon)d}/{3}} )$ work and
$O( \log^2 n + d\log n )$ depth ($d, m, n$ are the graph degeneracy, \#edges,
and \#vertices, respectively). This is the best work bound among
poly-logarithmic depth maximal clique listing algorithms, improving upon recent
schemes~\cite{eppstein2011listing, DBLP:conf/isaac/EppsteinLS10,
das2018shared}.

\iftr\begin{table*}[t]\fi
\ifconf\begin{table}[b]\fi
\vspaceSQ{-1em}
\ifsq\setlength{\tabcolsep}{0.6pt}\fi
\iftr\setlength{\tabcolsep}{3.0pt}\fi
\ifsq\renewcommand{\arraystretch}{0.5}\fi
\centering
\ssmall
\sf
\iftr
\begin{tabular}{llccccccccccccccccccccccc}
\fi
\ifconf
\begin{tabular}{lcccyyyycccccccccccccccc}
\fi
\toprule
\iftr
\multirow{2}{*}{\makecell[l]{\textbf{Reference /}\\ \textbf{Infrastructure}}} & \multirow{2}{*}{\makecell[c]{\textbf{Summary of focus}\\\textbf{(functionalities)}}} & 
\fi
\ifconf
\multirow{2}{*}{\makecell[l]{\textbf{Reference /}\\ \textbf{Infrastructure}}} & 
\fi
%
\multicolumn{3}{c}{\textbf{New Alg}} & \multicolumn{4}{c}{\hl{\textbf{Gen.~APIs}}} & \multicolumn{5}{c}{\textbf{Metrics}} & \multicolumn{4}{c}{\textbf{Storage}} & \multicolumn{5}{c}{\textbf{Compres.}} & \multicolumn{2}{c}{\textbf{Th.}}
\\
\iftr
\cmidrule(lr){3-5} \cmidrule(lr){6-9} \cmidrule(lr){10-14} \cmidrule(lr){15-18} \cmidrule(lr){19-23} \cmidrule(lr){24-25} 
\fi
\ifconf
\cmidrule(lr){2-4} \cmidrule(lr){5-8} \cmidrule(lr){9-13} \cmidrule(lr){14-17} \cmidrule(lr){18-22} \cmidrule(lr){23-24} 
\fi
\iftr
 & &  
\fi
\ifconf
 &
\fi 
\textbf{$\exists$} & \textbf{na} & \textbf{sp} &
\textbf{N} & \textbf{G} & \textbf{S} & \textbf{P} & 
\textbf{rt} & \textbf{me} & \textbf{fg} & \textbf{mf} & \textbf{af} & 
\textbf{ag} & \textbf{bg} & \textbf{aa} & \textbf{ba} &
\textbf{ad} & \textbf{of} & \textbf{fg} & \textbf{en} & \textbf{re} &
\textbf{$\exists$} & \textbf{nb}
\\  
\midrule
\iftr
\mbox{[B]} Cyclone~\cite{tang2016benchmarking} & Graph databases  
\fi
\ifconf
\mbox{[B]} Cyclone~\cite{tang2016benchmarking} 
\fi
& \faTimes & \faTimes & \faTimes  
& \faTimes & \VfaBatteryHalf & \faTimes & \faTimes 
& \faTimes & \faTimes & \faTimes & \faTimes & \faTimes 
& \faTimes & \faTimes & \faTimes & \faTimes 
& \faTimes & \faTimes & \faTimes & \faTimes & \faTimes 
& \faTimes & \faTimes  
\\
\iftr
\mbox{[B]} \makecell[l]{GBBS~\cite{dhulipala2020graph} + Ligra~\cite{shun2013ligra}} & General graph processing
\fi
\ifconf
\mbox{[B]} \makecell[l]{GBBS~\cite{dhulipala2020graph}\\ + Ligra~\cite{shun2013ligra}} 
\fi
& \faTimes  & \faTimes & \faTimes 
& \VfaBatteryHalf & \VfaBatteryFull & \VfaBatteryFull & \faTimes 
& \VfaBatteryFull & \VfaBatteryHalf & \faTimes & \faTimes & \faTimes 
& \VfaBatteryFull & \VfaBatteryFull & \VfaBatteryFull & \VfaBatteryFull 
& \VfaBatteryFull & \VfaBatteryHalf & \faTimes & \VfaBatteryHalf & \faTimes 
& \VfaBatteryFull & \VfaBatteryFull   
\\
\iftr
\mbox{[B]} GraphBIG~\cite{nai2015graphbig}  & General graph processing  
\fi
\ifconf
\mbox{[B]} GraphBIG~\cite{nai2015graphbig} 
\fi
& \faTimes & \faTimes & \faTimes 
& \VfaBatteryFull & \VfaBatteryHalf & \faTimes & \faTimes 
& \VfaBatteryFull & \VfaBatteryFull & \faTimes & \VfaBatteryFull & \faTimes 
& \VfaBatteryFull & \faTimes & \VfaBatteryFull & \faTimes 
& \faTimes & \faTimes & \faTimes & \faTimes & \faTimes 
& \faTimes & \faTimes   
\\ 
\iftr
\mbox{[B]} GAPBS~\cite{beamer2015gap}  & General graph processing
\fi
\ifconf
\mbox{[B]} GAPBS~\cite{beamer2015gap} 
\fi
& \faTimes & \faTimes & \faTimes 
& \faTimes & \VfaBatteryHalf & \faTimes & \faTimes 
& \VfaBatteryFull & \VfaBatteryHalf & \faTimes & \faTimes & \faTimes 
& \VfaBatteryFull & \faTimes & \faTimes & \faTimes 
& \faTimes & \faTimes & \faTimes & \faTimes & \faTimes 
& \faTimes & \faTimes   
\\ 
\iftr
\mbox{[B]} Graphalytics LDBC~\cite{boncz2013ldbc} & Graph databases
\fi
\ifconf
\mbox{[B]} LDBC~\cite{boncz2013ldbc} 
\fi
& \faTimes & \faTimes & \faTimes 
& \faTimes & \VfaBatteryHalf & \faTimes & \faTimes 
& \VfaBatteryFull$^*$ & \VfaBatteryFull$^*$ & \VfaBatteryHalf$^*$ & \VfaBatteryFull$^*$ & \VfaBatteryHalf$^*$ 
& \VfaBatteryFull & \faTimes & \VfaBatteryFull & \faTimes 
& \faTimes & \faTimes & \faTimes & \faTimes & \faTimes 
& \faTimes & \faTimes  
\\ 
\iftr
\mbox{[B]} WGB~\cite{ammar2013wgb} & General graph processing
\fi
\ifconf
\mbox{[B]} WGB~\cite{ammar2013wgb} 
\fi
& \faTimes & \faTimes & \faTimes 
& \faTimes & \VfaBatteryHalf & \faTimes & \faTimes 
& \VfaBatteryHalf & \VfaBatteryHalf & \faTimes & \faTimes & \faTimes 
& \faTimes & \faTimes & \faTimes & \faTimes 
& \faTimes & \faTimes & \faTimes & \faTimes & \faTimes 
& \faTimes & \faTimes  
\\
\iftr
\mbox{[B]} PBBS~\cite{blelloch2011problem}  & General graph processing
\fi
\ifconf
\mbox{[B]} PBBS~\cite{blelloch2011problem}
\fi
& \faTimes & \faTimes & \faTimes 
& \faTimes & \faTimes & \faTimes & \faTimes 
& \VfaBatteryHalf & \faTimes & \faTimes & \faTimes & \faTimes 
& \VfaBatteryHalf & \faTimes & \faTimes & \faTimes 
& \faTimes & \faTimes & \faTimes & \faTimes & \faTimes 
& \faTimes & \faTimes  
\\
\iftr
\mbox{[B]} Graph500~\cite{murphy2010introducing}  & Graph traversals
\fi
\ifconf
\mbox{[B]} Graph500~\cite{murphy2010introducing} 
\fi
& \VfaBatteryFull & \VfaBatteryHalf & \VfaBatteryFull 
& \faTimes & \VfaBatteryHalf & \faTimes & \faTimes 
& \VfaBatteryFull & \faTimes & \faTimes & \faTimes & \VfaBatteryHalf 
& \VfaBatteryFull & \faTimes & \faTimes & \faTimes 
& \faTimes & \faTimes & \faTimes & \faTimes & \faTimes 
& \faTimes & \faTimes  
\\ 
\iftr
\mbox{[B]} HPCS~\cite{bader2005design}  & General graph processing
& \faTimes & \faTimes & \faTimes 
& \faTimes & \VfaBatteryHalf & \faTimes & \faTimes 
& \VfaBatteryHalf & \VfaBatteryHalf & \faTimes & \faTimes & \faTimes 
& \VfaBatteryFull & \faTimes & \faTimes & \faTimes 
& \faTimes & \faTimes & \faTimes & \faTimes & \faTimes 
& \VfaBatteryHalf & \VfaBatteryHalf  
\\ 
\fi

\iftr
\mbox{[B]} Han et al. ~\cite{han2014experimental}  & Evaluation of graph processing systems 
& \faTimes & \faTimes & \faTimes 
& \VfaBatteryHalf & \faTimes & \faTimes & \faTimes 
& \VfaBatteryFull & \VfaBatteryFull & \faTimes & \VfaBatteryHalf & \faTimes 
& \VfaBatteryHalf & \VfaBatteryHalf & \faTimes & \faTimes 
& \faTimes & \faTimes & \faTimes & \faTimes & \faTimes 
& \faTimes & \faTimes  
\\ 
\fi

\iftr
\mbox{[B]} CRONO ~\cite{ahmad2015crono}  & Multicore systems  
\fi
\ifconf
\mbox{[B]} CRONO ~\cite{ahmad2015crono} 
\fi
& \faTimes  & \faTimes & \faTimes 
& \faTimes & \faTimes & \faTimes & \faTimes 
& \VfaBatteryFull & \VfaBatteryFull & \VfaBatteryHalf & \VfaBatteryFull & \faTimes 
& \faTimes & \faTimes & \faTimes & \faTimes 
& \faTimes & \faTimes & \faTimes & \faTimes & \faTimes 
& \faTimes & \faTimes 
\\ 
\iftr
\mbox{[B]} GARDENIA ~\cite{xu2019gardenia}  & Accelerators 
\fi
\ifconf
\mbox{[B]} GARDENIA ~\cite{xu2019gardenia}
\fi
& \faTimes & \faTimes & \faTimes 
& \faTimes & \VfaBatteryHalf & \faTimes & \faTimes 
& \VfaBatteryFull & \VfaBatteryHalf & \faTimes & \VfaBatteryFull & \faTimes 
& \VfaBatteryHalf & \faTimes & \VfaBatteryHalf & \faTimes 
& \faTimes & \faTimes & \faTimes & \faTimes & \faTimes 
& \faTimes & \faTimes  
\\ 
\iftr
\midrule
%
\iftr
\makecell[l]{\mbox{[F]} Arabesque~\cite{teixeira2015arabesque}}  & Graph pattern matching 
\fi
\ifconf
\mbox{[F]} Arabesque~\cite{teixeira2015arabesque}
\fi
& \VfaBatteryFull & \VfaBatteryHalf & \VfaBatteryHalf 
& \faTimes & \VfaBatteryFull & \faTimes & \faTimes & \VfaBatteryFull
& \VfaBatteryFull & \VfaBatteryFull & \faTimes & \faTimes 
& \VfaBatteryFull & \faTimes & \VfaBatteryFull & \faTimes 
& \VfaBatteryFull & \faTimes & \faTimes & \faTimes & \faTimes 
& \VfaBatteryFull & \faTimes   
\\
\iftr
\makecell[l]{\mbox{[F]} NScale~\cite{quamar2016nscale}}  & Ego-network analysis 
\fi
\ifconf
\mbox{[F]} NScale~\cite{quamar2016nscale}
\fi
& \VfaBatteryFull & \VfaBatteryHalf & \VfaBatteryFull 
& \faTimes & \VfaBatteryFull & \faTimes & \faTimes & \VfaBatteryFull
& \VfaBatteryFull & \VfaBatteryFull & \faTimes & \faTimes 
& \VfaBatteryFull & \faTimes & \faTimes & \VfaBatteryFull 
& \VfaBatteryFull & \faTimes & \faTimes & \faTimes & \faTimes 
& \VfaBatteryFull & \faTimes  
\\
\iftr
\makecell[l]{\mbox{[F]} G-Thinker~\cite{yan2017g}}  & Graph pattern matching 
\fi
\ifconf
\mbox{[F]} G-Thinker~\cite{yan2017g}
\fi
& \VfaBatteryFull & \faTimes & \VfaBatteryFull 
& \faTimes & \VfaBatteryFull & \faTimes & \VfaBatteryHalf & \VfaBatteryFull
& \VfaBatteryHalf & \VfaBatteryHalf & \faTimes & \faTimes 
& \VfaBatteryFull & \faTimes & \faTimes & \faTimes 
& \faTimes & \faTimes & \faTimes & \faTimes & \faTimes 
& \faTimes & \faTimes  
\\
\iftr
\makecell[l]{\mbox{[F]} G-Miner~\cite{chen2018g}}  & Graph pattern matching 
\fi
\ifconf
\mbox{[F]} G-Miner~\cite{chen2018g}
\fi
& \VfaBatteryFull & \VfaBatteryHalf & \VfaBatteryFull 
& \faTimes & \VfaBatteryHalf & \faTimes & \faTimes & \VfaBatteryFull
& \VfaBatteryFull & \VfaBatteryHalf$^*$ & \VfaBatteryHalf$^*$ & \faTimes 
& \VfaBatteryFull & \faTimes & \VfaBatteryFull & \faTimes 
& \faTimes & \faTimes & \faTimes & \faTimes & \faTimes 
& \VfaBatteryFull & \faTimes  
\\
\iftr
\makecell[l]{\mbox{[F]} Nuri~\cite{joshi2018efficient}}  & Graph pattern matching 
\fi
\ifconf
\mbox{[F]} Nuri~\cite{joshi2018efficient}
\fi
& \VfaBatteryFull & \faTimes & \VfaBatteryFull 
& \faTimes & \VfaBatteryFull & \faTimes & \faTimes & \VfaBatteryFull
& \VfaBatteryHalf$^*$ & \VfaBatteryHalf$^*$ & \faTimes & \faTimes 
& \VfaBatteryFull & \faTimes & \VfaBatteryFull & \faTimes 
& \faTimes & \faTimes & \faTimes & \faTimes & \faTimes 
& \faTimes & \faTimes  
\\
\iftr
\makecell[l]{\mbox{[F]} RStream~\cite{wang2018rstream}}  & Graph pattern matching
\fi
\ifconf
\mbox{[F]} RStream~\cite{wang2018rstream}
\fi
& \VfaBatteryFull & \faTimes & \VfaBatteryHalf 
& \faTimes & \VfaBatteryHalf & \faTimes & \faTimes & \VfaBatteryFull
& \VfaBatteryHalf$^*$ & \VfaBatteryHalf$^*$ & \faTimes & \faTimes 
& \VfaBatteryFull & \faTimes & \VfaBatteryFull & \faTimes 
& \faTimes & \faTimes & \faTimes & \faTimes & \faTimes 
& \faTimes & \faTimes  
\\
\iftr
\makecell[l]{\mbox{[F]} ASAP~\cite{iyer2018asap}}  & Graph pattern matching 
\fi
\ifconf
\makecell[l]{\mbox{[F]} ASAP~\cite{iyer2018asap}} 
\fi
& \VfaBatteryFull & \VfaBatteryHalf & \VfaBatteryHalf 
& \faTimes & \VfaBatteryHalf & \faTimes & \faTimes & \VfaBatteryFull
& \VfaBatteryHalf$^*$ & \VfaBatteryHalf$^*$ & \faTimes & \faTimes 
& \VfaBatteryHalf & \faTimes & \faTimes & \faTimes 
& \faTimes & \faTimes & \faTimes & \faTimes & \faTimes 
& \VfaBatteryFull & \faTimes  
\\
\iftr
\makecell[l]{\mbox{[F]} Fractal~\cite{dias2019fractal}}  & Graph pattern matching 
\fi
\ifconf
\mbox{[F]} Fractal~\cite{dias2019fractal}
\fi
& \VfaBatteryFull & \faTimes & \VfaBatteryFull 
& \faTimes & \VfaBatteryFull & \faTimes & \faTimes & \VfaBatteryFull
& \VfaBatteryFull & \VfaBatteryFull$^*$ & \faTimes & \faTimes 
& \VfaBatteryFull & \faTimes & \VfaBatteryFull & \faTimes 
& \faTimes & \faTimes & \faTimes & \faTimes & \faTimes 
& \faTimes & \faTimes  
\\
\iftr
\makecell[l]{\mbox{[F]} Kaleido~\cite{zhao2019kaleido}} & Graph pattern matching 
\fi
\ifconf
\mbox{[F]} Kaleido~\cite{zhao2019kaleido}
\fi
& \VfaBatteryFull  & \VfaBatteryHalf & \VfaBatteryHalf 
& \faTimes & \VfaBatteryHalf & \faTimes & \faTimes & \VfaBatteryFull
& \VfaBatteryFull & \VfaBatteryHalf$^*$ & \VfaBatteryHalf$^*$ & \faTimes 
& \VfaBatteryFull & \faTimes & \faTimes & \VfaBatteryFull 
& \VfaBatteryFull & \faTimes & \faTimes & \faTimes & \faTimes 
& \VfaBatteryFull & \faTimes 
\\
\iftr
\makecell[l]{\mbox{[F]} AutoMine+GraphZero~\cite{mawhirter2019automine,mawhirter2019graphzero}}  & Graph pattern matching 
\fi
\ifconf
\mbox{[F]} \makecell[l]{AutoMine~\cite{mawhirter2019automine} +\\ GraphZero~\cite{mawhirter2019graphzero}} 
\fi
& \VfaBatteryFull & \VfaBatteryHalf & \VfaBatteryHalf 
& \faTimes & \VfaBatteryHalf & \faTimes & \VfaBatteryFull & \VfaBatteryFull
& \VfaBatteryFull & \faTimes & \VfaBatteryHalf & \faTimes 
& \VfaBatteryFull & \faTimes & \faTimes & \faTimes 
& \faTimes & \faTimes & \faTimes & \faTimes & \faTimes 
& \VfaBatteryFull & \faTimes  
\\
\iftr
\makecell[l]{\mbox{[F]} Pangolin~\cite{chen2019pangolin}}  & Graph pattern matching 
\fi
\ifconf
\mbox{[F]} Pangolin~\cite{chen2019pangolin}
\fi
& \VfaBatteryFull & \faTimes & \VfaBatteryHalf 
& \faTimes & \VfaBatteryFull & \faTimes & \faTimes & \VfaBatteryFull
& \VfaBatteryFull & \VfaBatteryHalf$^*$ & \faTimes & \faTimes 
& \VfaBatteryFull & \faTimes & \VfaBatteryFull & \faTimes 
& \VfaBatteryFull & \faTimes & \faTimes & \faTimes & \faTimes 
& \faTimes & \faTimes  
\\
\iftr
\makecell[l]{\mbox{[F]} PrefixFPM~\cite{yan2020prefixfpm}}  & Graph Pattern Mining 
\fi
\ifconf
\mbox{[F]} PrefixFPM~\cite{yan2020prefixfpm}
\fi
& \VfaBatteryHalf & \faTimes & \VfaBatteryHalf 
& \faTimes & \VfaBatteryFull & \faTimes & \faTimes & \VfaBatteryHalf
& \faTimes & \faTimes & \faTimes & \faTimes 
& \VfaBatteryFull & \faTimes & \faTimes & \faTimes 
& \faTimes & \faTimes & \faTimes & \faTimes & \faTimes 
& \faTimes & \faTimes  
\\
\iftr
\makecell[l]{\mbox{[F]} Peregrine~\cite{jamshidi2020peregrine}}  & Graph Pattern Mining 
\fi
\ifconf
\mbox{[F]} Peregrine~\cite{jamshidi2020peregrine}
\fi
& \VfaBatteryFull & \faTimes & \VfaBatteryFull 
& \faTimes & \VfaBatteryFull & \faTimes & \faTimes & \VfaBatteryFull
& \VfaBatteryFull & \VfaBatteryFull & \VfaBatteryHalf & \faTimes 
& \VfaBatteryFull & \faTimes & \VfaBatteryFull & \VfaBatteryFull 
& \faTimes & \faTimes & \VfaBatteryHalf & \VfaBatteryHalf & \faTimes 
& \faTimes & \faTimes  
\\
%
%
%
%
\fi
\midrule
\iftr
\textbf{[B] GMS [This paper]} & \makecell[l]{Graph mining algorithms}  & \VfaBatteryFull & \VfaBatteryFull & \VfaBatteryFull & \VfaBatteryFull & \VfaBatteryFull & \VfaBatteryFull & \VfaBatteryFull & \VfaBatteryFull & \VfaBatteryFull & \VfaBatteryFull & \VfaBatteryFull & \VfaBatteryFull & \VfaBatteryFull & \VfaBatteryFull & \VfaBatteryFull & \VfaBatteryFull & \VfaBatteryFull & \VfaBatteryFull & \VfaBatteryFull & \VfaBatteryFull & \VfaBatteryFull & \VfaBatteryFull & \VfaBatteryFull  \\ 
\fi
\ifconf
\textbf{[B] GMS}  & \VfaBatteryFull & \VfaBatteryFull & \VfaBatteryFull & \VfaBatteryFull & \VfaBatteryFull & \VfaBatteryFull & \VfaBatteryFull & \VfaBatteryFull & \VfaBatteryFull & \VfaBatteryFull & \VfaBatteryFull & \VfaBatteryFull & \VfaBatteryFull & \VfaBatteryFull & \VfaBatteryFull & \VfaBatteryFull & \VfaBatteryFull & \VfaBatteryFull & \VfaBatteryFull & \VfaBatteryFull & \VfaBatteryFull & \VfaBatteryFull & \VfaBatteryFull  \\ 
\fi
\bottomrule
\end{tabular}
%
%
\caption{
\textmd{\textbf{\ul{Related work analysis, part~2}: a comparison of GMS to
graph \emph{benchmarks} (``[B]'') and graph pattern matching \emph{frameworks}
(``[F]'')}, focusing on \ul{supported functionalities important for developing
fast and simple graph mining algorithms}.
%
%
\iftr
We exclude benchmarks only partially related to graph
processing, with no focus on mining, such as
Lonestar~\cite{burtscher2012quantitative}, Rodinia~\cite{che2009rodinia},
Parboil~\cite{stratton2012parboil}, BigDataBench~\cite{wang2014bigdatabench},
BDGS~\cite{ming2013bdgs}, LinkBench~\cite{armstrong2013linkbench},
and SeBS~\cite{copik2020sebs}.
\fi
\textbf{New alg? ($\exists$):} Are there any new/enhanced algorithms offered?
\textbf{na:} do the new algorithms have provable performance properties?
\textbf{sp:} are there any speedups over tuned existing baselines?
\textbf{Modularity:} 
\tr{Is a given infrastructure modular, facilitating adding new
features?} 
The numbers
(\protect\includegraphics[scale=0.12,trim=0 16 0 0]{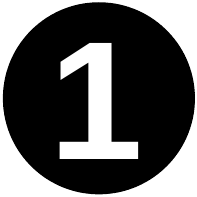} --
\protect\includegraphics[scale=0.12,trim=0 16 0 0]{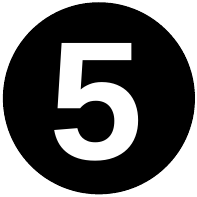},
\protect\includegraphics[scale=0.12,trim=0 16 0 0]{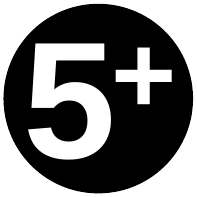}) indicate 
aspects of modularity, details in
Sections~\ref{sec:overview}--\ref{sec:bench_spec}. In general:
\hl{\textbf{Gen.~APIs:} Dedicated generic APIs for a seamless integration of
an arbitrary graph mining algorithm with:
\textbf{N} (an arbitrary vertex neighborhood),
\textbf{G} (an arbitrary graph representation),
\textbf{S} (arbitrary processing stages, such as preprocessing routines),
\textbf{P} (PAPI infrastructure).}
\textbf{Metrics:} Supported performance metrics.
\textbf{rt:} (plain) run-times.
\textbf{me:} (plain) memory consumption.
\textbf{fg:} support for fine-grained analysis (e.g., providing run-time fraction due to preprocessing).
\textbf{mf:} metrics for machine efficiency (details in~\cref{sec:metrics}).
\textbf{af:} metrics for algorithmic efficiency (details in~\cref{sec:metrics}).
\textbf{Storage:} Supported graph representations and auxiliary data structures.
\textbf{ag:} graph representations based on (sparse) integer arrays (e.g., CSR).
\textbf{bg:} graph representations based on (sparse or dense) bitvectors~\cite{han2018speeding, aberger2017emptyheaded}.
\textbf{aa:} auxiliary structures based on (sparse) integer arrays.
\textbf{ba:} auxiliary structures based on (sparse or dense) bitvectors.
\iftr\textbf{Compression:} Supported forms of compression or space-efficient data structures.\fi
\ifconf\textbf{Compression:} Supported forms of compression.\fi
\textbf{ad:} compression of adjacency data.
\textbf{of:} compression of offsets into the adjacency data.
\textbf{fg:} compression of fine-grained elements (e.g., single vertex IDs).
\textbf{en:} various forms of the encoding of the adjacency data (e.g.,
Varint~\cite{besta2018log}).
\textbf{re:} support for relabeling adjacency data (e.g., degree
minimizing~\cite{besta2018log}).
\textbf{Th.:} Theoretical analysis.
\textbf{$\exists$:} Any theoretical analysis is provided.
\iftr\textbf{Nb:} Whether any new bounds (or other new theoretical results) are derived.\fi
\ifconf\textbf{Nb:} Are there any new bounds? \fi
\faBatteryFull: Support. \faBatteryHalf: Partial support. 
\faBatteryHalf$^*$ / \faBatteryFull$^*$: A given metric is supported via an external profiler.
\faTimes: No support.
}}
\vspaceSQ{-2em}
\label{tab:comparison_functionalities}
\iftr\end{table*}\fi
\ifconf\end{table}\fi

\ifall
enables the clear design o GMS some simple code-wise optimizations in our
reimplementation of $k$-Clique Listing. When tested against the reference
implementation by Danisch et al.  \cite{danisch2018listing} they provide a
runtime gain of over 14\% when counting all 8-cliques of the orkut graph. When
utilizing the modularity of GMS and we change the preprocessing of the graph,
the runtime gain rises to 25\%. \yannick{There is no plot for this, but raw
data is available.}
\fi

\iftr


\vspace{0.5em}
\noindent
To summarize, we provide the specific contributions:
\vspaceSQ{-0.5em}

\begin{itemize}[noitemsep, leftmargin=0.5em]
\all{\item We propose GMS, the first \emph{benchmark specification} for general graph
mining. GMS crystallizes the results of analyzing more than 300 associated
research papers, and it aims to \emph{facilitate systematic and interpretable
design, implementation, analysis, and evaluation of high-performance graph
mining}.}
{\item We propose GMS, the first benchmark for graph mining, with
a specification based on more than 300 associated
research papers.}
{\item We deliver a GMS \emph{benchmarking
platform} that facilitates developing and tuning 
high-performance graph mining algorithms, with reference implementation of more
than 40 algorithms, and high modularity obtained with set algebra,
enabling experimenting with different fine- and coarse-grained algorithmic
elements.}
\all{\item We deliver a highly modular GMS design platform with
high-performance reference implementation of considered graph mining
algorithms. Our code is publicly available, it targets today's massively
parallel architectures, and it can be used as comparison baselines or as a
basis for anyone in the community to experiment with their own algorithmic
ideas.}
\item We propose a novel \emph{performance metric} for assessing the algorithmic 
throughput of graph mining algorithms.
\item We support GMS with the first extensive \emph{concurrency analysis} of
graph mining for performance insights that are portable and independent
of various implementation details.
\all{which can be used to quickly decide for or against implementing a
given algorithm.}
\item As an example of using GMS, we \emph{enhance 
state-of-the-art baselines} for core graph mining problems (degeneracy, maximal
clique listing, $k$-clique listing, and subgraph isomorphism), obtaining
respective speedups of $>$9$\times$, up to 10\%, $>$2$\times$, and
2.5$\times$. We also \emph{enhance their theoretical bounds}.
\all{Our results, enabled or much facilitated by the GMS infrastructure,
\emph{outperform state-of-the-art implementations}, e.g., ensuring speedups of
up to 25\% and $>$2.5$\times$ over a recent $k$-clique listing algorithm by
Danish et al.~\cite{danisch2018listing} and VF3-Light, a subgraph isomorphism
very recent baseline~\cite{carletti2018vf3}.}
\end{itemize}

\fi

\marginpar{\vspace{+4em}\colorbox{yellow}{\textbf{R-2}}}

\iftr
\subsection{GMS vs.~Graph-Related Benchmarks}
\label{sec:GMS_vs_B}
\else
\textbf{\ul{GMS vs.~Graph-Related Benchmarks}}
\fi
\ifall
We motivate GMS as {\textbf{the first benchmark for graph
mining problems}}.
There exist several graph processing benchmarks associated with different
areas. For example, the graph data management community recently introduced
(LDBC)~\cite{boncz2013ldbc, szarnyas2018early, erling2015ldbc, iosup2016ldbc},
a benchmark that (1) identifies and specifies important graph database
workloads, (2) presents relevant inputs, and (3) describes a methodology for
running experiments.
Moreover, the parallel programming community use Graph Algorithm Platform
Benchmark Suite (GAPBS)~\cite{beamer2015gap}, a benchmark specification with a
reference implementation of algorithms solving six graph problems that were
extensively studied in that area (BFS, PageRank, Betweenness Centrality,
Triangle Counting, Single Source Shortest Paths, and Connected Components).
There also exists Graph500~\cite{murphy2010introducing}, a benchmark and a
reference implementation for extreme-scale graph traversals on supercomputers.
Finally, different domains outside graph computations also proposed their
specific benchmarks, for example Deep500~\cite{ben2019modular} for large-scale
deep learning, MLPerf~\cite{reddi2019mlperf} for general machine learning, or
Top500~\cite{dongarra1997top500} for dense linear algebra computations.
GMS is an equivalent benchmark for graph mining.
\fi
We motivate GMS as {\textbf{the first benchmark for graph mining}}.
\hl{There exist graph processing benchmarks, but they do not 
focus on graph mining; we illustrate this in
Table~\mbox{\ref{tab:comparison_problems}} (``[B]'').}
They focus on graph \emph{database workloads}\tr{ (LDBC~\cite{boncz2013ldbc},
Cyclone~\cite{tang2016benchmarking}, LinkBench~\cite{armstrong2013linkbench})},
extreme-scale graph \emph{traversals}\tr{ (Graph500 and
GreenGraph500~\cite{murphy2010introducing})}, and different \emph{``low-complexity''}
(i.e., with run-times being low-degree polynomials in numbers of vertices or
edges) parallel graph algorithms such as PageRank, triangle counting, and
others, researched intensely in the parallel programming community (GAPBS~\cite{beamer2015gap}, GBBS \& Ligra~\cite{dhulipala2018theoretically},
WGB~\cite{ammar2013wgb}, PBBS~\cite{blelloch2011problem},
HPCS~\cite{bader2005design}, GraphBIG~\cite{nai2015graphbig},
Lonestar~\cite{burtscher2012quantitative}, Rodinia~\cite{che2009rodinia},
Parboil~\cite{stratton2012parboil}, BigDataBench~\cite{wang2014bigdatabench},
BDGS~\cite{ming2013bdgs}).
Despite some similarities (e.g., GBBS provides implementations of $k$-clique
listing), none of these benchmarks targets
general graph mining, and they do not offer novel performance metrics or
detailed control over graph representations, data layouts, and others.
We broadly analyze this in Table~\ref{tab:comparison_functionalities}, where we
compare GMS to other benchmarks in terms of the modularity of their software
infrastructures, offered metrics, control over storage schemes, support for
graph compression, provided theoretical analyses, and whether they improve
state-of-the-art algorithms.
Finally, GMS is the only benchmark that is used to directly enhance core
state-of-the-art graph mining algorithms, achieving both better bounds and
speedups in empirical evaluation.

\tr{Unlike other benchmarks, GMS proposes to
exploit \emph{set algebra} as a driving enabler for \emph{modularity},
\emph{simplicity}, but also \emph{high-performance}. This design decision
comes from our key observation that established formulations of many relevant
graph mining problems and algorithms heavily rely on set algebra.}

\all{\emph{Overall, we aim for GMS to be a \textbf{sweetspot} between graph processing
frameworks (that come with simplicity but are usually very limited in what
problems they solve and what features they enable~\cite{khan2016vertex}) and
plain specialized codes (that cover any scheme but are complex to use and
extend).}}


\marginpar{\vspace{30em}\colorbox{yellow}{\textbf{R-2}}}

\iftr
\subsection{GMS vs.~Pattern Matching {Frameworks}}
\label{sec:GMS_vs_F}
\else
\textbf{\ul{GMS vs.~Pattern Matching Frameworks}}
\fi
Many graph mining \emph{frameworks} have recently been proposed, for example
Peregrine~\cite{jamshidi2020peregrine}
and others~\cite{dias2019fractal, teixeira2015arabesque,
yan2020prefixfpm, mawhirter2019graphzero, chen2019pangolin, mawhirter2019automine, iyer2018asap,
zhao2019kaleido, joshi2018efficient, chen2018g, yan2017g}.
GMS does \emph{not} compete with such {frameworks}.
First, as Table~\ref{tab:comparison_problems} shows, such frameworks do not 
target broad graph mining.
Second, key offered functionalities also differ\tr{, see
Table~\ref{tab:comparison_functionalities}}. These frameworks
focus on \emph{programming models} and \emph{abstractions}, and on the
underlying \emph{runtime systems}\tr{\footnote{We do not include these aspects in
Table~\ref{tab:comparison_functionalities} due to space constraints -- these
aspects are \emph{not} in the focus of GMS and any associated columns would
have ``\faTimes'' for GMS}}.
Contrarily, GMS focuses on benchmarking and tuning \emph{specific parallel
algorithms}, with provable performance properties, to accelerate the most
competitive existing baselines.

\all{they prioritize expressiveness, focusing on \emph{programming models} or
\emph{abstractions} for \emph{pattern matching problems} \emph{only}, and
\emph{\ul{not}} on developing tuned \emph{high-performance and provably fast
parallel} \emph{algorithms} for specific mining problems.}

\all{As such, the GMS platform gives the user detailed control (e.g., with set
algebra based modularity) over high- and low-level aspects of implementation
and evaluation, for example graph representations and data structures,
algorithm optimizations, data layouts, compression schemes, or fine-grained
performance metrics.}

\all{Contrarily, mining frameworks do \emph{\ul{not}} enable (or facilitate)
\emph{experimenting} with different aspects of a graph problem to be solved,
Moreover, they usually do not provide \emph{infrastructure for insightful
benchmarking}, except for simple runtime measurements. 
Finally, while simplifying algorithm development, the framework abstractions
often introduce \emph{performance overheads}, may limit expressiveness when
developing different graph algorithms~\cite{kalavri2017high,
salihoglu2014optimizing, khan2016vertex, yan2014pregel}, and force their user
to depend on the provided runtime system, which may be inefficient.}

\ifall
(e.g., ``find all stars in a given graph'')
\fi

\ifall
Frameworks also focus almost solely on graph pattern matching problems (e.g.,
``find all stars in a given graph''). For simplicity, they are based
on paradigms such as the subgraph-centric paradigm~\cite{simmhan2014goffish}.
While simplifying algorithm development, such paradigms usually introduce
performance overheads and come with limited expressiveness when developing
different graph algorithms. Moreover, these framework usually do not allow for
effective utilization of parallel hardware and force their user to depend on
the provided runtime system, which may be inefficient~\cite{khan2016vertex}.
\fi


\section{NOTATION AND BASIC CONCEPTS}
\label{sec:back}
\vspaceSQ{-0.2em}

\tr{We first present the most basic used concepts.  However, GMS touches many
different areas, and -- for clarity -- \emph{we will present any other
background information later, when required}.}
\tr{Table~\ref{tab:symbols} lists the most important symbols used in this
work.}

\iftr

\begin{table}[h]
\centering
\footnotesize
\sf
\begin{tabular}{@{}ll@{}}
\toprule
$G = (V,E)$ & An graph $G$; $V, E$ are sets of vertices and edges.\\
$n, m$ & Numbers of vertices and edges in $G$; $|V| = n, |E| = m$.\\
$\Delta(v), N(v)$ & The degree and neighbors of $v \in V$.\\
$\Delta, \overline{d}$ & The maximum and the average degree in $G$ ($\overline{d} = m/n$).\\
\bottomrule
\end{tabular}
\caption{The most important symbols used in the paper.}
\label{tab:symbols}
\end{table}

\fi

\tr{\subsection{Graph Model}
\label{sec:models_int}}

We model an undirected graph $G$ as a tuple $(V,E)$; $V$ is a set of vertices
and $E \subseteq V \times V$ is a set of edges; $|V|=n$ and
$|E|=m$.  
%
%
The maximum degree of a graph is $\Delta$. The neighbors and the degree of a
given vertex~$v$ are denoted with $N(v)$ and $\Delta(v)$, respectively. \tr{The
vertices are identified by integer IDs: $V=\{1,\ldots,n\}$. }
\all{These IDs define a
total order $\succ$ on the vertices that is used to sort the neighborhoods.}
%

\tr{\subsection{Set Algebra Concepts}}

\tr{GMS uses basic \emph{set operations}: $A \cap B$, $A
\cup B$, $A \setminus B$, $|A|$, and $\in A$.
Any set operation can be implemented with various \emph{set algorithms}.
Sets usually contain vertices and at times edges. A set can be \emph{represented} 
differently, for example with a bitvector or an integer array.}

\tr{\subsection{Graph Representation}}

By default, we use a standard sorted \emph{Compressed Sparse Row
(CSR)} \emph{graph representation}. For an unweighted
graph, CSR consists of a contiguous array with IDs of neighbors of
each vertex ($2m$ words) and offsets to the neighbor data of each vertex ($n$
words). 
\tr{We also use more complex representations such as compressed bitvectors.}

\begin{figure*}[t]
\vspaceSQ{-1.5em}
\centering
\includegraphics[width=1.0\textwidth]{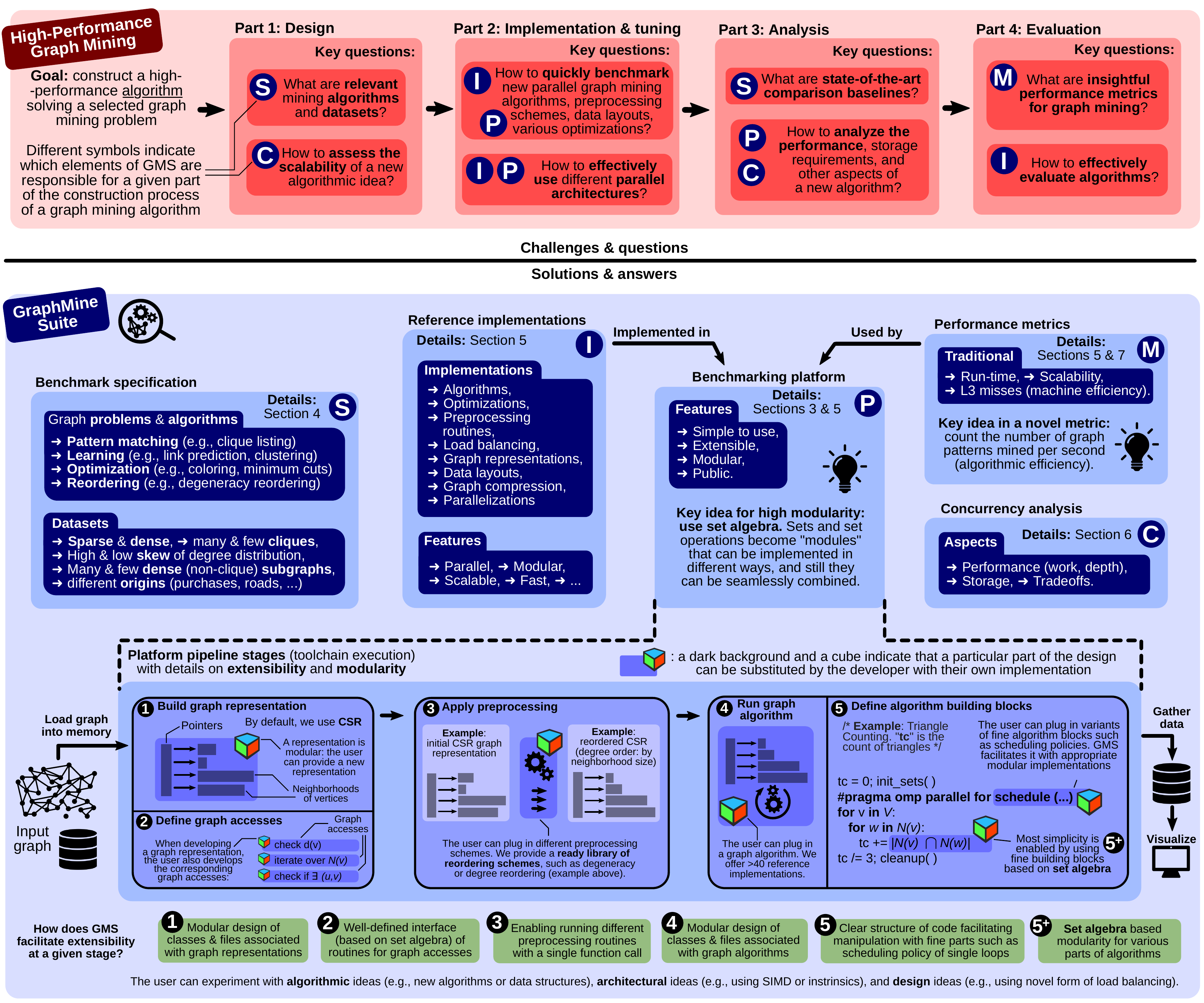}
\vspace{-2.25em}
\caption{\textmd{The overview of GMS and how it facilitates constructing,
tuning, and benchmarking graph mining algorithms. The {upper
red part} shows a process of constructing a graph mining algorithm,
and the associated research questions. The {middle blue part} shows the
corresponding different \textbf{elements} of the GMS suite
(\protect\includegraphics[scale=0.12,trim=0 16 0 0]{S.pdf} --
\protect\includegraphics[scale=0.12,trim=0 16 0 0]{M.pdf}). The {bottom blue
part} illustrates the details of the GMS design benchmarking, with the
\textbf{stages} of the GMS pipeline (execution toolchain) for running a given
graph mining algorithm (\protect\includegraphics[scale=0.12,trim=0 16 0
0]{1.pdf} -- \protect\includegraphics[scale=0.12,trim=0 16 0 0]{5.pdf},
\protect\includegraphics[scale=0.12,trim=0 16 0 0]{5p.pdf}).}}
\label{fig:design}
\vspaceSQ{-0.95em}
\end{figure*}

\ifall
\maciej{fix table}

\begin{table*}[bp]
\centering
\renewcommand{\arraystretch}{0.8}
\sf
\ssmall
\begin{tabular}{llll@{}}
\toprule
\textbf{Class of routines} & \textbf{Example routines} & \textbf{They enable extending GMS with...} & \textbf{Additional remarks} \\
\midrule
\includegraphics[scale=0.1,trim=0 16 0 0]{1.pdf} Pipeline management & load input graph, run graph algorithm, specify algorithm parameters & ...graph algorithms and representations \\ 
\includegraphics[scale=0.1,trim=0 16 0 0]{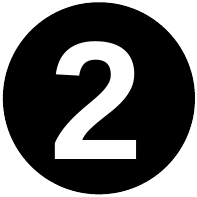} Preprocessing & apply degeneracy reordering, apply degree reordering,   & ...preprocessing routines & In the current version, we focus on vertex reordering. \\
\includegraphics[scale=0.1,trim=0 16 0 0]{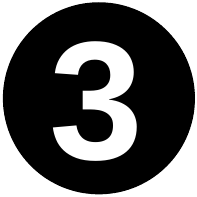} Graph accesses & Scan $N(v)$, check $d(v)$ & Accessing the graph structure &  \\
\includegraphics[scale=0.1,trim=0 16 0 0]{3.pdf} Set operations & $X \cap Y, X \cup Y, X \setminus Y, |X|, x \in X$  & Abstracting details of a graph algorithm \\
\bottomrule
\end{tabular}
%
%
\caption{GMS's interface: selected GMS routines.}
%
%
\label{tab:interface}
\vspaceSQ{-1.5em}
\end{table*}

\fi

\section{OVERVIEW OF GMS}
\label{sec:overview}
\vspaceSQ{-0.2em}

\all{GMS is a general publicly available design and benchmark suite for graph miming
that delivers a benchmark specification, fast reference implementation of
numerous algorithms, extensible and programmable design platform, concurrency
analysis, and metrics.} 
We start with an overview; see Figure~\ref{fig:design}.
%
%

The GMS \textbf{benchmark specification}~\includegraphics[scale=0.2,trim=0 16 0
0]{S.pdf} \hl{(details in Section~\mbox{\ref{sec:bench_spec}})} motivates representative graph mining problems and
state-of-the-art algorithms solving these problems, relevant datasets, 
performance
metrics~\includegraphics[scale=0.2,trim=0 16 0 0]{M.pdf},
and a
taxonomy that structures this information. 
%
%
The specification, in its entirety or in a selected subpart, enables choosing
relevant comparison baselines and important datasets that stress different
classes of algorithms. 
%

The specification is implemented in the \textbf{benchmarking
platform}~\includegraphics[scale=0.2,trim=0 16 0 0]{P.pdf} \hl{(details in Section~\mbox{\ref{sec:platform-design}})}. The platform facilitates
developing and evaluating high-performance graph mining algorithms.
\hl{The former is enabled by incorporating set algebra as the key driver for
modularity \emph{and} high performance.}
For the latter,
the platform forms a processing pipeline with
well-separated parts (see the bottom  of
Figure~\ref{fig:design}): loading the graph from I/O, constructing a
graph representation (\includegraphics[scale=0.2,trim=0 16 0 0]{1.pdf} --
\includegraphics[scale=0.2,trim=0 16 0 0]{2.pdf}), optional preprocessing
(\includegraphics[scale=0.2,trim=0 16 0 0]{3.pdf}) running selected graph
algorithms (\includegraphics[scale=0.2,trim=0 16 0 0]{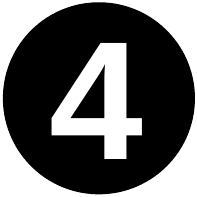} --
\includegraphics[scale=0.2,trim=0 16 0 0]{5.pdf},
\includegraphics[scale=0.2,trim=0 16 0 0]{5p.pdf}), and gathering data.
%

The \textbf{reference implementation of
algorithms}~\includegraphics[scale=0.2,trim=0 16 0 0]{I.pdf} \hl{(details in Section~\mbox{\ref{sec:hpc_algs}})} offers publicly
available, fast, and scalable baselines that effectively use massive
parallelism in today's architectures. 
\tr{We implement algorithms to make their design \emph{modular},
i.e., different building blocks of a given algorithm, such as a preprocessing
optimization, can be replaced with user-specified codes. }
\all{This, on one hand, enables users to not only use GMS as comparison
baselines, but also to \emph{easily} extend our implementation with their own
ideas, \emph{propelling research into high-performance graph mining}.}
\all{GMS enables experimenting with different algorithms and their variants,
optimizations, preprocessing routines, load balancing schemes, parallelization
strategies, data layouts and models, compression methods, and many others.}
 As data movement is dominating runtimes in irregular graph
computations, we also provide a large number of \emph{storage schemes}: graph
representations, data layout schemes, and graph compression.
\tr{We describe selected implementations, focusing on how they achieve
high performance and modularity, in
Section~\mbox{\ref{sec:hpc_algs}}.}


The \textbf{concurrency analysis}~\includegraphics[scale=0.2,trim=0 16 0
0]{C.pdf} \hl{(details in Section~\mbox{\ref{sec:concurrent}})} offers a theoretical framework to analyze performance, storage,
and the associated tradeoffs. We use work and
depth~\cite{Bilardi2011, blelloch2010parallel} that respectively describe the
total work done by all executing processors, and the length
of the associated longest execution path.
%

\all{Finally, GMS provides an \textbf{evaluation
methodology}~\includegraphics[scale=0.2,trim=0 16 0 0]{M.pdf} for
interpretability and reproducibility. The former \emph{``enables scientists to
understand the experiment, draw own conclusions, assess their certainty, and
possibly generalize results''}~\cite{hoefler2015scientific}.}


\ifall

\subsection{\mbox{GMS Platform: Modularity \& Extensibility}}
\label{sec:extensibility}

\fi

\ifall \maciej{fix, remove?}
enables extending GMS with whole graph algorithms,
level~\includegraphics[scale=0.2,trim=0 16 0 0]{2.pdf} facilitates providing
preprocessing routines, and level~\includegraphics[scale=0.2,trim=0 16 0
0]{3.pdf} allows for experimenting with small building blocks such as the
layout of some data structure, or the implementation of a specific function
used in a given graph algorithm. 

\includegraphics[scale=0.2,trim=0 16 0 0]{1.pdf}~\macb{Modularity of Pipeline Stages. } 
First, we clearly separate parts of the whole graph processing pipeline so that
\emph{the user can straightforwardly plug in their \textbf{new graph algorithm}
or a \textbf{new graph representation}}. For example, the user implements a new
algorithm in a separate file, using a rich GMS interface of functions for 
accessing the graph structure (see Table~\ref{tab:interface}). 
One can similarly add a new graph representation.

\includegraphics[scale=0.2,trim=0 16 0 0]{2.pdf}~\macb{Modularity of
Preprocessing Routines}
We also enable the user to \emph{\textbf{seamlessly add preprocessing
routines}}. We also provide a module with ready-to-go schemes such as
degeneracy reordering~\cite{DBLP:conf/latin/Farach-ColtonT14} (a key to find
$k$-cores). Their complexities are hidden beyond a well-defined interface, see
Table~\ref{tab:interface}.

\includegraphics[scale=0.2,trim=0 16 0 0]{3.pdf}~\macb{Modularity of Algorithm Building Blocks}
Finally, GMS facilitates \emph{\textbf{experimenting with different parts of an
algorithm}}, for example a data structure layout or an implementation of a
particular subroutine. For this, we incorporate the fact that many graph
algorithms, for example Bron-Kerbosch~\cite{bron1973algorithm}, \emph{are
formulated with set algebra and use a small number of well-defined
operations such as set intersection~$\cap$}. In GMS, we enable the user to
provide their own implementation of such operations and of sets themselves.
Table~\ref{tab:interface} illustrates selected set operations that are
currently offered by GMS.

In Section~\ref{sec:hpc_algs}, we will illustrate using
\includegraphics[scale=0.2,trim=0 16 0 0]{1.pdf} --
\includegraphics[scale=0.2,trim=0 16 0 0]{5p.pdf} to easily enhance
state-of-the-art graph mining algorithms. 

\fi

\ifall
In Section~\ref{sec:hpc_algs}, we describe details of using these routines with
several state-of-the-art graph mining algorithms, hat enabled us to quickly
experiment with.
\fi

\iftr
In the next sections, we detail the respective parts of GMS.  We will also
describe in more detail example use cases, in which we show how using GMS
ensures speedups over state-of-the-art baselines for $k$-clique
listing~\cite{danisch2018listing} and maximal clique
listing~\cite{DBLP:conf/isaac/EppsteinLS10}.
\fi

\begin{table*}[t]
\vspaceSQ{-1.25em}
\centering
\setlength{\tabcolsep}{4pt}
\ifsq\renewcommand{\arraystretch}{0.7}\fi
\sf
\ssmall
\begin{tabular}{llllll@{}}
\toprule
 & \textbf{Graph problem} & \textbf{Corresponding algorithms} & \textbf{E.?} & \textbf{P.?} & \textbf{Why included, what represents? (selected remarks)} \\
\midrule
\multirow{9}{*}{\makecell[c]{Graph\\Pattern\\Matching}}
 & $\bullet$ Maximal Clique Listing~\cite{diestel2018graph} & \makecell[l]{Bron-Kerbosch~\cite{bron1973algorithm} + optimizations (e.g., pivoting)~\cite{cazals2008note, DBLP:conf/isaac/EppsteinLS10, DBLP:journals/tcs/TomitaTT06}} & \faThumbsOUp\ \includegraphics[scale=0.13,trim=0 7 0 0]{5p.pdf} & \faThumbsDown   & Widely used, NP-complete, example of backtracking \\
\arrayrulecolor{lightgray}\cmidrule{3-6}\arrayrulecolor{black}
 & $\bullet$ $k$-Clique Listing~\cite{danisch2018listing} & \makecell[l]{Edge-Parallel and Vertex-Parallel general algorithms~\cite{danisch2018listing},\\ different variants of Triangle Counting~\cite{shun2015multicore, schank2007algorithmic}} & \faThumbsOUp\ \includegraphics[scale=0.13,trim=0 7 0 0]{5p.pdf} & \faThumbsDown  & P (high-degree polynomial), example of backtracking \\
\arrayrulecolor{lightgray}\cmidrule{3-6}\arrayrulecolor{black}
 & $\bullet$ Dense Subgraph Discovery~\cite{aggarwal2010managing} & Listing $k$-clique-stars~\cite{jabbour2018pushing} and $k$-cores~\cite{DBLP:conf/latin/Farach-ColtonT14} (exact \& approximate) &  \faThumbsOUp\ \includegraphics[scale=0.13,trim=0 7 0 0]{5p.pdf} & \faThumbsDown  & Different relaxations of clique mining \\
%
%
%
 & $\bullet$ Subgraph isomorphism~\cite{diestel2018graph} & VF2~\cite{cordella2004sub}, TurboISO~\cite{han2013turbo}, Glasgow~\cite{mccreesh2015parallel}, VF3~\cite{carletti2017introducing, carletti2019parallel}, VF3-Light~\cite{carletti2018vf3} & \faThumbsUp & \faThumbsDown  &  Induced vs.~non-induced, and backtracking vs.~indexing schemes  \\
%
 %
 & $\bullet$ Frequent Subgraph Mining~\cite{aggarwal2010managing} &  BFS and DFS exploration strategies, different isomorphism kernels & \faThumbsUp & \faThumbsDown  & Useful when one is interested in many different motifs \\
\midrule
\multirow{9}{*}{\makecell[c]{Graph\\Learning}}
 & $\bullet$ Vertex similarity~\cite{leicht2006vertex} & \makecell[l]{Jaccard, Overlap, Adamic Adar, Resource Allocation,\\ Common Neighbors, Preferential Attachment, Total Neighbors~\cite{robinson2013graph}} & \faThumbsOUp\ \includegraphics[scale=0.13,trim=0 7 0 0]{5p.pdf} & \faThumbsDown  &  \makecell[l]{A building block of many more comples schemes,\\ different methods have different performance properties} \\ 
\arrayrulecolor{lightgray}\cmidrule{3-6}\arrayrulecolor{black}
 & $\bullet$ Link Prediction~\cite{taskar2004link} & \makecell[l]{Variants based on vertex similarity (see above)~\cite{liben2007link, lu2011link, al2006link, taskar2004link},\\ a scheme for assessing link prediction accuracy~\cite{wang2014robustness}} & \faThumbsOUp\ \includegraphics[scale=0.13,trim=0 7 0 0]{5p.pdf} & \faThumbsDown  &  A very common problem in social network analysis \\ 
\arrayrulecolor{lightgray}\cmidrule{3-6}\arrayrulecolor{black}
 & $\bullet$ Clustering~\cite{schaeffer2007graph} & \makecell[l]{Jarvis-Patrick clustering~\cite{jarvis1973clustering} based on different\\ vertex similarity measures (see above)~\cite{liben2007link, lu2011link, al2006link, taskar2004link}} & \faThumbsOUp\ \includegraphics[scale=0.13,trim=0 7 0 0]{5p.pdf} & \faThumbsDown  &  \makecell[l]{A very common problem in general data mining; the selected\\scheme is an example of overlapping and single-level clustering} \\
\arrayrulecolor{lightgray}\cmidrule{3-6}\arrayrulecolor{black}
 & $\bullet$ Community detection & Label Propagation and Louvain Method~\cite{staudt2015engineering} & \faThumbsUp & \faThumbsDown  &  Examples of convergence-based on non-overlapping clustering\\
\midrule
\iftr
\multirow{4}{*}{\makecell[c]{Opti-\\mization\\problems}}
 & $\bullet$ Minimum Graph Coloring~\cite{papadimitriou1998combinatorial} & \makecell[l]{Jones and Plassmann's (JP)~\cite{jones1993parallel}, Hasenplaugh et al.'s (HS)~\cite{hasenplaugh2014ordering},\\ Johansson's (J)~\cite{johansson1999simple}, Barenboim's (B)~\cite{barenboim2016locality}, Elkin et al.'s (E)~\cite{elkin20142delta},\\ sparse-dense decomposition (SD)~\cite{harris2016distributed}} & \faThumbsUp & \faThumbsDown &  \makecell[l]{NP-complete; uses vertex prioritization (JP, HS),\\ random palettes (J, B), and adapted distributed schemes (E, SD)} \\
\arrayrulecolor{lightgray}\cmidrule{3-6}\arrayrulecolor{black}
 & $\bullet$ Minimum Spanning Tree~\cite{cormen2009introduction} & \makecell[l]{Boruvka~\cite{boruuvka1926jistem}} & \faThumbsUp & \faThumbsDown  & P (low complexity problem) \\
 & $\bullet$ Minimum Cut~\cite{cormen2009introduction} & A recent augmentation of Karger--Stein Algorithm~\cite{karger1996new} & \faThumbsUp & \faThumbsDown  & P (superlinear problem) \\
%
%
\midrule
\fi
\multirow{3}{*}{\makecell[c]{Vertex\\Ordering}}
 & $\bullet$ Degree reordering & A straightforward integer parallel sort & \faThumbsUp & \faThumbsOUp  & A simple scheme that was shown to bring speedups \\
 & $\bullet$ Triangle count ranking & Computing triangle counts per vertex & \faThumbsOUp\ \includegraphics[scale=0.13,trim=0 7 0 0]{5p.pdf} & \faThumbsOUp  & Ranking vertices based on their clustering coefficient \\
 & $\bullet$ Degenerecy reordering & Exact and approximate~\cite{DBLP:conf/latin/Farach-ColtonT14}~\cite{khaouid2015k} & \faThumbsOUp\ \includegraphics[scale=0.13,trim=0 7 0 0]{5p.pdf} & \faThumbsOUp  & Often used to accelerate Bron-Kerbosch and others \\
%
\bottomrule
\end{tabular}
%
%
\caption{\scriptsize\textmd{\textbf{Graph problems and algorithms considered in
GMS}.  ``\textbf{E.? (Extensibility)''} indicates how extensible given
implementations are in the GMS benchmarking platform: ``\faThumbsOUp'' indicates full
extensibility, including the possibility to provide new building blocks based
on set algebra (\protect\includegraphics[scale=0.12,trim=0 16 0 0]{1.pdf} --
\protect\includegraphics[scale=0.12,trim=0 16 0 0]{5.pdf},
\protect\includegraphics[scale=0.12,trim=0 16 0 0]{5p.pdf}).
``\faThumbsUp'': an
algorithm that does not straightforwardly (or extensively) use set algebra\tr{, offering
modularity levels \protect\includegraphics[scale=0.12,trim=0 16 0 0]{1.pdf} --
\protect\includegraphics[scale=0.12,trim=0 16 0 0]{5.pdf}''}}.
``\textbf{P.? (Preprocessing)} indicates whether a given algorithm can be
seamlessly used as a preprocessing routine; in the current GMS version, this
feature is reserved for the vertex reordering algorithms.
}
%
%
\label{tab:problems}
\vspace{-1.5em}
\end{table*}

%

\section{BENCHMARK SPECIFICATION}
\label{sec:bench_spec}


\tr{To construct a specification of graph mining algorithms, we extensively
reviewed related work~\cite{chakrabarti2006graph, washio2003state,
lee2010survey, rehman2012graph, gallagher2006matching, ramraj2015frequent,
jiang2013survey, aggarwal2010managing, tang2010graph, leicht2006vertex,
liben2007link, ribeiro2019survey, lu2011link, al2011survey}.}
The GMS specification has four parts: graph mining
\emph{\textbf{problems}}, \emph{\textbf{algorithms}}, \emph{\textbf{datasets}},
and \emph{\textbf{metrics}}\footnote{\scriptsize We encourage participation in
the GMS effort. If the reader would like to include some problem or algorithm
in the specification and the platform, the authors would welcome the input.}.
\cnf{
\emph{Due to space constraints, we only outline selected aspects. Full
specification is in the extended technical
report (the link is on page~1).}
}

\all{One large class is \emph{\textbf{graph pattern
matching}}~\cite{jiang2013survey}, which focuses on finding certain specific
subgraphs (also called \emph{motifs} or \emph{graphlets}). Examples of such
problems are $k$-clique listing~\cite{danisch2018listing}, maximal clique
listing~\cite{bron1973algorithm, cazals2008note, DBLP:conf/isaac/EppsteinLS10,
DBLP:journals/tcs/TomitaTT06}, $k$-star-clique
mining~\cite{jabbour2018pushing}, and many others~\cite{cook2006mining}.
Another class of graph mining problems is broadly referred to as
\emph{\textbf{graph learning}}~\cite{cook2006mining}, with problems such as
unsupervised learning or clustering~\cite{jarvis1973clustering}, link
prediction~\cite{liben2007link, lu2011link, al2006link, taskar2004link}, or
vertex similarity~\cite{leicht2006vertex}. 
Third, we also consider problems that are representative for a large family of
\emph{\textbf{optimization problems}}, such as deriving minimum spanning tree
or graph coloring~\cite{??}.
Finally, we separately consider schemes related to \emph{\textbf{reordering of
vertices}}. A well-known example is deriving graph degeneracy
ordering~\cite{??}.}

\vspaceSQ{-0.3em}
\subsection{Graph Problems and Algorithms}

We identify \textbf{four} major classes of graph mining problems and the
corresponding algorithms: \textbf{pattern matching, learning, reordering},
and (partially) \textbf{optimization}.
For each given class of problems, we aimed to cover a {wide} range of
problems and algorithms that {differ} in their design and performance
characteristics, for example P and NP problems, heuristics and exact schemes,
algorithms with time complexities described by low-degree and high-degree
polynomials, etc..
The specification is summarized in Table~\ref{tab:problems}.
Additional details are provided in the appendix, in Section~\ref{sec:app-details}.
\ifall
We also indicate (with~\includegraphics[scale=0.2,trim=0 16 0 0]{5p.pdf}) which
problems and algorithms are implemented in GMS using set operations. The user
has a fine level of control over the building blocks of these algorithms, as
explained in Section~\ref{sec:overview} and Figure~\ref{fig:design}.  The
problems not marked with~\includegraphics[scale=0.2,trim=0 16 0
0]{5p.pdf} \emph{cannot easily be implemented with set operations}: for
example, the power iteration scheme for computing PageRank does not use
sets~\cite{besta2017push}.  In all these cases, one can still benefit from
extensibility of GMS parts~\includegraphics[scale=0.2,trim=0 16 0 0]{1.pdf} --
\includegraphics[scale=0.2,trim=0 16 0 0]{5.pdf}.
\fi

\vspaceSQ{-0.3em}
\subsubsection{{Graph Pattern Matching}}
\label{sec:pattern-match}

\ One large class is {{graph pattern matching}}~\cite{jiang2013survey}, which
focuses on finding specific subgraphs (also called \emph{motifs} or
\emph{graphlets}) that are often (but not always) \emph{dense}. 
Most algorithms solving such problems consist of the \emph{searching
part} (finding candidate subgraphs) and the \emph{matching part}
(deciding whether a given candidate subgraph satisfies the search
criteria). The search criteria (the details of the searched subgraphs)
influence the time complexity of both searching and matching.
\all{Finding all maximal cliques has many applications in social network
analysis~\cite{wasserman1994social}, bioinformatics~\cite{day1986computational,
spirin2003protein}, and computational chemistry~\cite{rhodes2003clip}.}
First, we pick \textbf{listing all cliques} in a graph, as this
problem has a long and rich history in the graph mining domain, and numerous
applications. We consider both \textbf{maximal} cliques (an NP-hard problem)
and \textbf{$k$-cliques} (a problem with time complexity in $O(n^k)$), and the
established associated algorithms, most importantly
Bron-Kerbosch~\cite{bron1973algorithm},
Chiba-Nishizeki~\cite{chiba1985arboricity}, and their various
enhancements~\cite{manoussakis2018output, cazals2008note,
DBLP:conf/isaac/EppsteinLS10, DBLP:journals/tcs/TomitaTT06,
danisch2018listing}. 
Next, we cover a more general problem of listing \textbf{dense
subgraphs}~\cite{lee2010survey, jabbour2018pushing} such as $k$-cores,
$k$-star-cliques, and others. GMS also includes the Frequent Subgraph Mining
(FSM) problem~\cite{jiang2013survey}, in which one finds \textbf{all subgraphs}
(not just dense) that occur \emph{more often than a specified threshold}.
Finally, we include the established NP-complete \textbf{subgraph isomorphism}
(SI) problem, because of its prominence in both the theory and practice of
pattern matching, and because of a large number of variants that often have
different performance characteristics~\cite{cordella2004sub,
mccreesh2015parallel, carletti2018vf3, han2013turbo, ullmann1976algorithm}; SI
is also used as a subroutine in the matching part of FSM.

\vspaceSQ{-0.3em}
\subsubsection{{Graph Learning}}

\ We also consider various problems that can be loosely categorized as
{{graph learning}}.
These problems are mostly related to clustering, and they include
\textbf{vertex similarity}~\cite{robinson2013graph, leicht2006vertex,
robinson2013graph} (verifying how similar two vertices are), \textbf{link
prediction}~\cite{liben2007link, lu2011link, al2006link, taskar2004link,
wang2014robustness} (predicting whether two non-adjacent vertices can become
connected in the future, often based on vertex similarity scores), and
\textbf{Clustering and Community Detection}~\cite{jarvis1973clustering,
raghavan2007near, blondel2008fast} (finding various densely
connected groups of vertices, also often incorporating vertex similarity as a
subroutine).

\all{
\macb{Discussion: Clusters \& Communities vs.~Dense Subgraphs. }
\emph{Clustering and community detection (central problems in graph learning)
are {similar to dense subgraph discovery} (a central problem in graph pattern
matching)}. However, the latter use the notion of \emph{absolute density}: a
dense subgraph~$S$ is some relaxation of a clique (i.e., one does not consider
what is ``outside~$S$''). Contrarily, the former use a concept of
\emph{relative density}: one compares different subgraphs to decide which one
is dense~\cite{aggarwal2010managing}.}


\vspaceSQ{-0.3em}
\subsubsection{{Vertex Reordering}}
\label{sec:spec_reorder}

\ We also consider {{reordering of vertices}}.
Intuitively, the order in which vertices are processed in some algorithm may
impact the performance of this algorithm. For example, when counting
triangles, ordering vertices by degrees (prior to counting) minimizes the
number of times one triangle is (unnecessarily) counted more than once.
In GMS, we first consider the above-mentioned \textbf{degree ordering}.  
We also provide two algorithms for the \textbf{degeneracy
ordering}~\cite{DBLP:conf/latin/Farach-ColtonT14} (\textbf{exact} and
\textbf{approximate}), which was shown to improve the performance of maximal
clique listing or graph coloring~\cite{besta2020high, cazals2008note,
DBLP:conf/isaac/EppsteinLS10, DBLP:journals/tcs/TomitaTT06}.
\all{This ordering produces an \emph{orientation} of graph edges with
low out-degree of the vertices.}
\all{Second, we use {triangle ordering}, in which the number of triangles that
vertices belong to determines vertex order.}

\vspaceSQ{-0.3em}
\subsubsection{{Optimization}}

\ While GMS focuses less on optimization problems, we also include a
representative problem of graph coloring and selected other problems. 


\vspaceSQ{-0.3em}
\subsubsection{{Taxonomy and Discussion}}

\ Graph pattern matching, clustering, and optimization are related in that the
problems from these classes focus on \emph{finding certain subgraphs}.  In the
two former classes, such subgraphs are usually ``local'' groups of vertices,
most often dense (e.g., cliques, clusters)~\cite{aggarwal2010graph,
thomas2010margin, aggarwal2010survey, adedoyin2013survey, injadat2016data,
berkhin2006survey, parthasarathy2010survey}, but sometimes can also be sparse
(e.g., in FSM or SI). In optimization, a subgraph to be found can be
``global'', scattered over the whole graph (e.g., 
vertices with the same color).
Moreover, \emph{clustering and community detection (central problems in graph learning)
are {similar to dense subgraph discovery} (a central problem in graph pattern
matching)}. Yet, the latter use the notion of \emph{absolute density}: a
dense subgraph~$S$ is some relaxation of a clique (i.e., one does not consider
what is ``outside~$S$''). Contrarily, the former use a concept of
\emph{relative density}: one compares different subgraphs to decide which one
is dense~\cite{aggarwal2010managing}.

\ifall\m{old}
\vspaceSQ{-0.25em}
\subsection{Graph Learning}

Another class of graph mining problems is broadly referred to as
\emph{\textbf{graph learning}}~\cite{cook2006mining}, with problems such as
unsupervised learning or clustering~\cite{jarvis1973clustering}, link
prediction~\cite{liben2007link, lu2011link, al2006link, taskar2004link}, or
vertex similarity~\cite{leicht2006vertex}. 

\vspaceSQ{-0.3em}
\subsubsection{Vertex Similarity}
\label{sec:sets-similarity}

Vertex similarity measures can be used on their own, for example in graph
database queries~\cite{robinson2013graph}, or as a building block of more
complex algorithms such as clustering~\cite{jarvis1973clustering}. We consider
seven measures: \textbf{Jaccard}, \textbf{Overlap}, \textbf{Adamic Adar},
\textbf{Resource Allocation}, \textbf{Common Neighbors}, \textbf{Total
Neighbors}, and \textbf{Preferential Attachment}
measures~\cite{leicht2006vertex, robinson2013graph}. All these measures
associate (in different ways) the number of common neighbors of vertices $v$
and $u$ with the degree of similarity between $v$ and $u$.

\maciej{
Neo4j: ``We can use the Jaccard Similarity algorithm to work out the similarity
between two things. We might then use the computed similarity as part of a
recommendation query. For example, you can use the Jaccard Similarity algorithm
to show the products that were purchased by similar customers, in terms of
previous products purchased.''
Neo4j: ``We can use the Overlap Similarity algorithm to work out which things
are subsets of others. We might then use these computed subsets to learn a
taxonomy from tagged data.''
https://jbarrasa.com/2017/03/31/quickgraph5-learning-a-taxonomy-from-your-tagged-data/
``Adamic Adar is a measure used to compute the closeness of nodes based on
their shared neighbors.  The Adamic Adar algorithm was introduced in 2003 by
Lada Adamic and Eytan Adar to predict links in a social network.''
``Resource Allocation is a measure used to compute the closeness of nodes based
on their shared neighbors.  The Resource Allocation algorithm was introduced in
2009 by Tao Zhou, Linyuan Lü, and Yi-Cheng Zhang as part of a study to predict
links in various networks. ''
``Common neighbors captures the idea that two strangers who have a friend in
common are more likely to be introduced than those who don’t have any friends
in common.''
``Preferential Attachment is a measure used to compute the closeness of nodes,
based on their shared neighbors.  Preferential attachment means that the more
connected a node is, the more likely it is to receive new links. This algorithm
was popularised by Albert-László Barabási and Réka Albert through their work on
scale-free networks.''
``Total Neighbors computes the closeness of nodes, based on the number of
unique neighbors that they have. It is based on the idea that the more
connected a node is, the more likely it is to receive new links.  ''
Adamic Adar measure: "Friends and neighbors on the Web"
}

\vspaceSQ{-0.25em}
\subsubsection{Link Prediction}
\label{sec:lp}

Here, one is interested in developing schemes for predicting whether two
non-adjacent vertices can become connected in the future. There exist many
schemes for such prediction that are based on \textbf{variations of vertex
similarity} (\cref{sec:sets-similarity})~\cite{liben2007link, lu2011link,
al2006link, taskar2004link}. We provide them in GMS, as well as a simple
algorithm for \textbf{assessing the accuracy} of a specific link prediction
scheme~\cite{wang2014robustness}, which assesses how well a given prediction
scheme works.

\vspaceSQ{-0.25em}
\subsubsection{Clustering and Community Detection}
\label{sec:cl}

We consider graph clustering and community detection, a widely studied problem
used in a plethora of areas. We pick \textbf{Jarvis-Patrick clustering}
(JP)~\cite{jarvis1973clustering}, a scheme that uses similarity of two vertices
to determine whether these two vertices are in the same cluster. 
Moreover, we consider \textbf{Label Propagation} (LP)~\cite{raghavan2007near}
and the \textbf{Louvain method} (LM)~\cite{blondel2008fast}, two established
methods for detecting communities that, respectively, use the notions of
\emph{label dominance} and \emph{modularity} in assigning vertices to
communities. Together, our selection covers a wide range of schemes: single
iteration (JP) vs.~convergence based (LP, LM), multi-level (LM)
vs.~single-level (JP, LP), and overlapping (JP) vs.~non-overlapping (LM, LP).

\macb{Clusters \& Communities vs.~Dense Subgraphs. }
\emph{Clustering and community detection are \textbf{similar to
dense subgraph discovery} (\cref{sec:pattern-match})}. The difference is that
the latter use the notion of \emph{absolute density}: a dense subgraph~$S$ is
some relaxation of a clique (i.e., one does not consider what is ``outside~$S$'').
Contrarily, the former uses a concept of \emph{relative density}: one compares
different subgraphs to decide which one is dense~\cite{aggarwal2010managing}.

\vspaceSQ{-0.25em}
\subsection{Optimization Problems}

Third, we also consider problems that are representative for a large family of
\emph{\textbf{optimization problems}}, also deemed important in the
literature~\cite{aggarwal2010managing}. 
Here, we select the \textbf{graph coloring} (GC), the \textbf{mincut} (MC),
and the \textbf{minimum spanning tree} (MST) problem, all having a plethora of
well-known applications. Respectively, they represent an NP-Complete problem
and two P problems of different complexities.
We use \textbf{Boruvka's algorithm} for MST as it is easily parallelizable, a
very recent \textbf{parallel mincut} scheme~\cite{gianinazzi2018communication},
and several graph coloring algorithms that represent different approaches:
\textbf{Jones and Plassmann's}~\cite{jones1993parallel} and \textbf{Hasenplaugh
et al.'s}~\cite{hasenplaugh2014ordering} heuristics based on appropriate vertex
orderings and vertex prioritization,
\textbf{Johansson's}~\cite{johansson1999simple} and
\textbf{Barenboim's}~\cite{barenboim2016locality} randomized palette-based
heuristics that use conflict resolution, and \textbf{Elkin et
al.'s}~\cite{elkin20142delta} and \textbf{sparse-dense
decomposition}~\cite{harris2016distributed} that are examples of
state-of-the-art distributed algorithms.

\vspaceSQ{-0.25em} \subsection{Centralities}

Finally, GMS delivers \textbf{\emph{vertex centralities}} that -- intuitively
-- indicate the importance of vertices based on different properties. We
provide implementations of \textbf{Betweenness
Centrality}~\cite{brandes2001faster} where the importance of~$v$ is based on
numbers of shortest paths that go through~$v$ (we use a fast traversal-based
algorithm provided in GAPBS~\cite{beamer2015gap}), \textbf{degree centrality}
with the importance based on the degree, \textbf{triangle centrality}, where
more important vertices belong to more triangles, and
\textbf{PageRank}~\cite{page1999pagerank}, a popular website rank.
\fi

\subsection{Graph Datasets}
\label{sec:spec_datasets}

We aim at a dataset selection that is {computationally
challenging for all considered problems and algorithms},
cf.~Table~\ref{tab:problems}.
We list both \emph{large} and \emph{small} graphs, to indicate datasets that
can stress both low-complexity graph mining algorithms (e.g., centrality
schemes or clustering) \emph{and} high-complexity P, NP-complete,
and NP-hard ones such as subgraph isomorphism.

So far, existing performance analyses on parallel graph algorithms focused on
graphs with varying \emph{sparsities}~$m/n$ (sparse and dense), \emph{skews} in
degree distribution (high and low skew), \emph{diameters} (high and low), and
\emph{amounts of locality} that can be intuitively explained as the
\emph{number of inter-cluster} edges (many and few)~\cite{beamer2015gap}. In
GMS, we recommend to use such graphs as well, as the above properties influence
the runtimes of all described algorithms.

In Table~\ref{tab:problems}, graphs with high degree distribution skews are
indicated with large (relatively to $n$) maximum degrees~$\Delta$, which poses
challenges for load balancing and others.
\tr{Moreover, we list graphs with very high diameters (e.g., road networks) that
stress iterative algorithms where the runtime depends on the diameter.
Next, to
provide even more variability in the performance effects, we also consider
graphs with relatively high diameters \emph{and} with high skews in degree
distributions, such as the youtube social network.}

However, one of the insights that we gained with GMS is that \emph{\textbf{the
higher-order structure}, important for the performance of graph mining, can be
little related to the above properties}. For example,
in~\cref{sec:eval_higher}, we describe two graphs with almost identical sizes,
sparsities, and diameters, {but very different performance characteristics for
4-clique mining}. As we detail in~\cref{sec:eval_higher}, this is because {the
origin of these graphs determines whether a graph has many cliques \emph{or}
dense (but mostly non-clique) clusters}.
Thus, we also explicitly recommend to use graphs of \emph{different origins}.
We provide details of this particular case in~\cref{sec:eval_higher} (cf.~Livemocha
and Flickr). 

In addition, we explicitly consider the count of triangles~$T$, as (1) it
indicates clustering properties (and thus implies the
amount of locality), and it gives hints on different higher-order
characteristics (e.g., the more triangles per vertex, the higher a chance for
having $k$-cliques for $k > 3$).
Here, we also recommend using graphs that have \emph{large differences in counts
of triangles per vertex} (i.e., large $T$-skew). Specifically, a large
difference between the \emph{average} number of triangles per vertex~$T/n$
and the \emph{maximum} $T/n$ indicates that a graph may pose
additional load balancing problems for algorithms that list cliques
of possibly unbounded sizes, for example Bron-Kerbosch. We also consider
such graphs, see Table~\ref{tab:problems}. 

\marginpar{\vspace{5em}\colorbox{yellow}{\textbf{R-1}}}

\hl{Finally, GMS enables using synthetic graphs with the random uniform (the
\mbox{Erdős-Rényi} model~\mbox{\cite{erdHos1976evolution}}) and power-law (the
Kronecker model~\mbox{\cite{leskovec2010kronecker}}) degree distributions.
This is enabled by integrating the GMS platform with existing graph
generators~\mbox{\cite{beamer2015gap}}.
Using such synthetic graphs enables analyzing performance effects while
\emph{systematically} changing a specific \emph{single} graph property such as
\mbox{$n$}, \mbox{$m$}, or \mbox{$m/n$}, which is not possible with real-world
datasets.}

We stress that we refrain from prescribing {concrete datasets} as benchmarking
input (1) for flexibility, (2) because the datasets themselves evolve and (3)
the compute and memory capacities of architectures grow continually, making it
impractical to stick to a fixed-sized dataset. Instead, in GMS, we analyze and
discuss publicly available datasets in Section~\ref{sec:eval}, making
suggestions on their applicability for stressing performance of different
algorithms.

\enlargeSQ

\subsection{Metrics}
\label{sec:metrics}

In GMS, we first use simple \emph{running times} of algorithms (or their
specific parts, for a \emph{fine grained analysis}). Unless stated otherwise,
we use all available CPU cores, to maximize utilization of the underlying
system.
We also consider \emph{scalability analyses}, illustrating how the runtime
changes with the increasing amount of parallelism (\#threads). Comparison
between the measured scaling behavior and the ideal speedup helps to identify
potential scalability bottlenecks.
%
%
\sethlcolor{green}\hl{Finally, we consider \emph{memory consumption}.}

\marginpar{\vspace{-1em}\colorbox{green}{\textbf{R-1}}}

\sethlcolor{yellow}

We also assess the \textbf{machine-efficiency}, i.e., \emph{how well a
machine is utilized in terms of its memory bandwidth}. 
\hl{For this, we consider CPU core utilization, expressed with counts of
stalled CPU cycles.}
One can measure this number easily with, for example, the established PAPI
infrastructure~\cite{mucci1999papi} \hl{that enables gathering detailed
performance data from hardware counters}.
\hl{As we will discuss in detail in Section~\mbox{\ref{sec:platform-design}},
we seamlessly integrate GMS with PAPI, enabling gathering detailed data such as
stalled CPU cycles but also more than that, for example cache misses
and hits (L1, L2, L3, data vs.~instruction, TLB), memory reads/writes, and many
others.}

\marginpar{\vspace{-10em}\colorbox{yellow}{\textbf{R-1}}}

\marginpar{\vspace{-4em}\colorbox{yellow}{\textbf{R-1}}}

\all{For this, we count the number of L3 cache misses, which accounts for the
CPU-memory traffic. Note that -- while this metric does not consider concepts
such as work efficiency -- it does provide a raw measure of using the available
memory bandwidth.}

Finally, we propose a new metric for measuring the
``\textbf{algorithmic efficiency}'' (``\textbf{algorithmic throughput}'').
Specifically, we measure the \emph{number of mined graph patterns in a time
unit}. Intuitively, this metric indicates how efficient a given algorithm is in
finding respective graph elements.
An example such metric used in the past is \emph{processed edges per second}
(PEPS), used in the context of graph traversals and
PageRank~\cite{lin2018shentu}. Here, we extend it to graph mining and to
arbitrary graph patterns.
In graph pattern matching, this metric is the number of the respective
\emph{graph subgraphs found per second} (e.g., maximal cliques per second). In
graph learning, it is a count of vertex pairs with similarity derived per
second (vertex similarity, link prediction), or the number of
clusters/communities found per second (clustering, community detection).
\hl{The algorithmic efficiency facilitates deriving performance insights associated
with the structure of the processed graphs. By comparing relative
throughput differences between different algorithms \emph{for different input
graphs}, one can conclude whether these differences
consistently depend on \emph{pattern (e.g., clique) density}.}

\marginpar{\vspace{-4em}\colorbox{yellow}{\textbf{R-1}}}

\iftr

\sethlcolor{green}

\hl{The algorithmic efficiency metric may also be used to provide more
compact results. As an example, consider two datasets, one -- \mbox{$G_1$} -- with
many small cliques, the other -- \mbox{$G_2$} -- with few large cliques.  Bron
Kersbosch may be similar in both cases in its run-time, but its ``clique
efficiency'' would be high for \mbox{$G_1$} and low for \mbox{$G_2$}. Thus, one could deduce
based purely on the ``clique throughput'' that the best choice of algorithm
depends on the number of cliques in the graph, because BK's throughput suffers
more when there are few cliques, but it has a high throughput when there are
many of cliques.  This cannot be deduced based purely on the run-time, but only
using a combination of run-times \emph{and} total clique counts.}

\sethlcolor{yellow}

\fi


%

\subsection{Beyond The Scope of GMS}

\ifconf
GMS in its current shape does \emph{not} focus on ``low-complexity'' graph
algorithms such as BFS, traditionally not considered as a part of graph mining.
\hl{Many such algorithms can also use set algebra. For example,
the frontier in BFS, and its expansion, can be modeled with a set and several
set intersections, respectively. GMS currently does not offer
efficient implementations of set operations targeting such algorithms. We plan
to work on this in future GMS versions.}
%
%
\fi

\marginpar{\vspace{-5em}\colorbox{yellow}{\textbf{R-3}}}

\iftr
We fix GMB's scope to include problems and algorithms related to ``graph
mining'', often also referred to as ``graph analytics'', in the \emph{offline}
(\emph{static}) setting, with a \emph{single} input graph. Thus, we do
{not} focus on streaming or dynamic graphs (as they usually come with
vastly different design and implementation challenges~\cite{besta2019practice})
and we do {not} consider problems that operate on multiple
\emph{different} input graphs. We leave these two domains for future work.

GMS also does {not} aim to cover advanced statistical methods that -- for
example -- analyze power laws in input graphs.  For this, we recommend to use
specialized software, for example iGraph~\cite{csardi2006igraph}.

Finally, we also do not focus on many graph problems and algorithms
traditionally researched in the \emph{parallel programming} community and
usually do {not} considered as part of graph \emph{mining}. Examples are
PageRank~\cite{page1999pagerank}, Breadth-First
Search~\cite{beamer2013direction}, Betweenness
Centrality~\cite{brandes2001faster, madduri2009faster,
prountzos2013betweenness, solomonik2017scaling}, and others~\cite{demetrescu2009shortest,
besta2015accelerating, besta2017slimsell, gianinazzi2018communication, besta2019graph,
besta2020substream, besta2019demystifying, besta2020high}.
Many of these problems are addressed by abstractions such as
vertex-centric~\cite{malewicz2010pregel}, edge-centric~\cite{roy2013x},
GraphBLAS~\cite{kepner2016mathematical} and the associated linear algebraic
paradigm~\cite{kepner2016mathematical} with fundamental operations being matrix-matrix and matrix-vector
products~\cite{besta2020communication, kwasniewski2019red, kwasniewski2021parallel}.
These works were addressed in detail in past
analyses~\cite{besta2017push} and are included in existing suites such as
GAPBS~\cite{beamer2015gap}, Graph500~\cite{suzumura2011performance,
murphy2010introducing}, and GBBS~\cite{dhulipala2018theoretically}.
Still, all the GMS modularity levels (\includegraphics[scale=0.2,trim=0 16 0
0]{1.pdf} -- \includegraphics[scale=0.2,trim=0 16 0 0]{5p.pdf}) can be used to
extend the GMS platform with any of such algorithms.
\fi

\section{GMS PLATFORM \& SET ALGEBRA}
\label{sec:platform-design}
\vspaceSQ{-0.2em}

\marginpar{\vspace{-1em}\colorbox{yellow}{\textbf{R-2}}}

\enlargeSQ

We now detail the GMS platform and how it enables modularity,
extensibility, and high performance.
\tr{Details of using the platform are described in an 
extensive documentation (available at the provided link).}
%
%
%
\all{\maciej{fix} The groups and selected representative routines are in
Table~\ref{tab:interface}}
There are six main ways in which one can experiment with a graph mining
algorithm using the GMS platform, indicated in
Figure~\ref{fig:design} with \includegraphics[scale=0.2,trim=0 16 0 0]{1.pdf}
-- \includegraphics[scale=0.2,trim=0 16 0 0]{5p.pdf} and a
block~\includegraphics[scale=0.2,trim=0 16 0 0]{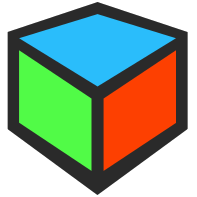}.
\all{\vspaceSQ{-0.3em}
\subsubsection{Graph Representation and Access Methods}}

{First}, the user can provide a \emph{new graph
representation}~\includegraphics[scale=0.2,trim=0 16 0 0]{1.pdf} and the
associated \emph{{routines for accessing the graph
structure}}~\includegraphics[scale=0.2,trim=0 16 0 0]{2.pdf}. By default, GMS
uses CSR. A seamless integration of a new graph representation is enabled by a
modular design of files and classes with the representation code, and a concise
interface (checking the degree~$d(v)$, loading neighbors~$N(v)$, iterating over
vertices~$V$ or edges~$E$, and verifying if an edge~$(u,v)$ exists) between a
representation and the rest of GMS.
The GMS platform also supports compressed graph representations. While many
compression schemes focus on minimizing the amount of used
storage~\cite{BoVWFI} and require expensive decompression, some graph
compression techniques entail \emph{mild} decompression overheads, and they can
even lead to overall \emph{speedups} due to lower pressure on the memory
subsystem~\cite{besta2018log}.
Here, we offer ready-to-go implementations of such schemes, including bit
packing, vertex relabeling, Log(Graph)~\cite{besta2018log}, and others.

\all{\vspaceSQ{-0.3em}
\subsubsection{Preprocessing Routines}}

Second, the user can seamlessly add \emph{{preprocessing
routines}}~\includegraphics[scale=0.2,trim=0 16 0 0]{3.pdf} such as the reordering
of vertices.
\all{; we provide different ready-to-go schemes such as degeneracy
reordering~\cite{DBLP:conf/latin/Farach-ColtonT14} (details in
Section~\ref{sec:bench_spec}).}
Here, the main motivation is that by applying a relevant vertex reordering
(relabeling), one can reduce the amount of work to be done in the actual
following graph mining algorithm. For example, the degeneracy order can
significantly reduce the work done when listing maximal
cliques~\cite{DBLP:conf/latin/Farach-ColtonT14}.
The user runs a selected preprocessing scheme with a single function call that
takes as its argument a graph to be processed.

\all{\vspaceSQ{-0.3em}
\subsubsection{Graph Algorithms}}

\enlargeSQ

Third, one can plug in a \emph{whole new graph
algorithm}~\includegraphics[scale=0.2,trim=0 16 0 0]{4.pdf}. 
\tr{thanks to a simple code structure and easy access to methods for loading a
graph from file, building representations, etc..}
%
%
\all{\vspaceSQ{-0.3em}
\subsubsection{Fine Parts of Graph Algorithms}}
GMS also facilitates modifying \emph{fine parts of an
algorithm}~\includegraphics[scale=0.2,trim=0 16 0 0]{5.pdf}, such as a
scheduling policy of a loop. For this, we ensure a \emph{modular structure of
the respective implementations, and annotate code}. 

\all{\vspaceSQ{-0.3em}
\subsubsection{Set Algebra Building Blocks}}

Finally, we use the fact that many graph algorithms, for example
Bron-Kerbosch~\cite{bron1973algorithm} and others~\cite{das2018shared,
han2018speeding, aberger2017emptyheaded, cazals2008note,
DBLP:conf/isaac/EppsteinLS10, DBLP:journals/tcs/TomitaTT06, carletti2018vf3,
carletti2019parallel, wang2014robustness, danisch2018listing,
DBLP:conf/isaac/EppsteinLS10}, are formulated with \emph{set algebra} and use a
{small} group of {well-defined} operations such as set intersection~$\cap$. In
GMS, we enable the user to provide their own implementation of such operations
and of the data layout of the associated sets. This facilitates controlling the
layout of a single auxiliary data structure or an implementation of a
particular subroutine (indicated with~\includegraphics[scale=0.2,trim=0 16 0
0]{5p.pdf}). 
\hl{Thus, one is able to break complex graph mining algorithms into simple
building blocks, and work on these building blocks independently.}
We already implemented a wide selection of routines for $\cap$, $\cup$,
$\setminus$, $|\cdot|$, and $\in$; we also offer different set layouts based on
integer arrays, bit vectors, and compressed variants of these two.

\marginpar{\vspace{-5em}\colorbox{yellow}{\textbf{R-2}}}

\all{\maciej{fix} Table~\ref{tab:interface} illustrates selected set operations
that are currently offered by GMS.}

\ifall
\hl{The central method for modularity and extensibility in GMS is to
incorporate set algebra operations. This way, one is able to break complex
graph mining algorithms into simple building blocks, and work on these building
blocks independently. We motivate this approach with an observation that many
graph mining algorithms contain a large number of set
operations~\mbox{\cite{??}}. We will illustrate this in
Section~\mbox{\ref{sec:hpc_algs}}, when describing implementing and tuning
several graph mining algorithms with sets and set operations.}
\fi

\hl{Set algebra building blocks in GMS are sets, set operations, set elements,
and set algebra based graph representations. The first three are grouped
together in the \texttt{Set} interface. The last one is a separate class that
appropriately combines the instances of a given \texttt{Set} implementation. We
now detail each of these parts.}

\marginpar{\vspace{-3em}\colorbox{yellow}{\textbf{R-2}}}

\marginpar{\vspace{1em}\colorbox{yellow}{\textbf{R-2}}}

\subsection{Set Interface}
\label{sec:set-interface}

\marginpar{\vspace{10em}\colorbox{yellow}{\textbf{R-2}}}

\hl{The \texttt{Set} interface, illustrated in
Listing~\mbox{\ref{lst:set-interface}}, encapsulates the representation of an
arbitrary set and its elements, and the corresponding set algorithms. By
default, set elements are vertex IDs (modeled as integers) but other elements
(i.e., integer tuples to model edges) can also be used.
Then, there are three types of methods in \texttt{Set}.}

\hl{First, there are methods implementing set basic set algebra operations,
i.e., ``union'' for $\cup$, ``intersect'' for $\cap$, and ``diff'' for
$\setminus$.  To enable performance tuning, they come in variants.
``\_inplace'' indicates that the calling object is being modified, as opposed
to the default method variant that returns a new set (avoiding excessive
data copying).  ``\_count'' indicates
that the result is the size of the resulting set, e.g., $|A \cap B|$ instead of
$A \cap B$ (avoiding creating unnecessary structures).
Then, \texttt{add} and \texttt{remove} enable devising optimized variants of
$\cup$ and $\setminus$ in which only one set element is inserted or removed
from a set; these methods always modify the calling set.}

\all{
\subsubsection{\colorbox{yellow}{Set Algebra Operations}}
}

\begin{lstlisting}[float=t,label=lst:set-interface,caption=\scriptsize\textmd{\hl{The set algebra interface provided by GMS.}}]
// Set: a type for arbitrary sets. 
// SetElement: a type for arbitrary set elements.
 
class Set {
public:
//In methods below, we denote "*this" pointer with $A$ 
//(1) Set algebra methods:
  Set diff(const Set &$B$) const; //Return a new set $C = A \setminus B$ 
  Set diff(SetElement $b$) const; //Return a new set $C = A \setminus \{b\}$
  void diff_inplace(const Set &$B$); //Update $A = A \setminus B$
  void diff_inplace(SetElement $b$); //Update $A = A \setminus \{b\}$
  Set intersect(const Set &$B$) const; //Return a new set $C = A \cap B$
  size_t intersect_count(const Set &$B$) const; //Return $|A \cap B|$
  void intersect_inplace(const Set &$B$); //Update $A = A \cap B$
  Set union(const Set &$B$) const; //Return a new set $C = A \cup B$
  Set union(SetElement $b$) const; //Return a new set $C = A \cup \{b\}$
  Set union_count(const Set &$B$) const; //Return $|A \cup B|$
  void union_inplace(const Set &$B$); //Update $A = A \cup B$
  void union_inplace(SetElement $b$); //Update $A = A \cup \{b\}$
  bool contains(SetElement $b$) const; //Return $b \in A$ ? true:false
  void add(SetElement $b$); //Update $A = A \cup \{b\}$
  void remove(SetElement $b$); //Update $A = A \setminus \{b\}$
  size_t cardinality() const; //Return set's cardinality

//(2) Constructors (selected):
  Set(const SetElement *start, size_t count); //From an array 
  Set(std::vector<SetElement> &vec); //From a vector
  //Set initialization with initializer list of elements:
  Set(std::initializer_list<SetElement> &data); 
  Set(); Set(Set &&); //Default and Move constructors
  Set(SetElement); //Constructor of a single-element set
  static Set Range(int $bound$); //Create set $\{0, 1, ..., bound - 1\}$

//(3) Other methods:
  begin() const; //Return iterators to set's start
  end() const; //Return iterators to set's end
  Set clone() const; //Return a copy of the set
  void toArray(int32_t *array) const; //Convert set to array
  operator==; operator!=; //Set equality/inequality comparison

private:
  using SetElement = GMS::NodeId; //(4) Define a set element
}
\end{lstlisting}

\hl{GMS offers other methods for
performance tuning. This includes constructors (e.g., a move
constructor, a constructor of a single-element set, or constructors from
an array, a vector, or an initializer list), and general methods such
as \texttt{clone}, which is used because -- by default -- the copy
constructor is disabled for sets to avoid accidental data copying.
GMS also offers conversion of a set to an integer
array to facilitate using established parallelization techniques.}

\marginpar{\vspace{1em}\colorbox{yellow}{\textbf{R-2}}}

\subsection{Implementations of Sets \& Set Algorithms}

On one hand, a set~$A$ can be represented as a contiguous sparse array with
integers modeling vertex IDs (``sparse'' indicates that only non-zero elements
are explicitly stored), of size $W \cdot |A|$, where $W$ is the memory word
size [bits]. This representation is commonly used to store vertex
neighborhoods.  However, one can \emph{also} represent $A$ with a dense
bitvector of size $n$ [bits], where the $i$-th set bit means that a vertex $i
\in A$ (``dense'' indicates that all zero bits are explicitly stored). While
being usually larger than a sparse array, a dense bitvector is more
space-efficient when $A$ is \emph{very} large, which happens when some vertex
connects to the majority of all vertices.
Now, depending on $A$'s and $B$'s representations, $A \cap B$ can itself be
implemented with different set algorithms. For example, if $A$ and $B$ are
sorted sparse arrays with similar sizes ($|A| \approx |B|$), one prefers the
``merge'' scheme where one simply iterates through $A$ and $B$, identifying
common elements (taking $O(|A|+|B|)$ time). If one set (e.g., $B$) is
represented as a bitvector, one may prefer a scheme where one iterates over the
elements of a sparse array~$A$ and checks if each element is in $B$, which
takes $O(1)$ time, giving the total of $O(|A|)$ time for the whole
intersection.

\marginpar{\vspace{-40em}\colorbox{yellow}{\textbf{R-2}}}

Moreover, a bitvector enables insertion or deletion of vertices into
a set in $O(1)$ time, which is useful in algorithms that rely on dynamic sets,
for example Bron-Kerbosch~\cite{das2018shared, cazals2008note,
DBLP:conf/isaac/EppsteinLS10, DBLP:journals/tcs/TomitaTT06}.
There are more set representations with other performance characteristics, such
as sparse~\cite{han2018speeding, aberger2017emptyheaded} or
compressed~\cite{besta2018survey} bitvectors, or hashtables, enabling further
performance/storage tradeoffs.

Importantly, using different \emph{\ul{set} representations} or \emph{\ul{set}
algorithms} does \emph{not} impact the formulations of \emph{\ul{graph}
algorithms}. GMS exploits this fact to facilitate development and experimentation.

\marginpar{\vspace{1em}\colorbox{yellow}{\textbf{R-2}}}

\hl{By default, GMS offers three implementations of \texttt{Set} interface:}

\marginpar{\vspace{5em}\colorbox{yellow}{\textbf{R-2}}}

\begin{itemize}[noitemsep, leftmargin=0.5em]
\item \hl{\textbf{RoaringSet} A set is implemented with a bitmap compressed
using recent ``roaring bitmaps''~\mbox{\cite{chambi2016better,
lemire2018roaring}}. A roaring bitmap offers diverse compression forms within
the same bitvector. They offer mild compression rates but do not
incur expensive decompression. As we later show, these structures result in
high performance of graph mining algorithms running on top of them.}
\item \hl{\textbf{SortedSet} GMS also offers sets stored as sorted vectors.
This reflects the established CSR graph representation design, where each
neighborhood is a sorted contiguous array of integers.}
\item \hl{\textbf{HashSet} Finally, GMS offers an implementation of Set with
a hashtable. By default, we use the Robin Hood library~\mbox{\cite{celis1986robin}}.}
\end{itemize}

\marginpar{\vspace{3em}\colorbox{yellow}{\textbf{R-2}}}

\vspaceSQ{-0.2em}
\subsection{Set-Centric Graph Representations}

\hl{Sets are building blocks for a graph representation: one
set implements one neighborhood. To enable using arbitrary set designs, GMS
harnesses templates, typed by the used set definition, see
Listing~\mbox{\ref{lst:setgraph}}.
GMS provides ready-to-go representations based on the RoaringSet,
SortedSet, and HashSet set representations.}

\marginpar{\vspace{5em}\colorbox{yellow}{\textbf{R-2}}}

\begin{lstlisting}[aboveskip=-1em, belowskip=-1em, float=h,label=lst:setgraph,caption=\scriptsize\textmd{\hl{A generic graph representation.}}]
template <class TSet>
class SetGraph {
public:
  using Set = TSet; int64_t num_nodes() const;
  const Set& out_neigh(NodeId node) const;
  int64_t out_degree(NodeId node) const;
  /* Some functions omitted */ };
\end{lstlisting}

\ifall
The following example from algorithms/set_based/maximal_clique_enum/parallel/eppsteinPAR.h shows how a function signature would look like in practice:

template <class SGraph, class Set = typename SGraph::Set>
    std::vector<Set> mceBench(const SGraph &rgraph, const pvector<NodeId> &ordering)
{
    // implementation omitted
}
This function is generic over SGraph and could be used for example with SetGraph<RoaringSet>. The second template parameter Set is set to default to the Set type associated with the SGraph. Inside of the function, new temporary sets are created by using this Set type. Also, in debug builds this function will return a vector of the found maximal cliques which also be represented by Set each.

For convenience we provide several default aliases for SetGraph with the default set implementations:

using RoaringGraph = SetGraph<RoaringSet>;
using SortedSetGraph = SetGraph<SortedSet>;
using RobinHoodGraph = SetGraph<RobinHoodSet>;
\fi


\marginpar{\vspace{5em}\colorbox{yellow}{\textbf{R-2}}}

\vspaceSQ{-0.5em}
\subsection{Pipeline Interface}
\label{sec:pipeline-interface}

\hl{Beyond the set algebra related interfaces, GMS also offers a dedicated API
for easy experimenting with other parts of the processing pipeline}
(\mbox{\includegraphics[scale=0.2,trim=0 16 0 0]{1.pdf}} --
\mbox{\includegraphics[scale=0.2,trim=0 16 0 0]{5.pdf}}).
\hl{This API is illustrated in Listing~\mbox{\ref{lst:pipeline-interface}}.  It
enables separate testing of each particular stage, but also enables the user to
define and analyze their own specific stages.}

\all{Defining your pipeline class To get started you should define a subclass of
GMS::Pipeline. Then add a function for each individual step, you can specify
the order of execution later at instantiation time. This class should be
stateful, i.e. provide all required data as members.}

\marginpar{\vspace{5em}\colorbox{yellow}{\textbf{R-2}}}

\begin{lstlisting}[aboveskip=-1em, belowskip=-1em, float=h,label=lst:pipeline-interface,caption=\scriptsize\textmd{\hl{A generic graph representation.}}]
class MyPipeline : public GMS::Pipeline {
public:
  //Any benchmark-specific arguments, including the input graph, 
  //are passed to the constructor 
  MyPipeline(const GMS::CLI::Args &a, const SortedSetGraph &g);
  //Functions for the individual steps.
  void convert(); //Potential conversion of g to another format
  void preprocess(); //Needed preprocessing
  void kernel(); //Desired graph mining algorithm
private:
  /* Any state variables that are shared between steps */ };
\end{lstlisting}


\marginpar{\vspace{4em}\colorbox{yellow}{\textbf{R-2}}}

\vspaceSQ{-0.5em}
\subsection{PAPI Interface}

\hl{GMS also uses the PAPI
library for easy access to hardware performance counters, cf.~\mbox{\cref{sec:metrics}}.
Importantly, we support seamless gathering of the performance data from
\emph{parallel} code regions}\footnote{\scriptsize \hl{We currently support OpenMP and plan to include other infrastructures such as Intel TBB.}}
\tr{\hl{An example usage of PAPI in GMS is in Listing~\mbox{\ref{lst:papi-interface}}.}
All the details on how to use the GMS PAPI support are also available in the online documentation.}

\iftr

\begin{lstlisting}[float=h,label=lst:papi-interface,caption=\scriptsize\textmd{\hl{Using PAPI for detailed performance measurements of a parallel region in GMS.}}]
//Init PAPI for parallel use, measure CPU cycles
//stalled on memory accesses, and on any resources
GMS::PAPIW::INIT_PARALLEL(PAPI_MEM_SCY, PAPI_RES_STL); 
GMS::PAPIW::START();
#pragma omp parallel
{
  //Benchmarked parallel region
}
GMS::PAPIW::STOP();
\end{lstlisting}

\fi

\section{\mbox{HIGH-PERFORMANCE \& SIMPLICITY}}
\label{sec:hpc_algs}

We now detail how using the GMS benchmarking platform leads to simple (i.e.,
programmable) \emph{\ul{and}} high-performance implementations of many 
graph mining algorithms.

\ifall


\subsection{Fine Modularity with Set Algebra}

To enable fine-grained modularity (``\includegraphics[scale=0.2,trim=0 16 0
0]{5p.pdf}'') \emph{without compromising on high performance}, GMS foremost relies
on set algebra. Our central observation is twofold. First,
\emph{large parts} of \emph{many} graph algorithms can be expressed with set
operations such as intersection $\cap$ or union $\cup$. Second, a set can be
represented with \emph{different data structures}, and similarly a set
operation can be instantiated with \emph{different algorithms}. 
As an example, consider a simple set intersection $A \cap B$ (where $A
\subseteq V$ and $B \subseteq V$ contain vertices), prevalent in many graph
mining algorithms~\cite{das2018shared, han2018speeding, aberger2017emptyheaded,
cazals2008note, DBLP:conf/isaac/EppsteinLS10, DBLP:journals/tcs/TomitaTT06,
carletti2018vf3, carletti2019parallel, wang2014robustness, danisch2018listing,
DBLP:conf/isaac/EppsteinLS10}.

On one hand, $A$ can be represented as a contiguous sparse array with integers
modeling vertex IDs (``sparse'' indicates that only non-zero elements are
explicitly stored), of size $W \cdot |A|$, where $W$ is the memory word size
[bits]. This representation is commonly used to store vertex neighborhoods.
However, one can \emph{also} represent $A$ with a dense bitvector of size $n$
[bits], where the $i$-th set bit means that a vertex $i \in A$ (``dense''
indicates that all zero bits are explicitly stored). While being usually larger
than a sparse array, a dense bitvector is more space-efficient when $A$ is
\emph{very} large, which happens when some vertex connects to the majority of
all vertices.
Now, depending on $A$'s and $B$'s representations, $A \cap B$ can itself be
implemented with different set algorithms. For example, if $A$ and $B$ are
sorted sparse arrays with similar sizes ($|A| \approx |B|$), one prefers the
``merge'' scheme where one simply iterates through $A$ and $B$, identifying
common elements (taking $O(|A|+|B|)$ time). If one set (e.g., $B$) is
represented as a bitvector, one may prefer a scheme where one iterates over the
elements of a sparse array~$A$ and checks if each element is in $B$, which
takes $O(1)$ time, giving the total of $O(|A|)$ time for the whole
intersection. 

Moreover, a bitvector enables insertion or deletion of vertices into
a set in $O(1)$ time, which is useful in algorithms that rely on dynamic sets,
for example Bron-Kerbosch~\cite{das2018shared, cazals2008note,
DBLP:conf/isaac/EppsteinLS10, DBLP:journals/tcs/TomitaTT06}. 
There are more
set representations with other performance characteristics, such as
sparse~\cite{han2018speeding, aberger2017emptyheaded} or
compressed~\cite{besta2018survey} bitvectors, or hashtables, enabling further
performance/storage tradeoffs.

\all{The set operation~$A \cap B$ can itself be implemented with different set
algorithms. For example, if $A$ and $B$ are sorted sparse arrays with similar
sizes ($|A| \approx |B|$), one prefers the ``merge'' scheme where one simply
iterates through $A$ and $B$, identifying common elements (taking $O(|A|+|B|)$
time). If one set is much smaller than the other ($|A| \ll |B|$), a
``galloping'' scheme~\cite{aberger2017emptyheaded} is preferred, where one
iterates over the elements of a smaller set and uses a binary search to check
if each element is in the bigger set (taking $O(|B| \log |A|)$ time).}

Importantly, using different \emph{\ul{set} representations} or \emph{\ul{set}
algorithms} does \emph{not} impact the formulations of \emph{\ul{graph}
algorithms}. GMS exploits this fact to facilitate development and experimentation.

\fi

\all{
\subsection{\mbox{Faster Graph Mining Algorithms in GMS}}
}

We now use the GMS benchmarking platform to {enhance existing graph mining
algorithms}. We provide consistent speedups (detailed in
Section~\ref{sec:eval}). Some new schemes also come with theoretical
advancements (detailed in Section~\ref{sec:concurrent}). 
The following descriptions focus on (1) how we ensure the
\emph{\textbf{modularity}} of GMS algorithms (for programmability), and (2)
what GMS design choices ensure \emph{\textbf{speedups}}.
Selected modular parts are marked with the $\ $
 ~~\tikzmarkin[set fill color=hlL, set border color=white, above offset=0.27,
right offset=2.2em, below
offset=-0.1]{mot1}\textcolor{black}{blue}\tikzmarkend{mot1}~ color and the
type of modularity (\includegraphics[scale=0.19,trim=0 16 0
0]{1.pdf} -- \includegraphics[scale=0.19,trim=0 16 0 0]{5p.pdf})\tr{ (explained
in~\cref{sec:overview} and Figure~\ref{fig:design})}.
\hl{Marked set operations are implemented using the \texttt{Set} interface, see Listing~\mbox{\ref{lst:set-interface}}.}
\tr{Whenever we use parallelization (``in parallel''), we ensure that it does
not involve conflicting memory accesses. For clarity, we focus on
\emph{formulations} and we discuss implementation details (e.g.,
parallelization) in the next sections.}

\marginpar{\vspace{-1em}\colorbox{yellow}{\textbf{R-2}}}

\iftr
\subsection{Use Case 1: Degeneracy Order \& $k$-Cores}
\label{sec:design_deg_order}
\else
\textbf{Use Case 1: Degeneracy Order \& $k$-Cores}
\fi
\all{\maciej{fix} Computing the \emph{degeneracy ordering} of the vertices is
an important routine used to accelerate many graph pattern matching algorithms,
and as a main part of computing the $k$-core
of~$G$~\cite{DBLP:conf/latin/Farach-ColtonT14}.
This ordering produces an orientation of the edge of the graph with low
out-degree of the vertices.}
A \emph{degeneracy} of a graph~$G$ is the smallest $d$ such that every subgraph
in~$G$ has a vertex of degree at most~$d$. Thus, degeneracy can serve as a way
to measure the graph sparsity that is ``closed under taking a graph subgraph''
(and thus more robust than, for example, the average degree). A degeneracy
\emph{ordering} (DGR) is an ``ordering of vertices of~$G$ such that each vertex
has $d$ or fewer neighbors that come later in this
ordering''~\cite{DBLP:conf/isaac/EppsteinLS10}. DGR can be obtained by
repeatedly removing a vertex of minimum degree in a graph. 
The derived DGR can be directly used to compute the $k$-core of $G$ (a maximal
connected subgraph of $G$ whose all vertices have degree at least $k$). This is
done by iterating over vertices in the DGR order, and removing vertices with
out-degree less than $k$.
\all{(in the oriented graph).}

DGR, when used as a preprocessing routine, has been shown to accelerate
different algorithms such as Bron-Kerbosch~\cite{DBLP:conf/isaac/EppsteinLS10}.
In the GMS benchmarking platform, we provide an implementation of DGR that is modular
and can be seamlessly used with other graph algorithms as preprocessing
(\includegraphics[scale=0.2,trim=0 16 0 0]{3.pdf}).
Moreover, we alleviate the fact that the default DGR is not easily
parallelizable and takes $O(n)$ iterations even in a parallel setting. For
this, GMS delivers a modular implementation of a recent
$(2+\epsilon)$-\emph{approximate} degeneracy order~\cite{besta2020high} (ADG), which has
$O(\log n)$ iterations for \emph{any} $\epsilon>0$. 
\all{Similarly to the standard exact degeneracy reordering, it loops over vertices,
with the difference that -- in each iteration -- a subset~$X$ of vertices is
selected (Line~3) and assigned the degeneracy order in parallel (Line~4). When
selecting, the usual criteria for the exact order are \emph{relaxed} based on
parameter~$\epsilon$ (Line~3), ensuring more parallelism in Line~4.}
\tr{Specifically, the strict degeneracy ordering can be relaxed by
introducing the approximation (multiplicative) factor~$k$ that determines, for
each vertex~$v$, \emph{the additional number of neighbors that can be ranked
higher in the order than $v$}. Formally, in a \emph{$k$-approximate degeneracy
ordering}, every vertex $v$ can have at most $k \cdot d$ neighbors ranked
higher in this order.}
Deriving ADG is in Algorithm~\ref{lst:apxcore}. It is
similar to computing the DGR, which iteratively removes vertices of the
smallest degree. The main difference is that one removes in parallel a
\emph{batch} of vertices with degrees smaller than $(1 +
\epsilon)\widehat{\delta_{U}}$ (cf.~set~$R$ and Line~\ref{ln:ADG_R}). The
parameter $\epsilon \ge 0$ controls the accuracy of the approximation;
$\widehat{\delta_U}$ is the average degree in the induced subgraph $G(U,
E[U])$, $U$ is a ``working set'' that tracks changes to $V$. ADG relies on set
cardinality and set difference, enabling the GMS set algebra modularity
(\includegraphics[scale=0.2,trim=0 16 0 0]{5p.pdf}).

\begin{lstlisting}[float=h,label=lst:apxcore,caption=\scriptsize\textmd{Deriving
the \textbf{approximate degeneracy order} (ADG) in GMS.
More than one number indicates that a given
snippet is associated with more than one modularity type.}]
|\vspace{0.15em}|//|\textbf{Input:}| |\hlLIR{6em}{ A graph $G$ }{1.pdf}|. |\textbf{Output:}| Approx. degeneracy order (ADG) $\eta$.
|\vspace{0.15em}|i = 1 // Iteration counter
$U$ = $V$ //$U$ is the induced subgraph used in each iteration $i$
while $U \neq \emptyset$ do:
  $\widehat{\delta_{U}}$ = $\Big(\sum_{v\in U}$ |\hlLIR{5em}{$\vert N_U(v)\vert$ }{2.pdf}| $\Big)$ / $\vert U\vert$ //Get the average degree in $U$
  //$R$ contains vertices assigned priority in this iteration:
  |\label{ln:ADG_R}||\vspace{0.35em}|$R$ = $\{v \in U :\ \ $|\hlLIR{5em}{$\vert N_U(v) \vert$ }{2.pdf}| $\leq (1+ \epsilon) \widehat{\delta_{U}} \}$
  |\vspace{0.15em}||\hlLIIR{13em}{\textbf{for} |$v \in R$ in parallel| }{2.pdf}{5.pdf}| do: $\eta(v)$ = i //assign the ADG order
  |\vspace{0.15em}|$U$ = |\hlLIR{3.5em}{$U \setminus R$ }{5p.pdf} |//Remove assigned vertices
  i = i+1
\end{lstlisting}

\iftr
\vspaceSQ{-0.3em}
\subsection{Use Case 2: Maximal Clique Listing}
\label{sec:design_max_clique}
\else
\textbf{Use Case 2: Maximal Clique Listing}
\fi
Maximal clique listing, in which one enumerates all \emph{maximal} cliques
(i.e., fully-connected subgraphs not contained in a larger such subgraph) in a
graph, is one of core graph mining problems~\cite{cazals2008note,
chiba1985arboricity, makino2004new, koch2001enumerating, tsukiyama1977new,
johnson1988generating, kose2001visualizing, xu2015distributed,
svendsen2015mining, lu2010dmaximalcliques, wu2009distributed,
lessley2017maximal, schmidt2009scalable, du2006parallel, zhang2005genome,
das2018shared, ottosen2010honour, stix2004finding, das2016change}.
The recursive backtracking algorithm by Bron and Kerbosch
(BK)~\cite{bron1973algorithm} together with a series of
enhancements~\cite{DBLP:journals/tcs/TomitaTT06, das2018shared,
eppstein2011listing, DBLP:conf/isaac/EppsteinLS10} (see Algorithm~\ref{alg:bk})
is an established and, in practice, the most efficient way of solving this
problem. Intuitively, in BK, one iteratively considers each vertex $v$ in a
given graph, and searches for all maximal cliques that contain $v$. The search
process is conducted \emph{recursively}, by starting with a single-vertex
clique $\{v\}$, and augmenting it with $v$'s neighbors, one at a time, until a
maximal clique is found.
\tr{Still, the number of maximal cliques in a general graph, and thus BK's
runtime, may be exponential~\cite{moon1965cliques}.}
\all{The original recursive backtracking MCE algorithm was developed by Bron and
Kerbosch~\cite{bron1973algorithm}. 
Intuitively, one first considers each vertex~$v$ as a ``starting point'' of any
maximal clique, and recursively searches for any maximal clique that
contains~$v$.}

\all{\subsubsection{Order of Processing Vertices}
\ }
Importantly, the order in which all the vertices are selected for processing
(at the \emph{outermost} level of recursion) may heavily impact the amount of
work in the following iterations~\cite{das2018shared,
eppstein2011listing, DBLP:conf/isaac/EppsteinLS10}. Thus, in GMS, we use
different vertex orderings, integrated using the GMS preprocessing modularity
(\includegraphics[scale=0.2,trim=0 16 0 0]{3.pdf}). One of our core
enhancements is to use the ADG order 
\iftr
(\cref{sec:design_deg_order}).
\else
(see above).
\fi
As we will
show, this brings theoretical (Section~\ref{sec:concurrent}) and empirical
(Section~\ref{sec:eval}) advancements.

\all{\subsubsection{Sets and Set Operations in BK}
\ }
A key part are vertex sets $P$, $X$, and $R$. They together navigate the
way in which the recursive search is conducted. $P$ (``Potential'') contains
candidate vertices that will be considered for belonging to the clique
currently being expanded. $X$ (``eXcluded'') are the vertices that are
definitely \emph{not} to be included in the current clique ($X$ is maintained
to avoid outputting the same clique more than once). $R$ is a currently
considered clique (may be non-maximal).
In GMS, we extensively experimented with different set representations for $P$,
$X$, and $R$, which was facilitated by the set algebra based modularity
(\includegraphics[scale=0.2,trim=0 16 0 0]{5p.pdf}). Our goal was to use
representations that enable fast ``bulk'' set operations such as intersecting
large sets (e.g., $X \cap N(v)$ in Line~\ref{ln:bk-bulk-set}) but also
efficient fine-grained modifications of such sets (e.g., $X = X \cup \{v\}$ in
Line~\ref{ln:bk-small-set}).  For this, we use roaring bitmaps.
As we will show (Section~\ref{sec:eval}), using such bitvectors as
representations of $P$, $X$, and $R$ brings overall speedups of even more than
9$\times$.

Now, at the outermost recursion level, for each vertex~$v_i$, we have $R =
\{v_i\}$ (Line~\ref{ln:bk-sets-outer}). This means that the considered clique
starts with $v_i$. Then, we have $P = N(v_i) \cap \{v_{i+1}, ..., v_n\}$ and
$X = N(v_i) \cap \{v_{1}, ..., v_{i-1}\}$. This removes unnecessary vertices
from $P$ and $X$. As we proceed in a fixed order of vertices in the main loop,
when starting a recursive search for $\{v_i\}$, we will definitely \emph{not}
include vertices $\{v_1, ..., v_{i-1}\}$ in $P$, and thus we can limit $P$ to
$N(v_i) \cap \{v_{i+1}, ..., v_n\}$ (a similar argument applies to $R$).
Note that these intersections may be implemented as simple splitting of the
neighbors $N(v_i)$ into two sets, based on the vertex order. This is another
example of the decoupling of general simple set algebraic formulations in GMS
and the underlying implementations (\includegraphics[scale=0.2,trim=0 16 0 0]{5p.pdf}).


In each recursive call of \texttt{BK-Pivot}, each vertex from $P$ is added to
$R$ to create a new clique candidate $R_{new}$ explored in the following
recursive call.  In this recursive call, $P$ and $X$ are respectively
restricted to $P \cap N(v)$ and $X \cap N(v)$ (any other vertices besides
$N(v)$ would not belong to the clique $R_{new}$ anyway).  After the recursive
call returns, $v$ is moved from $P$ (as it was already considered) to $X$ (to
avoid redundant work in the future).
The key condition for checking if $R$ is a \emph{maximal} clique is $P \cup X
== \emptyset$. If this is true, then no more vertices can be added to $R$
(including the ones from $X$ that were already considered in the past) and thus
$R$ is maximal.

\all{\subsubsection{Pivoting}}

The BK variant in GMS also includes an additional important optimization called
\emph{pivoting}~\cite{DBLP:journals/tcs/TomitaTT06}. Here, for any vertex $u
\in P \cup X$, only $u$ and its \emph{non} neighbors (i.e., $P \setminus N(u)$)
need to be tested as candidates to be added to $P$. This is because any
potential maximal clique must contain \emph{either} $u$ \emph{or} one of its
non-neighbors. Otherwise, a potential clique could be enlarged by adding $u$ to
it.  Thus, when selecting $u$ (Line~\ref{ln:bk-select-u}), one may use any
scheme that \emph{minimizes} $|P \setminus
N(u)|$~\cite{DBLP:journals/tcs/TomitaTT06}.
The advantage of pivoting is that it further prunes the search space and thus
limits the number of recursive calls.

\all{and no pivoting is used at this level. Then, for each vertex, the pivoting
based BK algorithm \texttt{BK-Pivot}~\cite{DBLP:journals/tcs/TomitaTT06} is
executed for each vertex $v_i$. To prune the search space, for each execution
of \texttt{BK-Pivot}, we initialize the set~$P$ of vertices potentially
included in the clique as $P = N(v_i) \cap \{v_{i+1}, ..., v_n\}$. This
enforces any vertices that \emph{follow} $v_i$ in the DG order to be already
considered as clique candidates. Similarly, we initialize the set~$X$ of
vertices that are not to be included in the maximal clique (in the recursive
call) as $X = N(v_i) \cap \{v_1, ..., v_{i-1}\}$. This enforces vertices that
\emph{precede} $v_i$ in the DG order to be excluded when considering candidate
vertices that could belong to a clique containing~$v_i$.}

For further performance improvements, we also use roaring bitmaps to implement
graph neighborhoods, exploiting the GMS modularity of representations and set
algebra (\includegraphics[scale=0.2,trim=0 16 0 0]{1.pdf},
\includegraphics[scale=0.2,trim=0 16 0 0]{2.pdf},
\includegraphics[scale=0.2,trim=0 16 0 0]{5p.pdf}).
\cnf{Finally, we also provide other optimizations based on set algebra that
further reduce work; they are described in the extended technical report.}

\iftr

An established way to derive the pivot vertex~$u \in P \cup X$, introduced by
Tomita et al.~\cite{DBLP:journals/tcs/TomitaTT06}, is to find $u =
\text{argmin}_{v \in P \cup X}|P \cap N(v)|$. This approach minimizes the size
of $P$ before the associated recursive \texttt{BK-Pivot} call.
Yet, it comes with a computational burden, because -- to select~$u$ --
one must conduct the set operation~$|P \cap N(v)|$ as many as $|P \cup X|$
times. This issue was addressed by proposing to derive $|P \cap N_H(v)|$
instead of $|P \cap N(v)|$, where $H$ is an induced subgraph of~$G$, with the
vertex set $P \cup X$ and the edge set $\{\{x,y\} \in E \mid x \in P \land y
\in P \cup X\}$~\cite{eppstein2011listing}. Using $N_H(v)$ reduces the amount
of work in each $|P \cap N_H(v)|$, because $N_H(v)$ is smaller than $N(v)$.
Such subgraph~$H$ is \emph{precomputed} before choosing~$u$, and is then passed
to the recursive \texttt{BK-Pivot} call, to accelerate precomputing
subgraphs~$H$ at deeper recursion levels.

We observe that the precomputed subgraph~$H$ can be used not only
to accelerate pivot selection, but also in several other set operations.
First, one can use $P \setminus N_H(u)$ 
%
%
instead of $P \setminus N(u)$ to reduce the cost of set difference;
note that this does not introduce more iterations in the
following loop because no vertex in $P$ is included in $N(u) \setminus N_H(u)$.
Second, we can also use $H$ to compute $P \cap N_H(v)$ and $X \cap N_H(v)$
instead of $P \cap N(v)$ and $X \cap N(v)$, also reducing the amount
of work in set intersections.

We also investigated the impact of constructing the $H$ subgraphs on each
recursion level, as initially advocated~\cite{eppstein2011listing}, versus only
at the outermost level. We observe that, while the latter always offers
performance advantages due to the large reductions in work, the former often
introduces overheads that outweight gains, due to the memory cost (from
maintaining many additional subgraphs) and the increase in runtimes (from
constructing many subgraphs). In our final BK-ADG version, we only derive $H$
at the outermost loop iteration, once for each vertex $v$, and use a given~$H$
at each level of the search tree associated with~$v$.

We also developed a variant of BK-ADG that, similarly to BK-DAS, uses
nested parallelism \emph{at each level of recursion}. This
approach proved consistently slower than the version without this feature.

\fi

\begin{lstlisting}[float=h,label=alg:bk,
caption=\textmd{\textbf{Enumeration of maximal cliques}, a Bron-Kerbosch variant by
Eppstein et al.~\cite{eppstein2011listing} with GMS enhancements.}]
/* |\textbf{Input:}| |\hlLIR{6em}{ A graph $G$ }{1.pdf}|. |\textbf{Output:}| all maximal cliques. */

//|\ul{Preprocessing}|: reorder vertices with DGR or ADG; see |\cref{sec:design_deg_order}|.
  $(v_1, v_2, ..., v_{n})$ = |\hlLIR{23em}{ preprocess$(V$, /* selected vertex order */$)$ }{3.pdf}|

//|\ul{Main part}|: conduct the actual clique enumeration.
|\label{ln:bk-main-loop}|for $v_i \in (v_1, v_2, ..., v_{n})$ do: //Iterate over $V$ in a specified order
  //For each vertex $v_i$, find maximal cliques containing $v_i$.
  //First, remove unnecessary vertices from $P$ (candidates
  //to be included in a clique) and $X$ (vertices definitely
  //not being in a clique) by intersecting $N(v_i)$ with vertices
  //that follow and precede $v_i$ in the applied order.
|\label{ln:bk-sets-outer}|  $P$ = |\hlLIR{12em}{ $N(v_i) \cap \{v_{i+1}, ..., v_n\}$ }{5p.pdf}|; $X$ = |\hlLIR{12em}{ $N(v_i) \cap \{v_{1}, ..., v_{i-1}\}$ }{5p.pdf}|; $R$ = $\{v_i\}$

  //Run the Bron-Kerbosch routine recursively for $P$ and $X$.
  BK-Pivot($P$, $\{v_i\}$, $X$)

BK-Pivot($P, R, X$) //Definition of the recursive BK scheme
|\vspace{0.25em}|  if |\hlLIR{7em}{ $P \cup X == 0$ }{5p.pdf}|: Output $R$ as a maximal clique
|\label{ln:bk-select-u}||\vspace{0.25em}|  $u$ = |\hlLIR{7em}{ pivot$(P \cup X)$ }{5p.pdf}| //Choose a "pivot" vertex $u \in P \cup X$
|\vspace{0.25em}|  for |\hlLIR{8em}{ $v \in P \setminus N(u)$ }{5p.pdf}|: // Use the pivot to prune search space
    //New candidates for the recursive search
|\label{ln:bk-bulk-set}||\vspace{0.25em}|    $P_{new}$ = |\hlLIR{6em}{ $P \cap N(v)$ }{5p.pdf}|; $X_{new}$ = |\hlLIR{6em}{ $X \cap N(v)$ }{5p.pdf}|; $R_{new}$ = |\hlLIR{5em}{ $R \cup \{v\}$ }{5p.pdf}|
    //Search recursively for a maximal clique that contains $v$
    BK-Pivot($P_{new}, R_{new}, X_{new}$)
    //After the recursive call, update $P$ and $X$ to reflect
    //the fact that $v$ was already considered
|\label{ln:bk-small-set}|    $P$ = |\hlLIR{5em}{ $P \setminus \{v\}$ }{5p.pdf}|; $X$ = |\hlLIR{5em}{ $X \cup \{v\}$ }{5p.pdf}|
\end{lstlisting}


\iftr
\subsection{Use Case 3: $k$-Clique Listing}\label{sec:kCliqueListing}
\else
\textbf{Use Case 3: $k$-Clique Listing}\label{sec:kCliqueListing}
\fi
GMS enabled us to enhance a state-of-the-art $k$-clique listing
algorithm~\cite{danisch2018listing}.
Our GMS formulation is shown in
Algorithm~\ref{lst:kcls}. We reformulated the original scheme (without changing
its time complexity) to expose the implicitly used set operations (e.g.,
Line~\ref{ln:kc-set-int}), to make the overall algorithm more modular.
The algorithm uses recursive backtracking. One starts with iterating
over edges (2-cliques), in Lines~\ref{ln:kc-edges-1}--\ref{ln:kc-edges-2}. In
each backtracking search step, the algorithm augments the considered cliques by
one vertex~$v$ and restricts the search to neighbors of~$v$ that come after $v$
in the used vertex order.

\all{As usual, the numbers indicate elements that are extensible. For example,
\includegraphics[scale=0.2,trim=0 16 0 0]{1.pdf} indicates that the user can
use their own graph representation (by default, we use CSR).}
Two schemes marked with~\includegraphics[scale=0.2,trim=0 16 0 0]{3.pdf}
indicate two preprocessing routines that appropriately reorder vertices and --
for the obtained order -- assign directions to the edges of the input
  graph~$G$. Both are well-known optimizations that reduce the search space
  size~\cite{danisch2018listing}. For such a modified~$G$, we denote
  \emph{out-neighbors} of any vertex~$u$ with $N^+(u)$.  Then, operations
  marked with \includegraphics[scale=0.2,trim=0 16 0 0]{5p.pdf} refer to
  accesses to the graph structure and different set operations that can be
  replaced with any implementation, as long as it preserves the semantics of
  set membership, set cardinality, and set intersection.

\begin{lstlisting}[float=h,label=lst:kcls,caption=\textmd{$k$-Clique Counting; see Listing~\ref{lst:apxcore} for the explanation of symbols.}]
|\vspace{0.5em}|/*|\textbf{Input:}| |\hlLIR{7em}{ A graph $G$ }{1.pdf}| , $k \in \mathbb{N}$ |\textbf{Output:}| Count of $k$-cliques $ck \in \mathbb{N}$. */

//|\ul{Preprocessing}|: reorder vertices with DGR or ADG; see |\cref{sec:design_deg_order}|.
//Here, we also record the actual ordering and denote it as $\eta$
  $(v_1, v_2, ..., v_{n}; \eta)$ = |\hlLIR{23em}{ preprocess$(V$, /* selected vertex order */$)$ }{3.pdf}|

//Construct a directed version of $G$ using $\eta$. This is an
|\vspace{0.25em}|//additional optimization to reduce the search space:
|\vspace{0.5em}|$G$ = |\hlLIR{5em}{ dir($G$) }{3.pdf}| //An edge goes from $v$ to $u\ \text{iff}\ \eta(v) < \eta(u)$
|\vspace{0.1em}|$ck$ = $0$ //We start with zero counted cliques.
|\label{ln:kc-edges-1}||\vspace{0.25em}|for |\hlLIR{4em}{ $u \in V$ in parallel \textbf{do:} }{2.pdf}| //Count u's neighboring $k$-cliques
|\label{ln:kc-edges-2}|  $C_2$ = $N^+(u)$; $ck$ += count(2, $G$, $C_2$)

|\vspace{0.15em}|function count($i$, $G$, $C_{i}$):
|\vspace{0.25em}|  if ($i$ == $k$): return |\hlLIR{4em}{ $\vert C_{k} \vert$ }{5p.pdf}| //Count $k$-cliques 
|\vspace{0.0em}|  else:
    ci = 0
|\label{ln:kc-set-int}|   |\vspace{0.25em}|for |\hlLIR{5em}{ $v \in C_{i}$ }{5p.pdf}| do: //search within neighborhood of v
|\vspace{0.25em}|      $C_{i+1}$ = |\hlLIR{7em}{ $N^+(v) \cap C_{i}$ }{5p.pdf}| // $C_i$ counts $i$-cliques. 
      $ci$ += count(i+1, $G$, $C_{i+1})$
    return ci
\end{lstlisting}

\begin{table*}[t]
\vspaceSQ{-1.25em}
\setlength{\tabcolsep}{0.8pt}
\renewcommand{\arraystretch}{2.4}
\centering
 \scriptsize
\sf
\begin{tabular}{lllllllllll}
\toprule
 & \makecell[c]{$k$-Clique Listing\\ \emph{Node Parallel}~\cite{danisch2018listing} } 
 & \makecell[c]{$k$-Clique Listing\\ \emph{Edge Parallel}~\cite{danisch2018listing} }
 & \makecell[c]{{\ding{72}} \textbf{$k$-Clique Listing} \\\textbf{with ADG (\cref{sec:kCliqueListing})}}  
 & \makecell[c]{ADG\\(Section~\ref{sec:hpc_algs})}
 & \makecell[c]{Max.~Cliques \\Eppstein et al.~\cite{DBLP:conf/isaac/EppsteinLS10} } 
 & \makecell[c]{Max.~Cliques \\Das et al.~\cite{das2018shared}} 
 & \makecell[c]{{\ding{72}} \textbf{Max.~Cliques} \\\textbf{with ADG (\cref{sec:theory-new-bounds})}}
 & \makecell[c]{Subgr.~Isomorphism\\ \emph{Node Parallel}~\cite{cordella2004sub,carletti2017introducing} } 
 & \makecell[c]{Link Prediction$^\text{\textdagger}$,\\ JP Clustering} \\
%
%
%
\midrule
\makecell[l]{\textbf{Work}} 
& $O\left(m k \left( \frac{d}{2}\right)^{k-2} \right)$ 
& $O\left(m k \left(\frac{d}{2}\right)^{k-2} \right)$  
& $O\left(mk\left(d+\frac{\epsilon}{2}\right)^{k-2}\right)$
& $O(m)$
& $O\left (d m 3^{d/3} \right )$ 
& $O\left(3^{n/3}\right)$
& $O\left(d m 3^{(2+\epsilon)d/3}\right)$
& $O\left(n \Delta^{k-1}\right)$
%
%
& $O(m\Delta)$ \\ 
\makecell[l]{\textbf{Depth}} 
& $O\left(n + k \left(\frac{d}{2}\right)^{k-1} \right)$ 
& $O\left(n + k \left(\frac{d}{2}\right)^{k-2} + d^2 \right)$
& $O\left( k\left(d+\frac{\epsilon}{2}\right)^{k-2} + \log^2 n + d^2\right)$
& $O\left(\log^2 n\right)$
& $O\left (d m 3^{d/3} \right)$
& $O\left(d \log n\right)$
& $O\left(\log^2 n + d \log n\right)$
& $O\left(\Delta^{k-1}\right)$ 
%
%
& $O(\Delta)$  \\

\makecell[l]{\textbf{Space}} 
& $O(n d^2 + K)$ 
& $O\left(m d^2 + K\right)$
& $O\left(m d^2 + K\right)$ 
& $O(m)$ 
& $O(m+nd + K)$
& $O(m + pd\Delta + K)$
& $O(m + pd\Delta + K)$
& $O(m+nk + K)$
%
%
& $O(m\Delta)$  \\ 
\bottomrule
\end{tabular}

\caption{\textmd{\textbf{Work, depth, and space for some graph mining
algorithms in GMS}. $d$ is the graph degeneracy, $K$ is the output size, 
$\Delta$ is the maximum degree, $p$ is the number of processors, $k$ is the number of vertices in the graph
that we are mining for, $n$ is the number of vertices in the graph that we are mining, and $m$ is the
number of edges in that graph. $^\text{\textdagger}$ Link prediction and the JP clustering complexities are valid for the Jaccard,
Overlap, Adamic Adar, Resource Allocation, and Common Neighbors vertex
similarity measures.
{\ding{72}}Algorithms derived in this work.
\tr{Additional bounds for BK are in Table~\ref{tab:bounds_more}}
%
%
%
}}
\label{tab:theory-table}
\vspaceSQ{-3.5em}
\end{table*}

The modular design and using set algebra enables us to easily experiment with
different implementations of $C_i$, $N^+(u) \cap C_i$, and others.  For
example, we successfully and rapidly redesigned the reordering scheme,
reducing the number of pointer chasing and the total amounts of communicated
data.  We investigated the generated assembly code of the respective part; it
has 22 x86 \texttt{mov} instructions, compared to \texttt{31} before the design
enhancement\footnote{We used ``compiler explorer'' (\url{https://godbolt.org/})
for assembly analysis\scriptsize}.
Moreover, we improved the memory consumption of the algorithm. The
space allocated per subgraph~$C_i$ (e.g., \includegraphics[scale=0.2,trim=0 16
0 0]{5p.pdf}) is now upper bounded by $|C_i|^2$ (counted in vertices)
instead of the default $\Delta^2$.
\ifall
the squared number of nodes in the
subgraph rather than the squared max degree inside the entire graph. 
\fi
When parallelizing over edges, this significantly reduces the required memory
(for large maximum degrees~$\Delta$, even up to $>$90\%).
\ifall\maciej{make sure by default it's really not that bad}
Example: h-hud: with maximal node degree around 1e5 -> requires appr 1TB of RAM
-> runs now with less than 64GB.
\fi
Finally, the modular approach taken by the GMS platform enables more concise
(and thus less complex) algorithm formulation. Specifically, the original
version had to use a separate routine for listing cliques for $k=3$, while
the GMS's reformulation enables all variants for $k \ge 3$. 

\ifall
\yannick{Other: Finding abstract description of algorithm allowes a single
implementation that covers all cases from 3 up to any clique size. (In danischs
case: 3 cliques don't work in default edgeparallel version, I had to add some
code to make it work). 2 and 1 cliques are treated special because its just the
number of edges, nodes respectively.} As a second example, we noticed that the
memory consumption when iterating over the edges can be improved. We exchanged
the auxiliary structure that contains the vertices and neighbours and provide
reading-access to them with a new structure. This allows - depending on the
input graph - to reduce the required memory by a huge factor. We could for
instance count $k$-cliques in h-hud with less than 64 GB of RAM, where before
there was almost 1 TB needed.
\fi

\ifall

\maciej{Greg, could you describe any implementation and design optimizations
that you used? In this section, we focus more on "how we make the code and the
algorithm fast", while in Section~4, we focused more on "this is the problem
and the algorithm that we use"}

\fi

\ifall

\subsection{Community Detection}

\maciej{Andreas, Dimitrios, Foteini, and Athina, could you describe any
implementation and design optimizations that you used? In this section, we
focus more on "how we make the code and the algorithm fast", while in
Section~4, we focused more on "this is the problem and the algorithm that we
use"}

\subsection{Graph Coloring}

\maciej{Emanuel, Alain, Roger, Severin, could you describe any implementation
and design optimizations that you used? In this section, we focus more on "how
we make the code and the algorithm fast", while in Section~4, we focused more
on "this is the problem and the algorithm that we use"}

\subsection{Minimum Cuts}

\maciej{Lukas, could you describe key implementation and design optimizations
from the mincut paper? In this section, we focus more on "how we make the code
and the algorithm fast", while in Section~4, we focused more on "this is the
problem and the algorithm that we use"}

\fi


\iftr
\vspaceSQ{-0.2em}
\subsection{Use Case 4: Subgraph Isomorphism}
\label{sec:hpc_iso}

GMS ensured speedups in the most recent parallel variant of the VF3 subgraph
isomorphism algorithm~\cite{carletti2018vf3, carletti2019parallel}.
Here, the GMS platform facilitates plugging in arbitrary variants of algorithms
without having to modify other parts of the toolchain
(\includegraphics[scale=0.2,trim=0 16 0 0]{4.pdf} --
\includegraphics[scale=0.2,trim=0 16 0 0]{5.pdf}) (Listing is in the extended
report).
First, example used optimizations in the baseline are \emph{work splitting}
combined with \emph{work stealing}. Specifically, threads receive lists of
vertices from which they start recursive backtracking. However, due to diverse
graph structure, this search can take a variable amount of time (because there
is more backtracking for some vertices) so some threads finish early. To combat
this, we use a lockfree queue, where idling threads steal work from other
threads. The queue element is the ID of a vertex from where to begin
backtracking. The thread performs a compare-and-swap (CAS) atomic to retrieve a
vertex from its queue. Idle threads select threads (that they steal from)
uniformly at random.  We also use a \emph{precompute} scheme: during runtime,
we gather information about possible mappings between vertices with their
neighborhoods, and certain specific query graphs. This can accelerate, for
example, searching through certain parts of the target graph.
\else

\textbf{Use Case 4: Subgraph Isomorphism}
We use GMS to enhance 
a recent parallel VF3 subgraph
isomorphism algorithm~\cite{carletti2018vf3, carletti2019parallel}.
Details are in the technical report, the key techniques are
work splitting combined with work stealing.

\fi

\ifall
Moreover, we also use \emph{precomputation}. During runtime, we check for a possible vertex
mapping if the previously matched neighbors of that vertex are feasible. The list
of previously matched neighbors is computed at runtime. However, since we check
the query nodes in a specific order (which is same heuristic used in the
baseline~\cite{carletti2018vf3, carletti2019parallel}) we can also precompute
for every query vertex what neighbors are already matched by the time it is
checked.
\fi

\ifall
By default, VF3 different split. The backtracking starts searching from every
node in the target graph. The way the work is split is changed by giving every
thread a list of starting nodes where each can independently do a backtracking
search. However, due to different graph structures this search can take a
variable amount of time (because some nodes have to search more), so some
threads finish early.
VF3 work stealing. To combat this, a global lockless queue is implemented,
where idling threads steal work from other threads. The queue containes one
integer per node, which is each node in the target graph from where to begin
backtracking. The thread performs a compare-and-swap (CAS) to retrieve a node
from its queue. Threads that run out of work have a uniformly at random chosen
order of other threads where they try to steal work from.
VF3 precomputation. During run-time we check for a possible node mapping if the
previously matched neighbors of that node are feasible. The list of previously
matched neighbors is computed at runtime. However, since we check the query
nodes in a specific order (which is same heuristic used in the sequential
version from [6]) we can also precompute for every query node what neighbors
are already matched by the time it is checked.
\fi

\iftr

\begin{figure*}[b]
\centering
\includegraphics[width=0.9\textwidth]{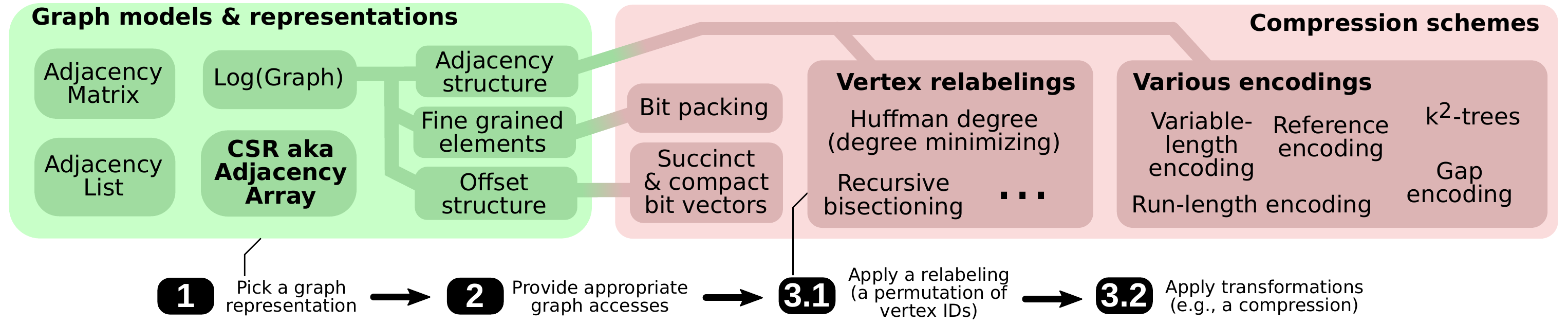}
\caption{\textmd{\textbf{Selected storage schemes (graph models, representations,
and graph compression methods) considered in the GMS platform}.} All
the schemes are outlined and pictured in more detail in
Figure~\ref{fig:reps-dets} (in the Appendix) and described in detail in a
recent survey~\cite{besta2018survey}. Developing and using a specific representation
in GMS corresponds to steps~\protect\includegraphics[scale=0.12,trim=0 16 0
0]{1.pdf} -- \protect\includegraphics[scale=0.12,trim=0 16 0 0]{3.pdf} in the
pipelined GMS design (cf.~Figure~\ref{fig:design}).} 
%
\label{fig:schemes_reps}
\end{figure*}

\fi

%

\iftr

\subsection{Use Case 5: Vertex Similarity \& Clustering}

We include vertex similarity and clustering in GMS. Vertex
similarity measures heavily use~$\cap$. For example, the well-known
Jaccard and overlap similarities of $u,v \in V$ are defined as $\frac{|N(u)
\cap N(v)|}{|N(u) \cup N(v)|}$ and $\frac{|N(u) \cap N(v)|}{\min(|N(u)|, |N(v)|)}$.
We provide a modular implementation in the GMS platform
(\includegraphics[scale=0.2,trim=0 16 0 0]{5p.pdf}), where one can use
different set representations (bitvectors, compressed bitvectors, integer
arrays, others) and two different routines for $\cap$: (1) simple
merging of sorted sets (taking $O(|N(v)| + |N(u)|)$ time) and a ``galloping'' variant
where, for each element~$x$ from a smaller set~$N(v)$, one uses binary search to check
if $x \in N(u)$ (taking $O(|N(v)| \log |N(u)|)$ time). This
enables fine tuning performance. 

\fi

\iftr

\subsection{Use Case 6: $k$-Clique-Star Listing}

A
$k$-clique-star is a $k$-clique with additional neighboring vertices that are
connected to all the vertices in the clique. $k$-clique-stars were proposed as
graph motifs that relax the restrictive nature of
$k$-cliques~\cite{jabbour2018pushing} (large cliques are expected to be rare
because \emph{every} vertex in a clique, regardless of the clique size, must be
connected to all other vertices in this clique). 
%
%
Our observation is that those extra vertices that are connected to the
$k$-clique actually form a $(k+1)$-clique (together with this $k$-clique). Thus,
to find $k$-clique-stars, we first mine $(k+1)$-cliques. Then, we find
$k$-clique-stars within each $(k+1)$-clique using set union, membership, and
difference.

\fi

\iftr

\subsection{Use Case 7: Link Prediction}

Here, one is interested in developing schemes for predicting whether two
non-adjacent vertices can become connected in the future. There exist many
schemes for such prediction~\cite{liben2007link, lu2011link, al2006link,
taskar2004link} and for \textbf{assessing the accuracy} of a specific link prediction
scheme~\cite{wang2014robustness}.
We start with some graph with \emph{known} links (edges). We derive $E_{sparse}
\subseteq E$, which is $E$ with random links removed; $E_{sparse} = E \setminus
E_{rndm}$.  $E_{rndm} \subseteq E$ are randomly selected \emph{missing} links
from $E$ (\emph{links to be predicted}).  We have $E_{sparse} \cup E_{rndm} =
E$ and $E_{sparse} \cap E_{rndm} = \emptyset$.
Now, we apply the link prediction scheme~$S$ (that we want to test) to each
edge $e \in (V \times V) \setminus E_{sparse}$. The higher a value $S(e)$, the
more probable $e$ is to appear in the future (according to $S$).  Now, the
effectiveness $eff$ of $S$ is computed by verifying how many of the edges with
highest prediction scores ($E_{predict}$) actually are present in the original
dataset~$E$: $eff = |E_{predict} \cap E_{rndm}|$.

\fi

\all{
\vspaceSQ{-0.2em}
\subsection{Other Use Cases}

We used GMS to experiment with other algorithms, for example we developed
different variants of Bron-Kerbosch for maximal cliques.  We 
detail these designs in the report and we provide most important results in the
Evaluation.
}

\ifall
\maciej{fix text}

\textbf{Bron-Kerbosch} is a recursive backtracking algorithm that heavily uses
different set operations. The main recursive function \texttt{BKPivot} has
three arguments that are dynamic sets containing vertices. $R$ is a partially
constructed, non-maximal clique~$c$, $P$ are candidate vertices that \emph{may}
belong to $c$ but are yet to be tried, and $X$ are vertices that definitely do
\emph{not} belong to $c$.  The algorithm recursively calls \texttt{BKPivot} for
each new candidate vertex, checks if this gives a clique, and updates
accordingly $P$ and $X$.
Several optimizations significantly reduce the search space of potential
cliques~\cite{cazals2008note, DBLP:conf/isaac/EppsteinLS10,
DBLP:journals/tcs/TomitaTT06}. First, the set of candidates is $P \setminus
N(u)$ instead of $P$, where $u$ is a vertex selected from $P \cup X$. Second,
the outermost loop iterates over $V$ using the degeneracy order and uses it to
prune $P$ and $X$, involving two additional set intersections in each
iteration.

\maciej{Zur, could you check if all the established optimizations that you use
are described above, and if not, describe them briefly? Thanks!}
\zur{The text above sums it quite nicely up, I'll add my addition down below, s.t. you can merge it easily}
In order to make Tomita et al. pivot selection fast, one can reduce the the graph into a subgraph.
Each recursive call limits the set of vertices in the graph to $P \cup X$ which are composed of the neighborhood of a chosen pivot $v$.
The next pivot $u$ is chosen s.t. it maximizes $\text{argmax}\;{|P \cap N(u)|}$. Thus for any recursive call we can shrink a given Graph $G$ to $G_{P, X}$.
By means, we keep only the vertices in $P \cup X$ as well all edges which are either in the cut of $P$ and $X$ or completely inside $P$.
This way successive intersection operations are thus cheaper to execute \cite{DBLP:conf/isaac/EppsteinLS10}.
\zur{It is too expensive to compute the subgraph in each recursive iteration. I address th problem in the section down below}

\maciej{Zur, could you describe any implementation and design optimizations
that you used? In this section, we focus more on "how we make the code and
the algorithm fast", while in Section~4, we focused more on "this is the
problem and the algorithm that we use"}

We provide multiple variants.
\zur{Maciej, I'll sum up all my implementations in subsubsection}
\subsubsection*{Caching/streaming}
After reordering the vertices according a degeneracy order, the function BkPivot is launched over a loop.
The parameter $P$ and $X$ are computed for each vertex $v$ by intersecting the already visited vertices in degeneracy order with $X$
and intersecting all the following vertices with $P$. Using a sequential algorithm the set of visited, resp. the following vertices
can be build iteratively, by adding, resp. removing from the sets after each iteration.  
This loop dependency burdens the possibility to parallelize. 
The obvious way, is to form the visited set as well as the remaining set fresh at each iteration. 
But for graphs with a lot of nodes, this could get the bottleneck, especially if the graph is sparse.
\zur{Claim is based on oberservations}
For sparse graph, it may be more efficient to construct the sets the other way around by iterating over each $u \in N(v)$
and adding $u$ to visited when ever $ord(u) < ord(v)$ and to remaining otherwise ($ord$ depicts any pre processing order, e.g. degeneracy ordering).
For $avg_{v\in V}(deg(v)) << |V|$ this is much more efficient.
\zur{Claim is based on oberservations}
\zur{Idea is from Apruba et. al which in turn got it from PECO }
To have a general way of dealing with this problem we provide a caching/streaming scheme to benefit from precomputed Sets.
Therefore we specify a chunksize and in that chunk we build the sets iteratively using one or two threads and consuming 
the prebuild sets as soon as they a ready with the other threads. Since Building the sets iteratively is reasonably fast and the recursive calls 
takes longer to execute, there should be only high contention and wait time in the beginning of a chunk. 
As for the division of the working sets in chunks, it holds: The higher the cache Size, the more work can be distributed over threads
as well as less synchronization overhead at the end. But It also burdens the memory and may introduce bad locality.
\zur{Claim is based on Assumptions, should we measure/proove this in depth?}
\zur{Maciej, see the chat with greg for more details and pseudocode}
Our test set consist of sparse graphs which makes this scheme nevertheless clearly inferior to Aprubas et. al variation.  
Therefore we reuse it for all the following variations
\subsubsection*{Subgraph}
We use a Subgraph class which has the same interface as our regular class graph but holds a hashmap internally to connect the
original label (in the original graph) with the new label index in the build subgraph. Therefore the function operating on the graphs may be agnostic 
about the graph type. After a subgraph was created for a recursive call, one can iterate through the vertices in the subgraph and choosing 
$\text{argmax}\{\max_{w \in X}deg(w), \max{w\in P}|P\cap N(w)|\}$. After finding such vertex in the subgraph, the original label has to be retrieved by
a scan of the mapping. This can optionally be further reduced by caching the original vertex label together with the number of outgoing edges to $P$.
But of course this also means more memory consumption.
\zur{See my implementation of RoaringSubGraph and FastRoaringSubGraph}
The pivot finding task ist the heaviest task in a recursive call using $O(|V|)$ intersections.
The use of subgraphing reduces the intersection complexity successiv recursive calls, as it scales down the neighbourhood of each 
vertex and thus the set size that is used in an intersection. But building the Subgraph introduces also some overhead.
The Idea of using a subgraph was discussed by Eppstein. Apurba et al. uses Subgraphing only in the outermost loop.  
Additionally, we have a variation, where a subgraph is also build in the first layer of recursive calls in any case 
and in the second layer only if it the operation reduces the number of vertices by more than a specified factor (e.g. say 0.9).
It could be interessting to analyse when a downscaling could be beneficial.
Further one could study ways to optimize the downscaling process.
\zur{Greg had a good comment on mattermost which could be nice to consider}

\maciej{Important - if any of your optimizations made this algorithm faster
than the competition, underline it here!}
\zur{Following is more or less a comment for Maciej}
Compared with apurba et al. provided c++ implementation, mine is worse on orkut, but is the same on wikitalk and as-skitter.
He uses nested parallelism based on tbb. I would have to switch to another set implementation to be able to do the same.
I haven't investigated on how they manage oversubscription, which will happen along the recursive call tree.

\fi

\iftr

\subsection{Developing Graph Representations}

The right data layout is one of key enablers of high performance.
There exists a plethora of graph representations, layouts, models, and
compression schemes~\cite{besta2018survey}.
Different compression schemes may vastly differ in the compression ratio as
well as the performance of accessing and mining a graph. For example, some
graphs compressed with a combination of techniques implemented in the WebGraph
framework~\cite{boldi2004webgraph} can use even below one bit per link. Yet,
decompression overheads may significantly impact the performance of graph
mining algorithms running on such compressed graphs. Then, a recent compressed
Log(Graph) representation can deliver 20-35\% space reductions with simple bit
packing, while eliminating decompression overheads or even \emph{delivering
speedups due to reduced amounts of transferred data}~\cite{besta2018log}.
Besides graph compression, there exist many other schemes related to
representations, for example NUMA-awareness in graph
storage~\cite{zhang2015numa}; they all impact performance of graph processing.

We consider the aspect of data layout and graph representation design in GMS
and we enable the user to analyze \emph{relationships between graph storage and
the performance of graph algorithms}. Specifically, the user can rapidly
develop or use an existing storage scheme and analyze its space utilization and
its impact on the performance of graph algorithms and graph queries. The
considered storage schemes are illustrated in Figure~\ref{fig:schemes_reps},
they include {\textbf{graph models and representations}}
(e.g., Log(Graph)~\cite{besta2018log}), and {\textbf{graph
compression schemes}} (e.g., difference encoding~\cite{besta2018survey},
$k^2$-trees~\cite{brisaboa2009k}, bit packing~\cite{besta2018survey}, or
succinct offsets~\cite{besta2018survey, gog2014theory}). All these schemes
offer different trade-offs between the required storage space and the delivered
performance. The current version of GMS implements many of these schemes, but
it also offers an intuitive and extensive interface that facilitates
constructing new ones.

We provide more details of these storage schemes, and how to use them, in the
Appendix (Section~\ref{sec:app-maze}). In general, in GMS, one first selects a
model to be used (by default, it is the Adjacency List Model) and its specific
implementation. GMS uses a simple Compressed Sparse Row (CSR) by default.
Other available schemes include Log(Graph) with its bit packing of vertex IDs
or succinct and compact offsets to neighborhoods~\cite{besta2018log}. 
Then, one must provide the implementation of graph accesses (fetching neighbors
or a given vertex, checking the degree of a given vertex, verifying if a given
edge exists). 
After that, one may apply additional preprocessing.  First, one
can relabel vertex IDs (i.e., apply a permutation of vertex IDs), for example
the Huffman degree relabeling.  Second, one may provide a transformation of
each (permuted) neighborhood, for example encoding neighborhoods using Varint
compression.

Further details on the permutations and transformations of vertex neighborhoods
can also be found in the Log(Graph) paper~\cite{besta2018log}.

\fi

\section{CONCURRENCY ANALYSIS}
\label{sec:concurrent}

\enlargeSQ

\ifall\maciej{Prev, with some stuff that I removed}
\begin{table*}[t]
\vspaceSQ{-1.25em}
\setlength{\tabcolsep}{3pt}
\renewcommand{\arraystretch}{1.3}
\centering
 \scriptsize
\sf
\begin{tabular}{llllllllll}
\toprule
 & \makecell[c]{$k$-Clique Listing\\ \emph{Edge Parallel}~\cite{danisch2018listing} }
 & \makecell[c]{$k$-Clique Listing\\ \emph{Node Parallel}~\cite{danisch2018listing} } 
 & \makecell[c]{$k$-Clique Listing\\ \emph{Edge Parallel + $\approx$DO}}  
 & \makecell[c]{\emph{Approximate Degeneracy}\\ Order ($\approx$DO) ~\cite{dhulipala2017julienne,DBLP:journals/pvldb/BahmaniKV12}}
 & \makecell[c]{Maximal Cliques \\ \emph{Subgraphing}~\cite{DBLP:conf/isaac/EppsteinLS10} } 
 & \makecell[c]{ Subgraph Isomorphism \\ \emph{Node Parallel Backtracking} } 
 & \makecell[c]{Link\\Prediction$^\text{\textdagger}$} 
 &  \makecell[c]{Link\\Prediction$^\text{\textsection}$} 
 & \makecell[c]{Jarvis-Patrick\\Clustering} \\ 
\midrule
\makecell[l]{\textbf{Work}} 
& $O\left(m k( \frac{c}{2})^{k-2} \right)$ 
& $O\left(m k (\frac{c}{2})^{k-2} \right)$  
& $\Omega\left(mk(c+\frac{\epsilon}{2})^{k-2}\right)$
& $O(m)$
& $O\left (c d n 3^{c/3} \right )$ 
& $O(nd^{k-1})$
& $O(m d)$ 
& $O(n^2 + md)$ 
& $O(md)$ \\ 
\makecell[l]{\textbf{Depth}} 
& $O\left(n + k (\frac{c}{2})^{k-2} \right)$ 
& $O\left(n + k (\frac{c}{2})^{k-1} \right)$
& $\Theta\left( k(c+\frac{\epsilon}{2})^{k-2} + \log^2 n\right)$
& $O(\log^2 n)$
& $O\left (c d 3^{c/3} \right)$
& $O(d^{k-1})$ 
& $O(d)$
& $O(d)$ 
& $O(d)$  \\

\makecell[l]{\textbf{Space}} 
& $O(m c^2 + K)$ 
& $O(m c + K)$
& $O(m c^2 + K)$ 
& $O(m)$ 
& $O(m+nc + K)$
& $O(m+nk + K)$
& $O(md)$
& $O(n^2 + md)$ 
& $O(md)$  \\ 
\bottomrule
\end{tabular}

\caption{
Work, depth, and space for some graph mining algorithms in GMS. $c$ is the
graph degeneracy, $K$ is the output size, $d$ is the maximum degree, $n$ is
the number of vertices, and $m$ is the number of edges.  Link prediction complexities are valid for the
following vertex similarity measures: 
$^\text{\textdagger}$Jaccard, Overlap, Adamic Adar, Resource Allocation, Common Neighbors;
%
%
$^\text{\textsection}$Preferential Attachment~\protect\cite{leicht2006vertex, neo4j_sim}.
``$\approx$DO'' means the approximate degeneracy ordering.
}
\label{tab:theory-table}
\vspaceSQ{-3.5em}
\end{table*}
\fi


%
In this part of GMS, we show how to assess a priori the properties of parallel
graph mining algorithms, \emph{reducing time spent on algorithm design and
development} and providing \emph{performance insights that are portable across
machines that differ in certain ways (e.g., in the sizes of their caches) and
independent of various implementation details}. 
\tr{We first broadly discuss the
approach and the associated trade-offs. Second, as use cases, we pick
$k$-clique and maximal clique listing, and we
\emph{enhance state-of-the-art algorithms addressing these problems}.}
\tr{
\Cref{tab:theory-table} summarizes the GMS theoretical results; {many of these bounds are novel}.
}

\subsection{Methodology, Models, Tools}

We use the established \emph{work-depth analysis} for bounding run-times of parallel algorithms.
Here, the total number of instructions performed by an algorithm
(over all number of processors for a given input size) is the \emph{work} of
the algorithm. The longest
chain of sequential dependencies (for a given input size) is the \emph{depth}
of an algorithm~\cite{Bilardi2011, blelloch2010parallel}.  This approach is
used in most recent formal analyses of parallel algorithms in the shared-memory
setting~\cite{DBLP:conf/spaa/HasenplaughKSL14, dhulipala2018theoretically}. 
Overall, we consider \emph{four} aspects of a parallel algorithm: (1) the
overhead compared to a sequential counterpart, quantified with
work, (2) the scalability, which is illustrated by depth, (3) the space usage,
and -- when applicable -- (4) the approximation ratio. 
\tr{These four aspects often enable different tradeoffs.}

\all{In the following, we will focus on $k$-clique listing to illustrate in
practice how one can achieve and reason about tradeoffs between performance,
space, and approximation.}

\subsection{Discussion On Trade-Offs}

%
For many problems, there is a \textbf{tradeoff between work, depth, space}, and
sometimes \textbf{approximation ratio}~\cite{miller2015improved, karger1996new,
danisch2018listing}. Which algorithm is the best choice hence depends on the
available number of processors and the available main memory. For today's
shared memory machines, typically the number of processors/cores is relatively
small (e.g., 18 on our machines) and main memory is not much bigger than the
graphs we would like to process (e.g., 64GiB or 768GiB on our machines, see
Section~\ref{sec:eval}). Thus, \emph{reducing work (and maintaining close
to linear space in the input plus output)} is a high priority to obtain good
performance in practice~\cite{dhulipala2018theoretically}.

An algorithm with a work that is much larger than the best sequential algorithm
will require many processors to be faster than the latter. An algorithm with
large depth will stop scaling for a small number of processors.
An estimate of the runtime of an algorithm with work $W$ and depth $D$ on $p$
processors is $W/p+D$. This estimate is optimistic as it neglects the cost for
scheduling threads and caching issues (e.g., false sharing). Yet, it has proven
a useful model in developing efficient graph algorithms in
practice~\cite{dhulipala2018theoretically}.

%
The {space} used by a parallel algorithm limits the largest problem that can be
solved on a fixed machine. This is crucial for graph mining
problems with exponential time complexities where we want the \emph{space to be
close to the input size plus the output size}.


We illustrate a work / depth / space tradeoff with $k$-clique
listing~\cite{danisch2018listing} (\cref{sec:kCliqueListing}). All following
designs are pareto-optimal in terms of the work / depth / space tradeoff and
they are useful in different circumstances (for different machines).
\ifconf
\else

\fi
First, consider a \textbf{naive} algorithm variant. Starting from every vertex,
one spawns parallel recursive searches to complete the current clique.
The advantage of this approach is that is has low depth $O(k)$, but the work
and space is $\Theta(n \Delta^{k-1})$, which can be
prohibitive.

This approach can be enhanced by using the DGR order to guide the search
as described in~\cref{sec:kCliqueListing} (the ``\textbf{Node Parallel}'' variant).
Here, one invokes a parallel search starting from each vertex for cliques that
contain this vertex as the first vertex in the order. This reduces the space to
almost linear $\Theta(n d^2)$, where $d$ is the degeneracy of the
graph. The depth is increased to $\Theta(n + k (d/2)^{k-1})$.  This design was
reported to have poor scalability in practice~\cite{danisch2018listing}. 

One can also invoke a parallel search for every \emph{edge} (``\textbf{Edge
Parallel}'')  and try to find a clique that contains it (and follows
the DGR order). The depth decreases by a factor of $d$ to $\Theta(n +
k (d/2)^{k-2} + d^2)$, but the space increases by a factor of $\frac{m}{n}$ to $O(m d^2)$. This
approach has a good work / depth / space tradeoff in
practice~\cite{danisch2018listing}.

\marginpar{\vspace{1em}\colorbox{yellow}{\textbf{R-2}}}

\subsection{Bounds for Graph Mining Algorithms}
\label{sec:theory-new-bounds}

\marginpar{\vspace{5em}\colorbox{yellow}{\textbf{R-2}}}

\hl{Table~\mbox{\ref{tab:theory-table}} presents work-depth and space bounds for considered graph
mining algorithms. 
%
%
Here, we obtain \emph{new better} bounds for maximal
clique listing. The main idea is to combine existing corresponding
algorithms~\mbox{\cite{eppstein2011listing, das2018shared}}
with the ADG ordering. 
Specifically, the new maximal clique listing improves upon the
Eppstein et al.~\mbox{\cite{eppstein2011listing}} and Das et
al.~\mbox{\cite{das2018shared}}: our depth is better than both while work is
better than that of~\mbox{\cite{eppstein2011listing}} and adds only a small
factor to work in~\mbox{\cite{eppstein2011listing}}.}
\cnf{We provide detailed proofs in the technical report.}
We also provide a new \mbox{$k$}-clique listing variant, again by using ADG.
The variant scales better than Danisch et
al.~\mbox{\cite{danisch2018listing}} (column~2) if \mbox{$n$} is much bigger
than \mbox{$k d^{k-2}$}. This variant matches a recent scheme by Shi et al.~\cite{shi2020parallel},
which uses a similar approach.


\subsection{Improving $k$-Clique Listing} 

Finally, one can use the \textbf{approximate degeneracy order} (ADG,
cf.~\cref{sec:design_deg_order}) instead of DGR, which results in \ul{new performance bounds}.
Proceed as for the Edge Parallel variant, but use the
$(2+\epsilon)$-approximate parallel degeneracy order. This is easy to implement in
the GMS benchmarking platform, as all one has to do is to change the
preprocessing reordering routine from DGR to ADG. The depth becomes
$\Theta(k (d+\frac{\epsilon}{2})^{k-2} + \log^2 n)$ and the work is increased
to $\Omega(m k (d+\frac{\epsilon}{2})^{k-2})$. This design scales better if 
\tr{$n$ is much bigger than $k d^{k-2}$}
\cnf{$n \gg k d^{k-2}$}
and outperforms other variants in practice
(see Section~\ref{sec:eval}).

\subsection{Improving Maximal Clique Listing}

We now analyze our parallel maximal clique listing algorithm (cf.~\cref{sec:design_max_clique})
\all{It offers \emph{strong theoretical bounds} on work and depth, and is
\emph{high-performance} and \emph{scalable} in practice.}
The key idea is to use ADG, the \emph{relaxation} of the strict degeneracy
order when processing vertices iteratively in the highest level of recursion in
the BK algorithm. 
As in $k$-cliques, this is easy to implement with the GMS platform.
This improves upon the Eppstein et
al.~\cite{eppstein2011listing} (BK-E) and Das et al.~\cite{das2018shared}
(BK-DAS): our depth is better than both while work is better than that of
BK-DAS and adds only a small factor to the work amount in BK-E.



\all{\subsubsection{Analysis \& Results}
\ }
For constant degeneracy graphs, our algorithm has linear work and
poly-logarithmic depth, see Table~\ref{tab:theory-table} for a comparison with
previous work. We note that, for many classes of sparse graphs, such as
scale-free networks~\cite{barabasi1999scaleFree} and 
planar graphs~\cite{DBLP:conf/isaac/ZhouN94}, $\Delta \gg d$.
Moreover, we often also have $\log n
\ll \Delta$. Thus, the depth of BK-ADG is in such cases lower than that of
BK-DAS. 

We provide our bounds for the case where \emph{nested parallelism} is employed. 
If only the outer loop which launches the calls to BK-Pivot and the construction
of the arguments to BK-Pivot is parallelized, 
the depth is still $O( (d\Delta)^{(2+\epsilon)d/3})$ and the space is only $O(m + nd + K)$ (where $K$ is the output size).

\tr{\begin{table}[h]
\renewcommand{\arraystretch}{1.7}
  \centering
  \small
  \tabcolsep=1pt
  \begin{tabular}{lccc} %
    \toprule
    & Work & Depth \\
    \midrule
    Chiba and Nishizeki~\cite{chiba1985arboricity} & $O\left( d^2 n (n-d) 3^{d/3} \right)$ & $O\left( d^2 n (n-d) 3^{d/3} \right)$. \\
    Chiba and Nishizeki~\cite{chiba1985arboricity} & $O\left( n d ^{d+1} \right)$ & $O\left( n d ^{d+1} \right)$. \\
    Chrobak and Eppstein~\cite{chrobak1991planar} & $O\left( n d^2 2^d \right)$ & $O\left( n d^2 2^d \right)$. \\
    Eppstein et al.~\cite{eppstein2011listing} & $O\left(d m \ 3^{ \frac{d}{3}}\right)$ & $O\left(d m \ 3^{ \frac{d}{ 3}}\right)$. \\
    Das et al.~\cite{das2018shared} & $O\left(3^{ \frac{n }{ 3}}\right)$ & $O\left(d \log n\right)$. \\
    \emph{\textbf{This Paper}} & $O\left(d m \ 3^{ \frac{(2+\epsilon) d }{ 3}}\right)$ & $O\left(\log^2 n + d \log n\right)$.
  \\
    \bottomrule
  \end{tabular}
  \caption{Additional bounds for enumerating all maximal cliques.}
  \label{tab:bounds_more}
\end{table}}

%

We first state the cost of computing the ADG order (cf.~\cref{sec:design_deg_order}), which is the key
difference to the algorithm by Das et al.~\cite{das2018shared}.
\vspaceSQ{-0.3em}
\begin{lemma}\label{lem:apx}
Computing a $(2+\epsilon)$-approximate degeneracy order takes $O(m)$ work and
$O(\log^2 n)$ depth, for any constant $\epsilon$.
\end{lemma}
\vspaceSQ{-0.3em}

\all{\begin{proof}
Every iteration removes at least a constant $\frac{1}{1+\epsilon}$ fraction of
the vertices (because at most a $\frac{1}{1+\epsilon}$ fraction of vertices can
have a degree larger than $(1+\epsilon)$ times the average degree). Hence,
$O(\log n)$ iterations suffice. Each iteration can be implemented with $O(\log
n)$ depth to find the average degree and update the representation of the
graph. Each iteration takes work proportional to the number of edges which have
at least one endpoint in the removed vertices. Hence, this is $O(m)$ over all
iterations.

The average degree is at most twice as big as the degeneracy (because we can
repeatedly remove a degree $d$ vertex until no vertices are left, meaning the
sum of the vertex degrees is at most $nd$). Therefore, the algorithm computes a
$2(1+\epsilon)$-approximate degeneracy order.
\end{proof}}

Eppstein gave a generic work bound for an invocation of BK-Pivot($P$, $v_i$,
$X$) that we can use for our setting.

\vspaceSQ{-0.3em}
\begin{lemma}[Eppstein~\cite{eppstein2011listing}]\label{lem:bk-pivot-work}
Excluding the work to report
the found maximal cliques, BK-Pivot($P$, $v_i$, $X$) takes $O( (d |X|) 3^{ |P| / 3})$ work.  
\end{lemma}
\vspaceSQ{-0.3em}

We combine Eppstein's bound with the bounds on BK-Pivot and ADG to obtain work
and depth bounds for BK-ADG.

\vspaceSQ{-0.3em}
\begin{lemma}
\sloppy
Finding all maximal cliques with BK-ADG takes $O(d m \ 3^{
  {(2+\epsilon) d}/{ 3}})$ work and $O(\log^2 n + d \log
  n)$ depth.
\end{lemma}
\vspaceSQ{-0.3em}

\vspaceSQ{-0.3em}
\begin{proof}

\tr{We first sketch the used {parallel compute primitives}.
\emph{Intersecting} two sets $A$ and $B$ takes $O(|A| \ |B|)$ work and $O(1)$
depth. Performing a \emph{Reduction} over an array of $n$ values (for example
to compute their sum) takes $O(n)$ work and $O(\log n)$ depth. }

Computing ADG 
takes $O(m)$ work and $O(\log ^2 n)$ depth; see Lemma~\ref{lem:apx}.
Next, for all invocations of BK-Pivot, $|P| \leq (2+\epsilon) d$, by 
the properties of the ADG order.
Moreover, the size of the set $|X|$ in the invocation of BK-Pivot($P$, $v_i$, $X$)
is at most $\Delta(v_i)$. Hence, by using Lemma~\ref{lem:bk-pivot-work} for 
each invocation of BK-Pivot, we conclude that the work is $O(d m \ 3^{ (2+\epsilon) d / 3})$, 
excluding the cost to output the maximal cliques.
Because there are  $(n-d)3^{d/3}$ maximal cliques~\cite{eppstein2011listing}, the work
 is not dominated by the cost to report the maximal cliques.

The depth of BK-Pivot is $O(M \log n)$, where $M$ is the size of the maximum
clique~\cite{das2018shared}. As the size of the largest clique is bounded by
the degeneracy (i.e., $M < d$), this is $O(d\log n)$. All the calls to BK-Pivot
from BK-ADG can be launched simultaneously. 
%
%
%
\end{proof}


\ifall

\subsection{Improving Subgraph Isomorphism}

To test if a graph of $k$ vertices is isomorphic to a subgraph of a graph of $n$ vertices,
 no work bound better than the trivial $O(n \Delta^{k-1})$ is known. In practice, 
backtracking can reduce the work in many cases~\cite{cordella2004sub,carletti2017introducing}, 
but introduces highly imbalances search trees. The order in which the search visits 
the vertices becomes crucial for performance and many
 heuristics have been proposed~\cite{cordella2004sub}. 



\fi

\ifall

\paragraph{Case Study: $k$-clique listing} We illustrate a work / depth / space
tradeoff on the example of $k$-clique listing. Designs 2-4 are variants of
Danish et al.'s algorithm~\cite{danisch2018listing} (see also \Cref{sec:kcls}).
These designs are all pareto-optimal in terms of the work / depth / space
tradeoff and be useful in different circumstances (for different machines).

\emph{Design 1. "Naive"} Starting from every vertex, spawn parallel recursive
searches in order to complete the current clique. The advantage of this
approach is that is has low depth $O(k)$, but the work and space is $\Theta(n
d^k)$, which can be prohibitive.

\emph{Design 2. "Node Parallel".} Use the degeneracy order to guide the search
as described in \Cref{sec:kcls}. Invoke a parallel search starting from each
node for cliques that contains this node as the first vertex in the order. This
reduces the space to almost linear $\Theta(n c^2))=O(m c)$, where $c$ is the
degeneracy of the graph. The depth is increased to $\Theta(n + k (c/2)^{k-1})$.
This design was reported to have poor scalability in
practice~\cite{danisch2018listing}. 

\emph{Design 3: "Edge Parallel".} Instead of parallelizing only over the
vertices, invoke a parallel search for every edge and try to find a clique that
contains it (and follows the degeneracy order). The depth is decreased by a
factor of $c$ to $\Theta(n + k (c/2)^{k-2})$, but space is increased by a
factor $c$ to $O(m c^2)$. This approach represents a good work / depth / space
tradeoff in practice~\cite{danisch2018listing}.

\emph{Design 4: "Approximate Degeneracy".} Proceed as for the Edge Parallel
variant, but use the (2+$\epsilon$) approximate parallel degeneracy order (See
\cref{sec:spec_reorder}, \cref{sec:design_deg_order}, and
App.~\ref{sec:app_impl}). This changes the depth to $\Theta( k c^{k-2})$ and
the work is increased to $\Omega(m k c^{k-2})$. This design scales better if
$n$ is much bigger than $k c^{k-2}$.

\fi

\ifall
\begin{table*}
\vspaceSQ{-1.25em}
\centering
\scriptsize
\sf
\setlength{\tabcolsep}{1pt}
\renewcommand{\arraystretch}{0.5}
\begin{tabular}{lllrrrrrrrrrl@{}}
\toprule
\textbf{Type} & 
\textbf{Graph and its source} $\dagger$ & 
\textbf{ID} &
\#vertices $n$ & 
\#edges $m$ & 
\makecell[c]{\#edges $m$\\(symmetrized)} &
\makecell[c]{sparsity\\${m}/{n}$} & 
\makecell[c]{Maximum\\in-degree} &
\makecell[c]{Maximum\\out-degree} &
\makecell[c]{$D$} & 
\#triangles $T$ & 
\makecell[c]{{$T$ per}\\{vertex}} &
\makecell[c]{Remarks (special features, etc.)} \\
\midrule
\multirow{12}{*}{\makecell[c]{Social\\networks\\(friend-\\ships)}} 
%
 & (K) Friendster & s-frs & 64,185,720 & 2,147,483,645 & 1,584,590,812 & 24.7 & 3124 & 5214 & 38 & {3,554,133,311} & 55.4 & \makecell[l]{Common. Relatively large $D$} \\
 & (K) Orkut & s-ork & 3,072,441 & 117,184,899 & *117,184,899 & 38.1 & 33,313 & 33,313 & 10 & {627,577,371} & 204.3 &Common  \\
 & (W) LiveJournal & s-ljn & 5,363,260 & 99,028,542 & 79,023,142  & 18.4 & 19,409 & 2,469 & $^H$7.36 & 285,730,264 & 53.2 & Common  \\
 & (K) Flickr & s-flc & 2,302,925 & 33,140,017 & 22,838,276 & 9.9& 21,001 & 26,367 & 23 & {837,605,842} & 363.7 & Large $T$ but low $m/n$. \\
 & (K) Pokec & s-pok & 1,632,803 & 30,622,564 & 22,301,964 & 13.7 & 13,733 & 8,763 & 14 & {32,557,458} & 19.9 & Low $T$ \\
 & (K) Libimseti.cz & s-lib & 220,970 & 17,359,346 & 17,233,144 & 78 & 33,389 & 25,042 & 6 & {69,115,603} & 312.8 & Low $D$ and large $m/n$ \\
 & (K) Catster/Dogster & s-cds & 623,766 & 15,699,276 & *15,699,276 & 25.2 & 80,637 & 80,637 & 15 & {656,390,451} & 1052.3 & Large $T$ \\
 & (K) Youtube & s-you & 3,223,589 & 9,375,374 & *9,375,374 & 2.9 & 91,751 & 91,751 & 31 & {12,226,580} & 3.8 & \makecell[l]{Common; very low $m/n$ and $T$} \\
 & (K) Flixster & s-flx & 2,523,386 & 7,918,801 & *7,918,801 & 3.1 & 1,474 & 1,474 & 8 & {7,897,122} & 3.1 & Very low $m/n$ and $T$ \\
 & (K) Livemocha & s-lmc & 104,103 & 2,193,083 & *2,193,083 & 21.1 & 2,980 & 2,980 & 6 &  3,361,651 & 32.3  & \makecell[l]{Very similar to v-flr, yet vastly different \\4-clique counts (4,359,646 in s-lmc)} \\
%
%
%
%
\midrule
\multirow{12}{*}{\makecell[c]{Hyperlink\\graphs}} 
 & (C) Web Data Commons 2012 & h-wd2 &  3,563,602,789 & 225,840,663,232 & 128,736,914,167  & 63.4 & $\approx$95,000,000 & $\approx$56,000 & 331 & 9,648,842,110,027 & 2707.6 & \makecell[l]{Very large $m, n, m/n, D, T$} \\
 & (C) Web Data Commons 2014 & h-wd4 &  1,724,573,718 & 124,141,874,032 &  64,422,807,961  & 72 & $\approx$45,000,000 & $\approx$32,000 & 207 & 4,587,563,913,535 & 2660.1 & \makecell[l]{Very large $m, n, m/n, D, T$} \\
 \iftr
 & (W) EU domains (2015) & h-deu & 1,070,557,254 & 91,792,261,600 &  & & 20,252,239 & 35,340 & $^H$14.18 & & \\
 & (W) UK domains (2014) & h-duk & 787,801,471 & 47,614,527,250  & & & 8,605,490 & 16,635 & $^H$24.63 & & \\
\fi
 & (W) ClueWeb12 & h-clu & 978,408,098 & 74,774,358,622 & 42,574,107,469 & 76.4 & 75,611,690 & 7,447 & $^H$15.71 &  1,995,295,290,765 & 2039.3 & Large $m/n$ and $T$ \\
 & (W) GSH domains (2015) & h-dgh & 988,490,691 & 33,274,090,228 & 33,877,399,152  & 33.7 & 58,860,299 & 32,114 & $^H$14.91 & 1,784,146,861,411 & 1804.9 & Large $T$ \\
 \iftr
 & (W) SK domains (2005) & h-dsk & 50,636,154 & 1,949,412,601 & 1,810,063,330 & & 8,563,808 & 12,870 & $^H$17.56 & & \\
 & (W) IT domains (2004) & h-dit & 41,291,594 & 1,150,725,436 & & & 1,326,745 & 9,964 & $^H$19.16 & & \\
 & (W) Arabic domains (2005) & h-dar & 22,744,080 & 639,999,458 & & & 575,618 & 9,905 & $^H$22.39 & & \\
\fi
 & (K) Wikipedia/DBpedia (en) & h-wdb & 12,150,976 & 378,142,420 & 288,257,813 & 23.7 &963,032 & 963,032 & 10 & {11,686,212,734} & 961.8 & Relatively low $m/n$ but high $T$ \\
 \iftr
 & (W) Indochina domains (2004) & h-din & 7,414,866 & 194,109,311 & & & 256,425 & 6,985 & $^H$28.12 & & \\
\fi
 & (K) Wikipedia (en) & h-wen & 18,268,992 & 172,183,984 & 126,890,209 & 6.9 & 632,558 & 632,558 & 12 & {328,743,411} & 18.0 & Common \\
 & \makecell[l]{(K) Wikipedia (it)} & h-wit & 1,865,965 & 91,555,008 & 68,022,541 & 36.5 & 237,151 & 237,151 & 10 & {3,139,806,683} & 1682.7 & Very large $T$ \\
 & (K) Hudong & h-hud & 2,452,715 & 18,854,882 & 18,690,759 & 7.6 & 204,277 & 21 & 108 & {12,028,700} & 4.9 & Low $T$ \\
 & (K) Baidu & h-bai & 2,141,300 & 17,794,839 & 17,014,946 & 7.9 & 97,950 & 2,596 & 20 & {25,207,196} & 11.8& Low $m/n$ \\
 & (K) DBpedia & h-dbp & 3,966,924 & 13,820,853 & 12,610,982 & 3.2 & 472,799 & 1,006 & 67 & {8,329,548} & 2.1 & Very low $m/n$ and $T$ \\
\midrule
\multirow{1}{*}{\makecell[c]{Commu-\\nication}} 
 & (K) Twitter follows & m-twt & 52,579,682 & 1,963,263,821 & 1,614,106,187 & 30.7 & 3,503,656 & 779,958 & 18 & {55,428,217,664} & 1054.2& Very large $T$ \\
 \iftr
 & (S) Stack Overflow interactions & m-stk & 2,601,977 & 63,497,050 & & & & & & & \\
 \fi
 & (K) Wikipedia talk (en) & m-wta & 2,394,385 & 5,021,410 & 4,659,565 & 1.9 & 3,311 & 100,022 & 11 & {9,203,519} & 3.8 & Very low $m/n$ and $T$ \\
\midrule
\multirow{2}{*}{\makecell[c]{Collabo-\\ration}} 
\iftr
\fi
 & (K) DBLP co-authorship & l-dbl &  1,314,050  &  18,986,618  &  10,724,828 & 8.16 &   5,260 &   5,260 & 24 & {12,184,090} & 9.27 & Low $m/n$ and $T$ \\
 & (S) Citation network (patents) & l-cit & 3,774,768 & 16,518,948 & 16,518,947 & 4.36 & 793 & 793 & 22 & {7,515,023} & 2.0 & Very low $m/n$ and $T$ \\
%
\midrule
 & (N) Chebyshev4 & & 68.1K & 5.3M & 5.3M & 77.8 & 68.1K & 68.1K & & 444.6M & 6.5K & Very large $T$ \emph{and} $T/n$, $T_{max} = 5.8\text{M}$, $C_{lb} = 33$ \\
 & (N) Bio-CE-PG Gene functional associations & & 1.9K & 47.8K & 47.8K & 25.1 & 913 & 913 & & 2.4M & 1.3K & Very large $T$ \emph{and} $T/n$ \emph{and} relatively low $m/n$, $T_{max} = 29.5\text{k}$, $C_{lb} = 28$ \\
 & (N) Copresence-INVS15, economic & & 219 & 1.3M & 1.3M & 6K & 39.9K & 39.9K & & 18.3B & 83.7M & \emph{Extremely} large $T$, $T/n$, $m/n$, $T_{max} = 392.4\text{M}$, $C_{lb} = 66$ \\
 & (N) econ-beacxc economic & & 497 & 50K & 50K & 100 & 497 & 497 & & 9.1M & 18.3K & Very large $T$ \emph{and} $T/n$ \\ 
 & (N) edit-frwiki, wikipedia & & 94.3K & 5.7M & 5.7M & 60.4 & 107.3K & 107.3K & & 834.6M & 8.9K & \\
 & (N) \\
\midrule
\multirow{5}{*}{\makecell[c]{Various}} 
\iftr
\fi
 & (K) Wikipedia evolution (de) & v-ewk & 2,166,669 & 86,337,879 & 39,705,981 & 18.3 & 175,906 & 218,465 & 10 & {169,876,249} & 78.4 & Time-aware structure \\
\iftr
 & (??) EU road network & v-eur & & & & & & & & & & The largest available road network \\
\fi
 & (D) USA road network & v-usa & 23,947,347 & 58,333,344 & 28,854,312 & 1.2 & 9 & 9 & & $\approx$1,300,000 &$\approx$0.1 & Very low $m/n$ and $T$ \\
 & (K) Internet topology (Skitter) & v-skt & 1,696,415 & 11,095,298 & *11,095,298 & 6.5 & 35,455 & 35,455 & 31 & {28,769,868} & 17.0 & Low $m/n$ and $T$ \\
%
 & (K) Flickr (photo relations) & v-flr & 105,938 & 2,316,948 & *2,316,948 & 21.9 & 5,425 & 5,425 & 9 & 107,987,357 & 1019.3 & \makecell[l]{Very similar to s-lmc, yet vastly different \\4-clique counts (9,578,965,096 in v-flr)} \\
\bottomrule
\end{tabular}
\caption{\textmd{The characteristics of various publicly available real-world
graphs. Graphs are sorted by $m$ in each class separately. ``*'' indicates that
the graph is undirected and thus the number of edges remains identical after
symmetrization. ``$^H$'' indicates the harmonic diameter. (W), (S), (K), (D),
and (C) refer to the publicly available datasets, explained
in~\cref{sec:eval_m}. For more details on the selection of the graphs,
see~\cref{sec:spec_datasets}.
\maciej{Zur - we should probably add some graphs from the Network Repository,
that we used because of their clique counts... Any suggestions? :)}
}}
\vspaceSQ{-2em}
\label{tab:graphs}
\vspaceSQ{-1.5em}
\end{table*}
\fi

\begin{figure*}[t]
\vspaceSQ{-1.25em}
\centering
\includegraphics[width=1.0\textwidth]{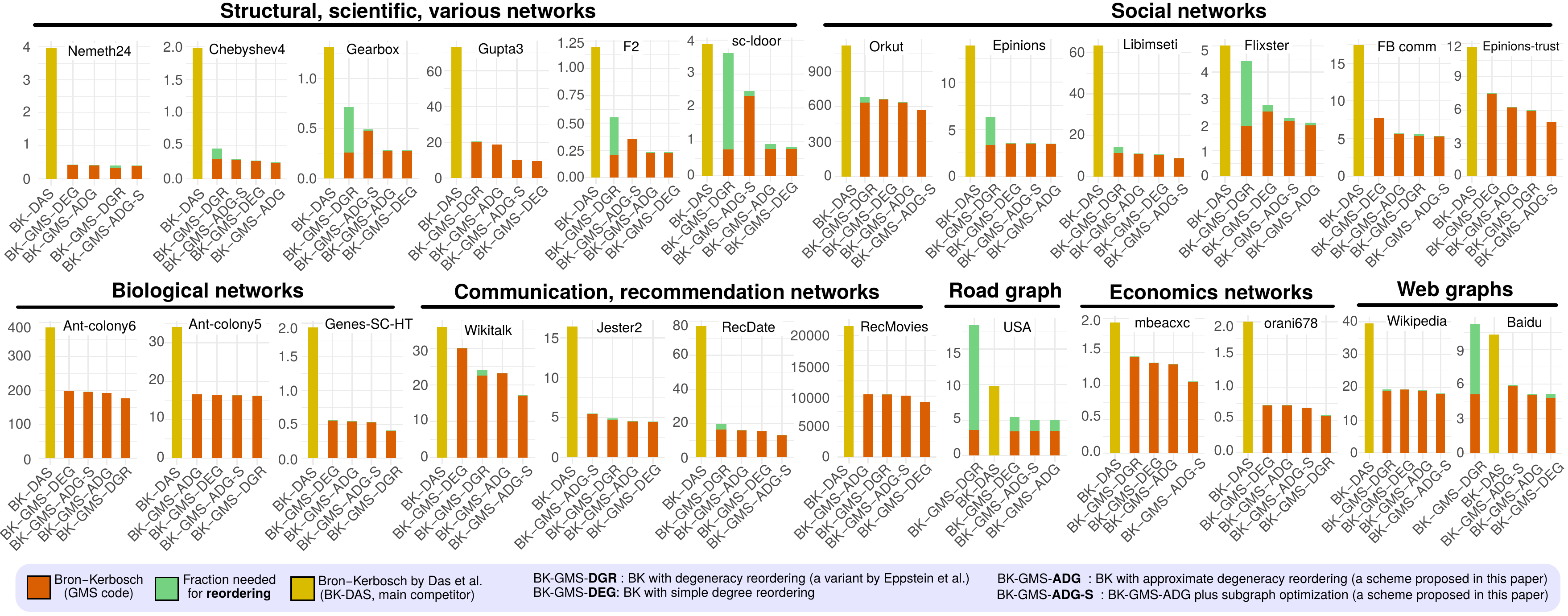}
\vspaceSQ{-2.25em}
\caption{\textmd{\textbf{Speedups of the parallel GMS BK algorithm} over a
state-of-the-art implementation by Das et al.~\protect\cite{das2018shared}
(BK-DAS) and a recent algorithm by Eppstein et
al.~\protect\cite{DBLP:conf/isaac/EppsteinLS10} (BK-GMS-DGR). System: Daint.}}
\label{fig:bk-faster}
\vspaceSQ{-1em}
\end{figure*}


\section{EVALUATION}
\label{sec:eval}
\vspaceSQ{-0.2em}


We describe how GMS facilitates performance analysis of various aspects
of graph mining, and accelerates the state of the art.
\tr{\emph{We focus on accelerating the four core mining problems from
Section~\ref{sec:hpc_algs}}.}
%

\vspaceSQ{-0.3em}
\subsection{Datasets, Methodology, Architectures}
\label{sec:eval_m}

We first sketch the {evaluation methodology}.
{For measurements, we omit the first 1\% of performance data as warmup. We derive
enough data for the mean and 95\% non-parametric confidence intervals.
We use arithmetic means as summaries.}

\all{
\subsubsection{Measurements}
\ 
}
\all{To ensure \emph{interpretability} and \emph{reproducibility}, we prescribe
using the ``12 rules for scientific benchmarking of parallel
codes''~\cite{hoefler2015scientific}.
The former \emph{``enables scientists to understand the experiment, draw own
conclusions, assess their certainty, and possibly generalize
results''}~\cite{hoefler2015scientific}. The latter ensures that there is
enough information to \emph{reproduce} the outcomes of experiments.
The considered set of rules prescribes a methodology for obtaining performance
results that are meaningful. We employ the most important rules in GMS, such as
``when reporting speedups, clearly specify the comparison baseline and provide
absolute runtime of this baseline''. Following other rules, in our
measurements,}
\all{
We omit the first 1\% of performance data as warmup. We derive
enough data for the mean and 95\% non-parametric confidence intervals.
We use arithmetic means as summaries.
}
\all{More details on methodology are found in the associated
publication~\cite{hoefler2015scientific}.}

\vspaceSQ{-0.3em}
\subsubsection{Datasets}
\ 
{We consider {SNAP (S)}~\cite{snapnets}, {KONECT
(K)}~\cite{kunegis2013konect}, {DIMACS (D)}~\cite{demetrescu2009shortest},
Network Repository (N)~\cite{nr}, and {WebGraph (W)}~\cite{BoVWFI} datasets.
As explained in~\cref{sec:spec_datasets}, for flexibility, we do not fix
specific datasets.  Instead, we illustrate a wide selection of public datasets
in Table~\ref{tab:graphs}, arguing \emph{which parameters make them useful or
challenging}.  Details of these parameters are in~\cref{sec:spec_datasets}.}
\all{As explained in~\cref{sec:spec_datasets}, we do not prescribe specific datasets
(1) for flexibility, (2) because the datasets themselves evolve and (3) the
compute and memory capacities of architectures grow continually, making it
impractical to stick to a fixed-sized dataset.
Instead, we illustrate a wide selection of public datasets in
Table~\ref{tab:graphs}, arguing their usefulness. To ensure that our discussion
is useful for a future line of works, we select and describe large graphs with
$m > 10$M from {SNAP (S)}~\cite{snapnets}, {KONECT
(K)}~\cite{kunegis2013konect}, {DIMACS (D)}~\cite{demetrescu2009shortest},
Network Repository (N)~\cite{nr}, {Web Data Commons (C)}~\cite{wdc}, and
{WebGraph (W)}~\cite{BoVWFI} datasets.}
\all{If there are multiple related graphs, for example different snapshots of
the .eu domain, we select the largest one. One exception is the additional
Italian Wikipedia snapshot selected due to its interestingly high density.}
\all{Our selection covers graphs with different properties with respect to $n$,
$m$, and -- especially -- $m/n$, which results in different performance
properties of virtually all the algorithms in Table~\ref{tab:problems}.}

\ifconf\begin{table}[b]\fi
\iftr\begin{table*}[t]\fi
\vspaceSQ{-1.25em}
\centering
\scriptsize
\ifsq\ssmall\fi
\iftr\footnotesize\fi
\sf
\ifsq\setlength{\tabcolsep}{1pt}\fi
\ifsq
\renewcommand{\arraystretch}{0.5}
\else
\renewcommand{\arraystretch}{1.1}
\fi
\begin{tabular}{lrrrrrrrl@{}}
\toprule
\textbf{Graph} $\dagger$ & 
$n$ & 
$m$ & 
\makecell[c]{$\frac{m}{n}$} & 
\makecell[c]{$\widehat{d_{i}}$} &
\makecell[c]{$\widehat{d_{o}}$} &
$T$ & 
\makecell[c]{{$\frac{T}{n}$}} &
\makecell[c]{\textbf{Why selected/special?}} \\
\midrule
%
%
\textbf{[so]} (K) Orkut & 3M & 117M & 38.1 & 33.3k & 33.3k & 628M & 204.3 & Common, relatively large \\
\textbf{[so]} (K) Flickr & 2.3M & 22.8M & 9.9 & 21k & 26.3k & 838M & 363.7 & Large $T$ but low $m/n$. \\
\textbf{[so]} (K) Libimseti & 221k & 17.2M & 78 & 33.3k & 25k & 69M & 312.8 & Large $m/n$ \\
\textbf{[so]} (K) Youtube & 3.2M & 9.3M & 2.9 & 91.7k & 91.7k & 12.2M & 3.8 & \makecell[l]{Very low $m/n$ and $T$} \\
\textbf{[so]} (K) Flixster & 2.5M & 7.91M & 3.1 & 1.4k & 1.4k & 7.89M & 3.1 & Very low $m/n$ and $T$ \\
\textbf{[so]} (K) Livemocha & 104k & 2.19M & 21.1 & 2.98k & 2.98k & 3.36M & 32.3  & \makecell[l]{Similar to Flickr, but\\ a lot fewer 4-cliques (4.36M)} \\
\textbf{[so]} (N) Ep-trust & 132k & 841k & 6 & 3.6k & 3.6k & 27.9M & 212 & Huge $T$-skew ($\widehat{T} = 108\text{k}$) \\
\textbf{[so]} (N) FB comm. & 35.1k & 1.5M & 41.5 & 8.2k & 8.2k & 36.4M & 1k & Large $T$-skew ($\widehat{T} = 159\text{k}$) \\
%
%
\textbf{[wb]} (K) DBpedia & 12.1M & 288M & 23.7 & 963k & 963k & 11.68B & 961.8 & Rather low $m/n$ but high $T$ \\
\textbf{[wb]} (K) Wikipedia & 18.2M & 127M & 6.9 & 632k & 632k & 328M & 18.0 & Common, very sparse \\
\textbf{[wb]} (K) Baidu & 2.14M & 17M & 7.9 & 97.9k & 2.5k & 25.2M & 11.8& Very sparse \\
\textbf{[wb]} (N) WikiEdit & 94.3k & 5.7M & 60.4 & 107k & 107k & 835M & 8.9k & Large $T$-skew ($\widehat{T} = 15.7\text{M}$) \\
%
%
%
%
%
%
%
%
\textbf{[st]} (N) Chebyshev4 & 68.1k & 5.3M & 77.8 & 68.1k & 68.1k & 445M & 6.5k & \makecell[l]{Very large $T$ \emph{and} $T/n$\\ \emph{and} $T$-skew ($\widehat{T} = 5.8\text{M}$)} \\
\textbf{[st]} (N) Gearbox & 154k & 4.5M & 29.2 & 98 & 98 & 141M & 915 & \makecell[l]{Low $\widehat{d}$ but large $T$;\\ low $T$-skew ($\widehat{T}=1.7\text{k}$)} \\
\textbf{[st]} (N) Nemeth25 & 10k & 751k & 75.1 & 192 & 192 & 87M & 9k & Huge $T$ \emph{but} low $\widehat{T} = 12\text{k}$ \\
\textbf{[st]} (N) F2 & 71.5k & 2.6M & 36.5 & 344 & 344 & 110M & 1.5k & Medium $T$-skew ($\widehat{T} = 9.6\text{k}$) \\
\textbf{[sc]} (N) Gupta3 & 16.8k & 4.7M & 280 & 14.7k & 14.7k & 696M & 41.5k & Huge $T$-skew ($\widehat{T}=1.5\text{M}$) \\
\textbf{[sc]} (N) ldoor & 952k & 20.8M & 21.5 & 76 & 76 & 567M & 595 & Very low $T$-skew ($\widehat{T} = 1.1\text{k}$) \\
\textbf{[re]} (N) MovieRec & 70.2k & 10M & 142.4 & 35.3k & 35.3k & 983M & 14k & Huge $T$ and $\widehat{T} = 4.9\text{M}$  \\
\textbf{[re]} (N) RecDate & 169k & 17.4M & 102.5 & 33.4k & 33.4k & 286M & 1.7k & Enormous $T$-skew ($\widehat{T} = 1.6\text{M}$) \\
\textbf{[bi]} (N) sc-ht (gene) & 2.1k & 63k & 30 & 472 & 472 & 4.2M & 2k & Large $T$-skew ($\widehat{T} = 27.7\text{k}$) \\
\textbf{[bi]} (N) AntColony6 & 164 & 10.3k & 62.8 & 157 & 157 & 1.1M & 6.6k & Very low $T$-skew ($\widehat{T} = 9.7\text{k}$) \\
\textbf{[bi]} (N) AntColony5 & 152 & 9.1k & 59.8 & 150 & 150 & 897k & 5.9k & Very low $T$-skew ($\widehat{T} = 8.8\text{k}$) \\
\textbf{[co]} (N) Jester2 & 50.7k & 1.7M & 33.5 & 50.8k & 50.8k & 127M & 2.5k & Enormous $T$-skew ($\widehat{T} = 2.3\text{M}$) \\
\makecell[l]{\textbf{[co]} (K) Flickr\\ (photo relations)} & 106k & 2.31M & 21.9 & 5.4k & 5.4k & 108M & 1019 & \makecell[l]{Similar to Livemocha, but\\ many more 4-cliques (9.58B)} \\
\textbf{[ec]} (N) mbeacxc & 492 & 49.5k & 100.5 & 679 & 679 & 9M & 18.2k & Large $T$, low $\widehat{T} = 77.7\text{k}$ \\
\textbf{[ec]} (N) orani678 & 2.5k & 89.9k & 35.5 & 1.7k & 1.7k & 8.7M & 3.4k & Large $T$, low $\widehat{T} = 80.8\text{k}$ \\
%
%
%
\ifall
\fi
\textbf{[ro]} (D) USA roads & 23.9M & 28.8M & 1.2 & 9 & 9 & 1.3M & 0.1 & Extremely low $m/n$ and $T$ \\
%
%
\bottomrule
\end{tabular}
\caption{\textmd{Some considered real-world
graphs. 
\textbf{\ul{Graph class/origin:}}
[so]: social network,
[wb]: web graph,
[st]: structural network,
[sc]: scientific computing,
[re]: recommendation network,
[bi]: biological network,
[co]: communication network,
[ec]: economics network,
[ro]: road graph.
\textbf{\ul{Structural features:}}
$m/n$: graph sparsity,
$\widehat{d_i}$: maximum in-degree,
$\widehat{d_o}$: maximum out-degree,
$T$: number of triangles,
$T/n$: average triangle count per vertex,
$T$-skew: a skew of triangle counts per vertex (i.e., the difference between
the smallest and the largest number of triangles per vertex). Here,
$\widehat{T}$ is the maximum number of triangles per vertex in a given graph.
\textbf{\ul{Dataset:}} (W), (S), (K), (D), (C), and (N) refer to the publicly available
datasets, explained in~\cref{sec:eval_m}. 
For more details, see~\cref{sec:spec_datasets}.
}}
\vspaceSQ{-2em}
\label{tab:graphs}
\ifconf\end{table}\fi
\iftr\end{table*}\fi

\vspaceSQ{-0.3em}
\subsubsection{Comparison Baselines}
\ 
For each considered graph mining problem, we compare different
GMS variants to \emph{the most optimized state-of-the-art
algorithms available}. We compare to the original existing
implementations. Details are stated in
the following sections.

\vspaceSQ{-0.3em}
\subsubsection{Parallelism}
\ 
Unless stated otherwise, we use \emph{full parallelism}, i.e., we run
algorithms on \emph{the maximum number of cores available on a given system}.
We also analyzed \emph{scaling} (how performance changes when varying number of
used cores), the results show consistent advantages of GMS variants over other
baselines.

\vspaceSQ{-0.3em}
\subsubsection{Architectures}
\ 
We used different systems for a \emph{broad evaluation and to analyze and
ensure performance portability} of our implementations.
First, we use an in-house Einstein and Euler servers.
Einstein is a Dell PowerEdge R910 with an Intel Xeon X7550 CPUs @ 2.00GHz with
18MB L3 cache, 1TiB RAM, and 32 cores per CPU (grouped in four sockets).
Euler has an HT-enabled Intel Xeon Gold 6150 CPUs @ 2.70GHz with 24.75MB L3
cache, 64 GiB RAM, and 36 cores per CPU (grouped in two sockets).
We also use servers from the CSCS supercomputing center, most importantly a
compute server with Intel Xeon Gold 6140 CPU @ 2.30GHz, 768 GiB RAM, 18 cores,
and 24.75MB L3. Finally, we also used XC50 compute nodes in the Piz Daint Cray
supercomputer (one such node comes with 12-core Intel Xeon E5-2690 HT-enabled
CPU 64 GiB RAM).

\vspaceSQ{-0.3em}
\subsection{Faster Maximal Clique Listing}

\emph{We start with our key result: GMS enabled us to outperform a
state-of-the-art fastest available algorithm for maximal clique listing} by Das et
al.~\cite{das2018shared} (BK-DAS) \emph{by nearly an order of magnitude}.
The results are in Figure~\ref{fig:bk-faster}. We compare BK-DAS with several
variants of BK developed in GMS as described in~\cref{sec:hpc_algs}.
BK-GMS-DGR uses the degeneracy order and is a variant of the Eppstein's
scheme~\cite{DBLP:conf/isaac/EppsteinLS10}, enhanced in
GMS.  BK-GMS-DEG uses the simple degree ordering. BK-GMS-ADG and BK-GMS-ADG-S
are two variants of a new BK algorithm proposed in this work, combining BK with
the ADG ordering; the latter also uses the subgraph caching optimization
(~\cref{sec:hpc_algs}).
We also compare to the original Eppstein scheme, it was always slower.
GMS also enabled us to experiment with Intel Thread Building Blocks 
vs.~OpenMP for threading in both the outermost loop and in inner loops (we
exploit nested parallelism), we only show the OpenMP variants as they
always outperform TBB.

Figure~\ref{fig:bk-faster} shows consistent speedups of GMS variants over BK-DAS.  We
could quickly deliver these speedups by being able to plug in different set
operations and optimizations in BK. Moreover, many plots show the large
preprocessing overhead when using DGR. It sometimes helps to reduce the actual
clique listing time (compared to ADG), but in most cases, ``ADG plus clique
listing'' are faster than ``DGR plus clique listing'': ADG is very fast and it
reduces the BK runtime to the level comparable to that achieved by DGR. This
confirms the theoretical predictions of the benefits of BK-GMS-ADG over
BK-GMS-DGR or BK-DAS.
Finally, the comparably high performance (for many graphs) of BK-GMS-ADG,
BK-GMS-ADG-S, and BK-GMS-DEG is due to the optimizations based on set algebra,
for example using fast \emph{and} compressed roaring bitmaps to implement
neighborhoods and auxiliary sets $P$, $X$, and $R$
(cf.~\cref{sec:hpc_algs}), which enables fast set operations heavily
used in BK.
\emph{Overall, BK-GMS is often faster than BK-DAS by $>$50\%, in some cases even
$>$9$\times$.}

We stress that the speedups of the implementations included in the GMS
benchmarking platform are consistent over many graphs of \emph{different
structural characteristics} (cf.~Table~\ref{tab:graphs}) that entail deeply
varying load balancing properties. For example, some graphs are \emph{very}
sparse, with virtually no cliques larger than triangles (e.g., the USA road
network) while others are \emph{relatively} sparse with \emph{many} triangles
(and higher cliques), with low or moderate skews in triangle counts per vertex
(e.g., Gearbox or F2). Finally, some graphs have \emph{large} or even
\emph{huge} skews in triangle counts per vertex (e.g., Gupta3 or RecDate),
which gives significant differences in the depths of the backtracking trees and
thus load imbalance.

We also derived the \textbf{algorithmic efficiency} results, i.e., the number
of maximal cliques found per second;
selected data is in Figure~\ref{fig:posterchild}. The
results follow the run-times; the GMS schemes consistently outperform
BK-DAS (the plots are in the technical report). These results show more
distinctively that BK-GMS finds maximal cliques consistently
better than BK-DAS, even if input graphs have vastly different
clustering properties. For example, BK-GMS-ADG outperforms BK-DAS for Gupta3
(huge $T$-skew), F2 (medium $T$-skew), and ldoor (low $T$-skew).

\vspaceSQ{-0.3em}
\subsection{Faster $k$-Clique Listing}

\emph{GMS also enabled us to accelerate a very recent $k$-clique listing
algorithm~\cite{danisch2018listing}}, see Figure~\ref{fig:k-clique-faster}.
\ifall
Overall, GMS's modular design allows scientists and programmers to quickly
develop and test different configurations and combinations of methods,
algorithms, and others.
\fi
We were able to rapidly experiment with different variants, such as node
parallel and edge parallel schemes, described in~\cref{sec:kCliqueListing} and
in Section~\ref{sec:concurrent}. Our optimizations
from~\cref{sec:kCliqueListing} (e.g., a memory-efficient layout of $C_i$)
ensure consistent speedups of up to 10\% for different parameters (e.g., clique
size~$k$), input graphs, and reordering routines.  Additionally, we show that
using the ADG order brings further speedups over DEG or DGR.
\ifall
Specifically, we observed effects caused by the tradeoff between performance
and accuracy, described in detail in Section~\ref{sec:concurrent}.
\fi

\ifall
The generated plots and data allow to accelerate
graph mining with relatively little effort. 
This is shown here exemplary on
$k$-clique Listing, which was tested using different parallezation strategies
and preprocessing methods: When inspecting the speedup of $k$-Clique Listing,
one can observe different results: 

(1) We observe an almost perfect linear speedup, as has also been predicted and
reported by Danisch et al. \cite{danisch2018listing}. \yannick{Eg: s-frs,
s-ork, h-wdg, v-skt (edgeparallel)} Depending on the machine, the speedup's
slope places between $0.6$ and $1$. \yannick{see orkut cscs vs orkut einstein}
This good speedup is easily explainable when noting that the code spends only a
tiny fraction of its runtime in building the subgraphs on which the
computations happen. Thus, the overhead introduced by the parallelization is
minimal and linear speedup is the consequence.

(2) We observe that parallelization over the nodes of the graph has a limit on
its speedup, whereas the parallelization over the edges still provides an
almost perfect linear speedup. This phenomen was also addressed by Danisch et
al. \cite{danisch2018listing} and attributed to the parallelization over the
edges yielding smaller tasks, such that all cpus can be equally utilized until
the graph is completely processed.\yannick{see v-skt} This phenomena is
dependent on the geometry of the processed graph. 

(3) We see very different runtimes between the parallelization strategies over
the edges and over the nodes.\yannick{see h-hud} This is a phenomena which was
not reported by Danisch et al. and has its origin in the structure of the
underlying graphs. In the h-hud graph for instance, which exhibits this
behaviour, a small percentage of the nodes have a very high degree (ranging in
$~10^5$). And while the overall graph is still sparse, this concentration of
high-degree nodes has the consequence that the code spends a huge part of its
runtime in building subgraphs, and its runtime behaviour is therefore almost
completely determined by it. Apart from cases like h-hud, the runtime of the
two parallelization strategies on a single core are of comparable size.

These three cases show that the parallelization over the edges is superior or
equal to the parallelization over the nodes, except for case (3). But case (3)
can be avoided with some basic knowledge of the graph's statistics.
\fi

\ifall
(3) We see very different runtimes between the parallelization strategies over
the edges and over the nodes.\yannick{see h-hud} This is a phenomena which was
not reported by Danisch et al. and has its origin in the structure of the
underlying graphs. In the h-hud graph for instance, which exhibits this
behaviour, a small percentage of the nodes have a very high degree (ranging in
$~10^5$). And while the overall graph is still sparse, this concentration of
high-degree nodes has the consequence that the code spends a huge part of its
runtime in building subgraphs, and its runtime behaviour is therefore almost
completely determined by it. Apart from cases like h-hud, the runtime of the
two parallelization strategies on a single core are of comparable size.
\fi

\ifall
\macb{Insights \& Summary}
When comparing our implementation to the reference implementation, we usually
see that for small graphs, ours is slightly slower. This is most likely
introduced through overhead introduced by object oriented aspects and aspects
of C++ (which matters, seeing that the reference implementation is plain C
code). But for larger graphs our implementation regularly outperforms the
reference. This is owed to the graph structures handling their own memory as
described in section \ref{sec:kCliqueListing} and leads to runtime gains as
high as 14\% on the s-ork graph when counting 8 cliques in parallel. In
sequential the gain in runtime is almost 18\%. When swapping the preprocessing
method, the runtime gain rises to 25\% (resp. 34\%).

\fi

\begin{figure}[t]
\vspaceSQ{-1.25em}
\centering
\includegraphics[width=1.0\columnwidth]{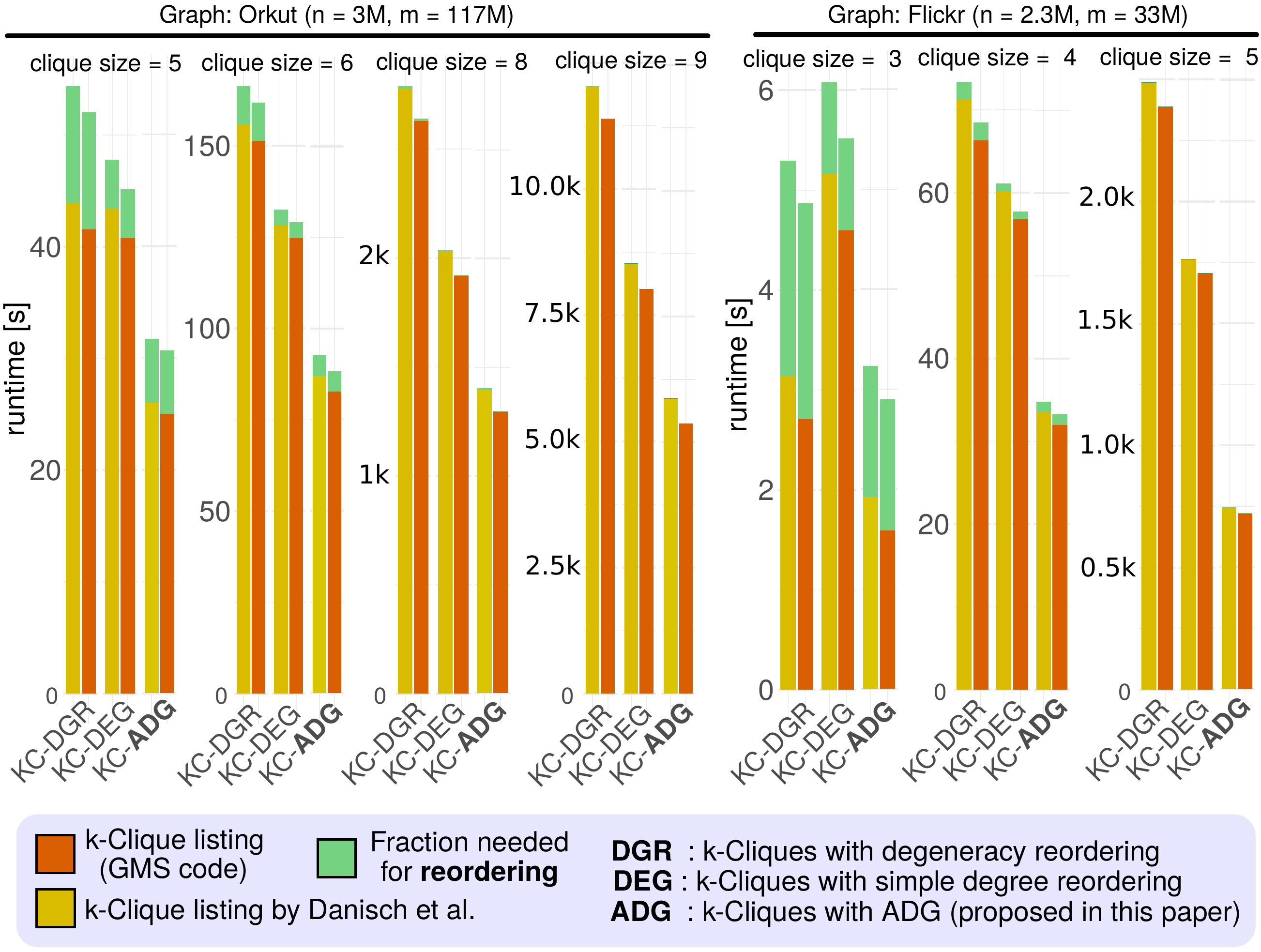}
\vspaceSQ{-2.25em}
\caption{\textmd{\textbf{Speedups of (1) the GMS implementation of $k$-clique
listing} over a state-of-the-art algorithm~\protect\cite{danisch2018listing},
and \textbf{(2) of the ADG reordering in $k$-clique listing} over DGR/DEG.
System: Daint.}} \label{fig:k-clique-faster}
\vspaceSQ{-1.25em}
\end{figure}

\vspaceSQ{-0.3em}
\subsection{Faster Degeneracy Reordering and $k$-Cores}
\label{sec:eval_deg}

We also analyze in more detail the performance of different reordering routines
(DEG, DGR, and ADG) and their impact on graph mining algorithms in GMS
(cf.~\cref{sec:hpc_algs}). We also show their impact on the run-time of
BK maximal clique listing by Eppstein et
al.~\cite{DBLP:conf/isaac/EppsteinLS10} (BK-E). The results are in
Figure~\ref{fig:results_reorder}.
ADG, due to its beneficial scalability properties
(cf.~Section~\ref{sec:concurrent}), \emph{outperforms the exact DGR}. At the
same time, it \emph{similarly} reduces the runtime of
BK-E~\cite{DBLP:conf/isaac/EppsteinLS10} (cf.~leftmost and rightmost bars).
The $2+\epsilon$ approximation ratio has mild influence on performance.
Specifically, the lower $\epsilon$ is,
the more (mild) speedup is observed. This is because
larger $\epsilon$ enables more parallelism, but then less
accurate degeneracy ordering may incur more work when listing cliques.
Moreover, ADG combined with BK-E cumulatively \emph{outperforms the simple DEG
reordering}: the latter is also fast, but its impact on the Bron-Kerbosch
run-time is lower, ultimately failing to provide comparable speedups. 
\tr{We note that the results of BK+DGR being slower than BK+DEG are consistent with
independent results in a recent BK paper~\cite{das2018shared}. This
additionally highlights our insight that using ADG over DGR and DEG is the
fastest of all three variants.}
We were able to rapidly experiment with different reorderings as
-- \emph{thanks to GMS's modularity} -- we could seamlessly integrate them with
BK-E~\cite{DBLP:conf/isaac/EppsteinLS10}.

\begin{figure}
    \centering
    \begin{minipage}{0.32\columnwidth}
        \centering
        \includegraphics[width=1.0\textwidth]{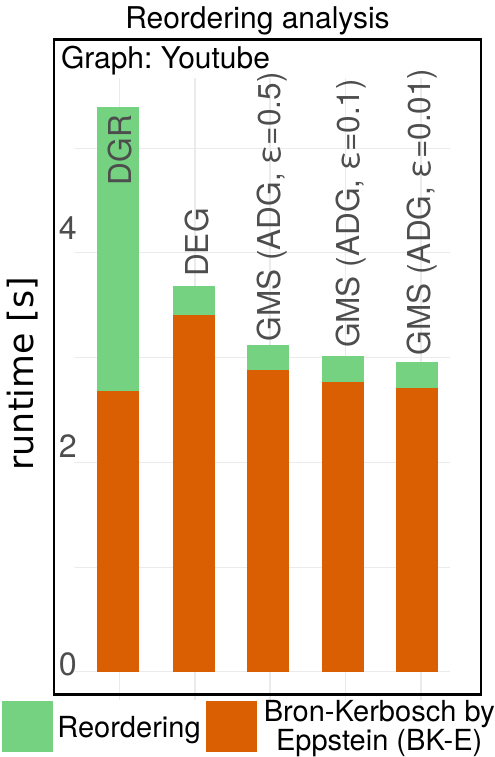} 
\vspaceSQ{-2em}
        \caption{\textmd{\textbf{Speedups of ADG
for different $\epsilon$} over 
DEG/DGR, details in~\cref{sec:eval_deg}. System: Ault.}}
\label{fig:results_reorder}
    \end{minipage}\hfill 
    \begin{minipage}{0.57\columnwidth}
        \centering
        \includegraphics[width=1.0\textwidth]{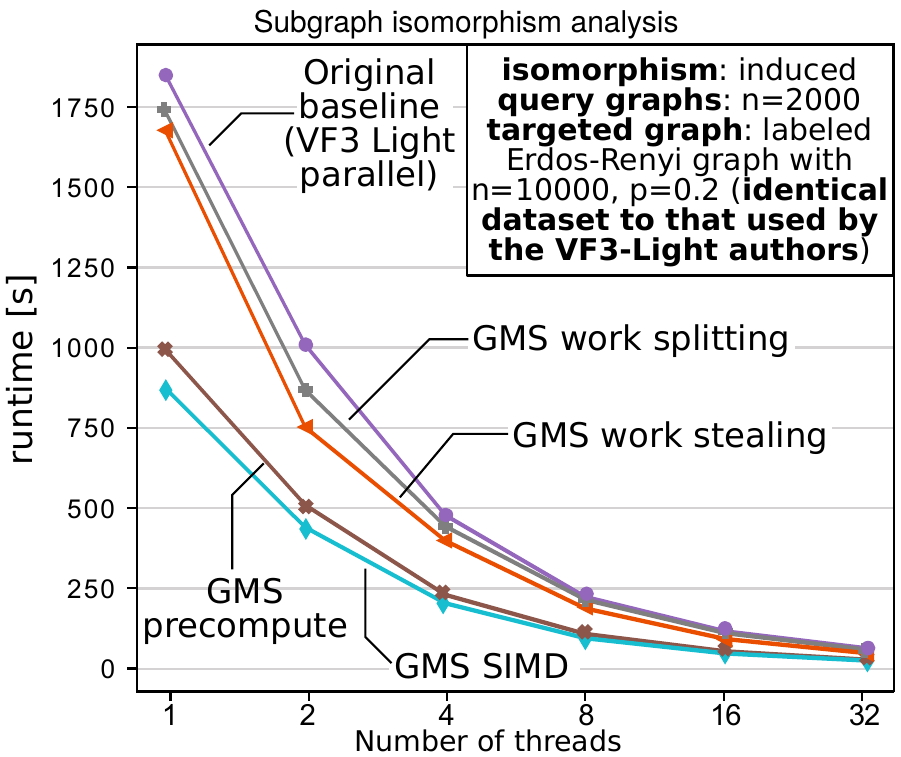} 
\vspaceSQ{-2em}
        \caption{\textmd{\textbf{Speedups of different GMS variants of subgraph isomorphism} over
the state-of-the-art parallel VF3-Light baseline~\protect\cite{carletti2019parallel}. Details in~\cref{sec:eval_sub}.
System: Euler.}}
\label{fig:results_sub}
    \end{minipage}
\vspaceSQ{-1.5em}
\end{figure}

\all{
\begin{figure}[h]
\centering
\includegraphics[width=1.0\columnwidth]{vf3-faster__reorder.pdf}
\vspaceSQ{-2.25em}
\caption{\textmd{\textbf{On the left:} 
Performance advantages of our approximate degeneracy reordering (ADG,
for different parameters of $\epsilon \in \{0.5, 0.1, 0.01\}$) over other
orderings (degree DEG, exact degeneracy DGR), and its positive impact on the runtime
of a recent Bron-Kerbosch algorithm for maximal clique
listing~\protect\cite{DBLP:conf/isaac/EppsteinLS10}. System: Ault.
\textbf{On the right:} 
Performance advantages of different GMS variants of subgraph isomorphism over
the state-of-the-art parallel VF3-Light baseline~\protect\cite{carletti2019parallel}.
System: Euler.}}
\label{fig:reorder-vf3-faster}
\vspaceSQ{-1.25em}
\end{figure}
}

\ifall\maciej{OLD data}
\begin{figure}[h]
\centering
\includegraphics[width=1.0\columnwidth]{BK_Paper_Daint___e.pdf}
\vspaceSQ{-2.25em}
\caption{\textmd{Performance advantages of two Bron-Kerbosch algorithm variants
delivered in GMS, compared to  a state-of-the-art implementation by Das et
al.~\protect\cite{das2018shared} (BK-D) and a recent algorithm by Eppstein et
al.~\protect\cite{DBLP:conf/isaac/EppsteinLS10} (BK-E-DGR). System: Daint.}}
\label{fig:bk-faster}
\vspaceSQ{-2em}
\end{figure}
\fi

\ifall
\begin{figure*}[t]
\vspaceSQ{-1.25em}
\centering
\includegraphics[width=1.0\textwidth]{kclique-faster-all_final_2.pdf}
\vspaceSQ{-2.25em}
\caption{\textmd{Performance advantages of GMS over a state-of-the-art parallel
$k$-clique listing algorithm~\protect\cite{danisch2018listing}. System: Daint.}}
\label{fig:k-clique-faster}
\vspaceSQ{-1.25em}
\end{figure*}
\fi

\begin{figure*}[t]
\vspaceSQ{-1.25em}
\centering
\subfloat[\hl{Analysis of synthetic graphs.}]{
\includegraphics[width=0.19\textwidth]{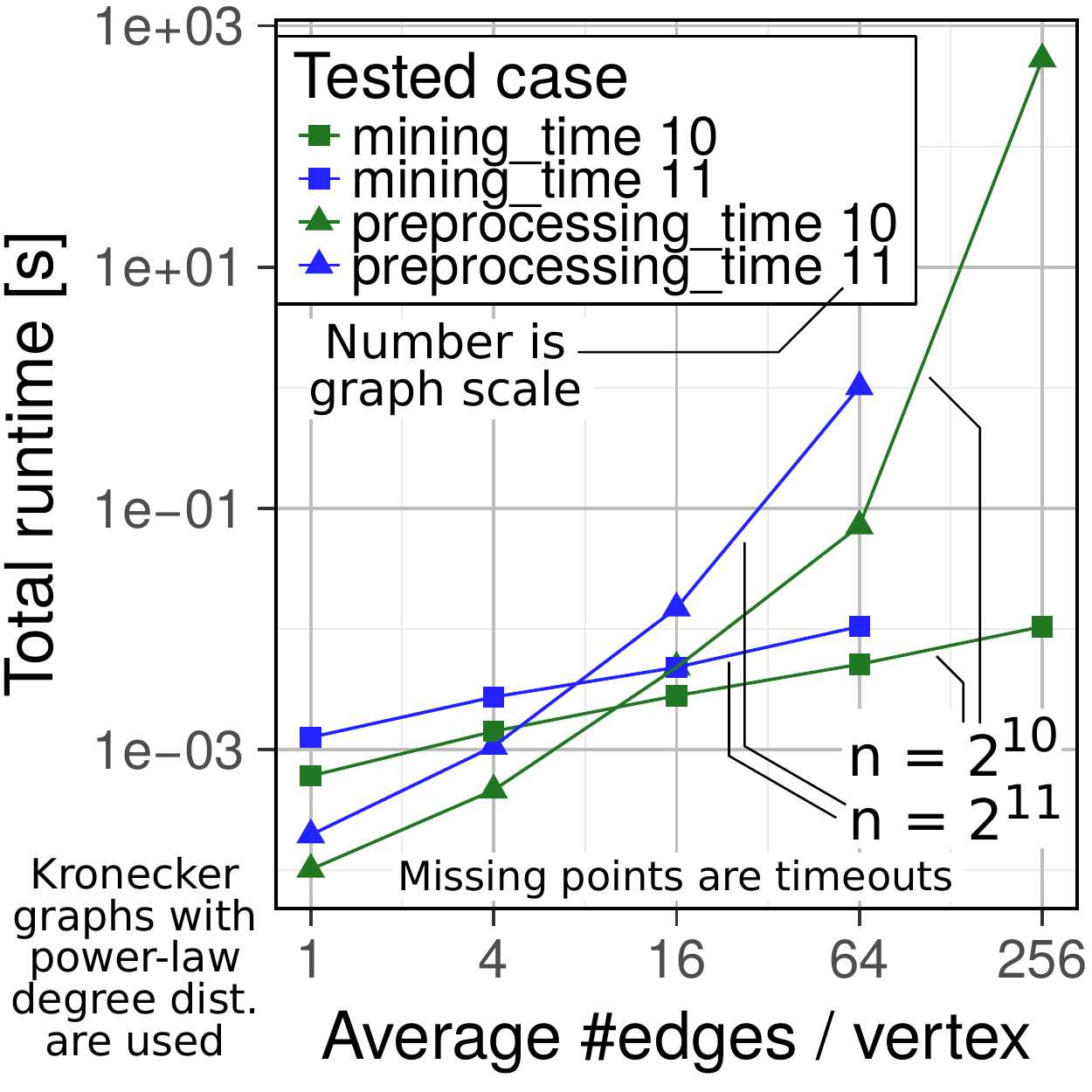}
\label{fig:eval-synt}
}
%
\subfloat[\hl{Analysis of machine efficiency.}]{
\includegraphics[width=0.6\textwidth]{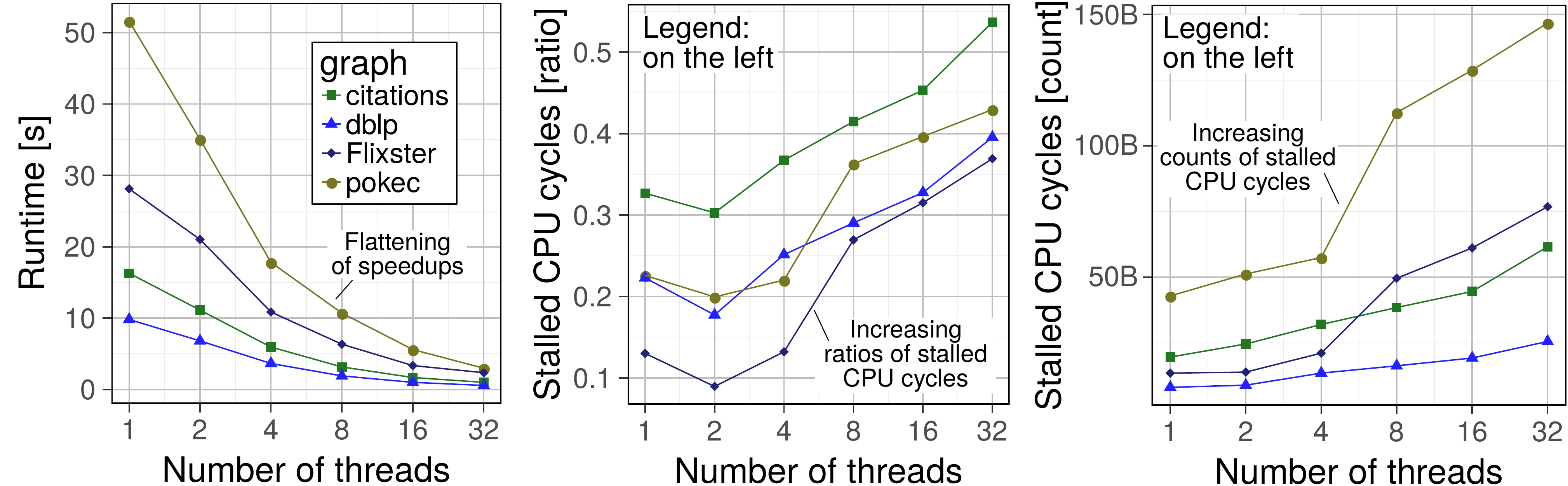}
\label{fig:eval-mem-eff}
}
\subfloat[\hl{Sizes of GMS graph representations.}]{
\includegraphics[width=0.19\textwidth]{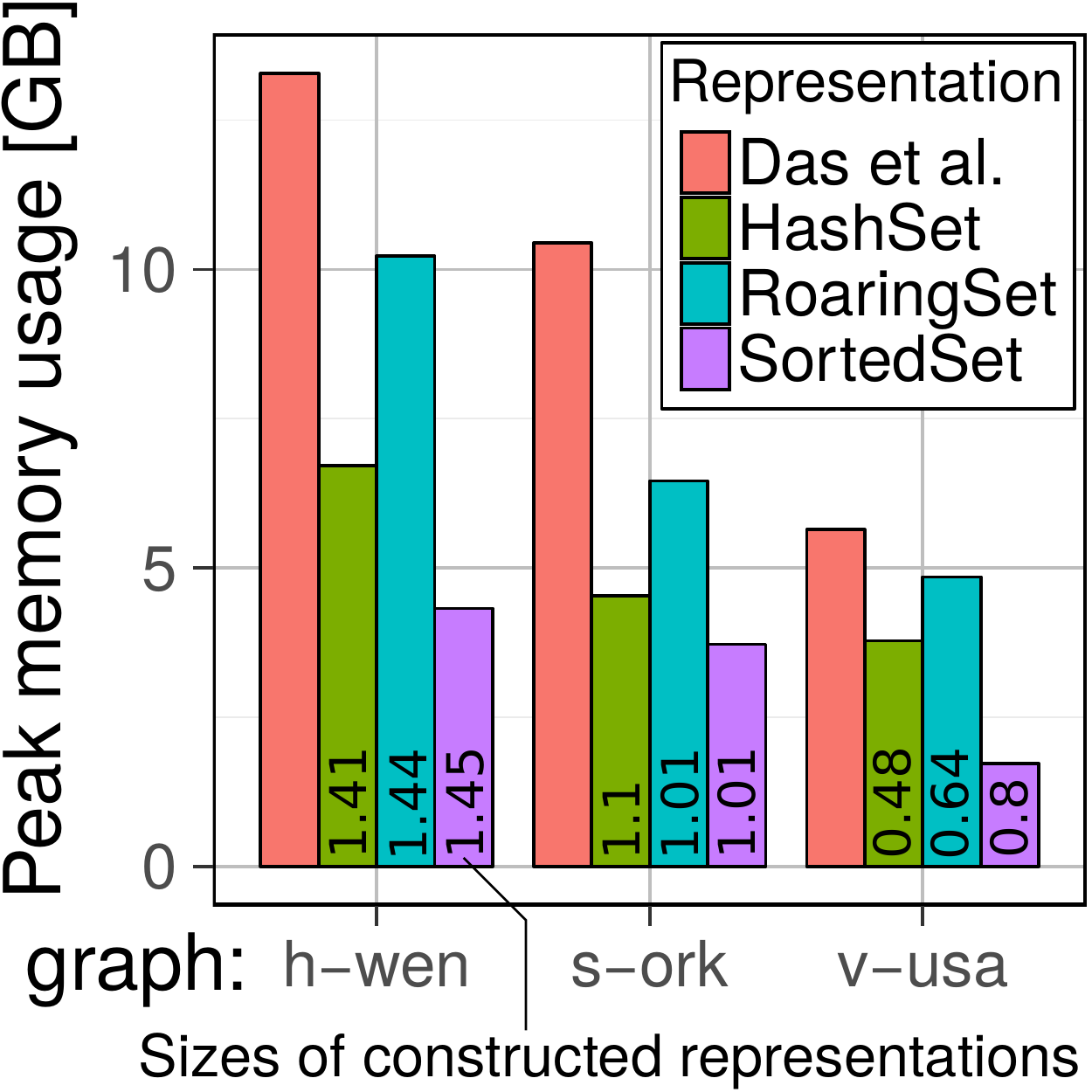}
\label{fig:eval-sizes}
}
\vspaceSQ{-1em}
\caption{\hl{\textmd{\textbf{Additional analyses for the parallel GMS BK algorithm (BK-GMS-DGR)}. System: Daint.}}}
\label{fig:eval-more-stuff}
\vspaceSQ{-1em}
\end{figure*}


%

%

\vspaceSQ{-0.3em}
\subsection{Faster Subgraph Isomorphism}
\label{sec:eval_sub}

\emph{GMS enabled us to accelerate a very recent parallel VF3-Light subgraph
isomorphism baseline by 2.5$\times$.} The results are in
Figure~\ref{fig:results_sub} (we use the same dataset as in the original
work~\cite{carletti2019parallel}). We illustrate the impact from different
optimizations outlined in~\cref{sec:hpc_algs}. 
\all{The gains are due to the combination of work splitting and work stealing,
and precompute optimizations, ultimately ensuring better load balancing.} 
We were also able to use SIMD vectorization in the binary search part of the
algorithms, leading to additional 1.1$\times$ speedup. 

\iftr
\vspaceSQ{-0.3em}
\subsection{Subtleties of Higher-Order Structure}
\label{sec:eval_higher}
\else
\vspaceSQ{-0.3em}
\subsection{Additional Analyses}
\label{sec:eval_others}
\fi

\textbf{Subtleties of Higher-Order Structure}
One of the insights that we gained with GMS is that graphs similar in
terms of $n$, $m$, sparsity~$m/n$, and degree distributions,
may have very different characteristics in their higher-order structure.
For example, a graph of photo relations in Flickr and a Livemocha social
network (see Table~\ref{tab:graphs} for details) are similar in
the above properties, but the former has 9,578,965,096 4-cliques while the
latter has only 4,359,646 4-cliques. This is because, while a in a social
network 4-cliques of friendships may be only \emph{relatively common}, they
should \emph{occur very often} in a network where photos are related if they
share \emph{some} metadata (e.g., location).  
\ifall
\emph{Importantly, this vastly
impacts the performance of algorithms for listing cliques and dense subgraphs.}
\fi
Thus, one should \emph{carefully select input datasets} to properly
evaluate respective graph mining algorithms, as seemingly similar graphs
may have very different higher-order characteristics, which may vastly
impact performance and conclusions when developing a new algorithm.

\all{
\vspaceSQ{-0.3em}
\subsection{Analysis of Performance Metrics}

\maciej{Zur, we would need some discussion on how these metrics look like...
Maybe we could just use existing data to derive the algorithmic efficiency, and
provide some discussion}
}

\iftr
\subsection{{Analysis of Synthetic Graphs}}

\else
\textbf{\hl{Analysis of Synthetic Graphs}}
\fi
\hl{We illustrate example results for synthetic graphs, see
Figure~\mbox{\ref{fig:eval-synt}} (with BK-GMS-DGR).  Using power-law Kronecker
graphs enable us to study the performance impact from varying the graph
sparsity~\mbox{$m/n$} while fixing all other parameters. For very sparse
graphs, the cost of mining cliques is much lower than that of vertex reordering
during preprocessing. However, as \mbox{$m/n$} increases, reordering begins to
dominate. This is because Kronecker graphs in general do not have large
cliques, which makes the mining process finish relatively fast, while
reordering costs grow proportionally to \mbox{$m/n$}.}

\marginpar{\vspace{-25em}\colorbox{yellow}{\textbf{R-1}}\\\colorbox{yellow}{\textbf{R-2}}}

\marginpar{\vspace{-4em}\colorbox{yellow}{\textbf{R-1}}\\\colorbox{yellow}{\textbf{R-2}}}

\iftr
\subsection{{Machine Efficiency Analysis}}

\else
\textbf{\hl{Machine Efficiency Analysis}}
\fi
\hl{We show example analysis of CPU utilization, using the PAPI interface
in GMS, see Figure~\mbox{\ref{fig:eval-mem-eff}}. The plots
illustrate the flattening of speedups with the increasing \#threads,
accompanied by the steady growth of stalled CPU cycles (both total counts and
ratios), showing that maximal clique listing is memory
bound~\mbox{\cite{cheng2012fast, yao2020locality, jamshidi2020peregrine,
zhang2005genome, eblen2012maximum}}.}

\marginpar{\vspace{-4em}\colorbox{yellow}{\textbf{R-1}}\\\colorbox{yellow}{\textbf{R-2}}}

\iftr
\subsection{{Memory Consumption Analysis}}

\else
\textbf{\hl{Memory Consumption Analysis}}
\fi
\hl{We illustrate example memory consumption results in Figure~\mbox{\ref{fig:eval-sizes}};
we compare the size of three GMS set-centric graph representations,
showing both \emph{peak} memory usage when constructing a representation (bars) and sizes of ready
representations (all in GB).
Interestingly, while the latter are similar (except for v-usa), peak memory usage
is visibly highest for RoaringSet.
We also compare to the representation used by Das et al.~\mbox{\cite{das2018shared}},
it always comes with the highest peak storage costs.}

\marginpar{\vspace{-4em}\colorbox{yellow}{\textbf{R-1}}\\\colorbox{yellow}{\textbf{R-2}}}


\iftr
\subsection{{Algorithmic Throughput Analysis}}

\else
\textbf{\hl{Algorithmic Throughput Analysis}}
\fi
\hl{The advantages of using algorithmic throughput can be seen by comparing
Figure~\mbox{\ref{fig:posterchild}} and~\mbox{\ref{fig:bk-faster}}.
While plain runtimes illustrate which algorithm is faster for which graph, the
algorithmic throughput also \emph{enables combining this outcome with the input
graph structure}. For example, the GMS variants of BK have relatively lower 
benefits over BK by Das et al.~\mbox{\cite{das2018shared}} \emph{whenever 
the input graph has a higher density of maximal cliques.}
This motivates using the GMS BK especially for very sparse graphs
without large dense clusters.
One can derive analogous insights for any other patterns such a \mbox{$k$}-cliques.
}

\marginpar{\vspace{-4em}\colorbox{yellow}{\textbf{R-1}}\\\colorbox{yellow}{\textbf{R-2}}}

\marginpar{\vspace{4em}\colorbox{yellow}{\textbf{R-2}}\\\colorbox{yellow}{\textbf{R-3}}}

\iftr
\subsection{{GMS and Graph Processing Benchmarks}}

\else
\textbf{\hl{GMS and Graph Processing Benchmarks}}
\fi
\hl{There is very little overlap with GMS and existing graph processing
benchmarks, see Section~\mbox{\ref{sec:intro}} and
Table~\mbox{\ref{tab:comparison_problems}}.  The closest one is
GBBS~\mbox{\cite{dhulipala2018theoretically}}, which supports the exact
same variant of mining \mbox{$k$}-cliques. We compare GBBS to GMS in
Figure~\mbox{\ref{fig:k-clique-more}}; we also consider the edge-based very
recent implementation by Danisch et al.~\mbox{\cite{danisch2018listing}}. GMS
offers consistent advantages for different graphs and large clique sizes.}


\begin{figure}[h]
\vspaceSQ{-0.25em}
\centering
\includegraphics[width=0.42\textwidth]{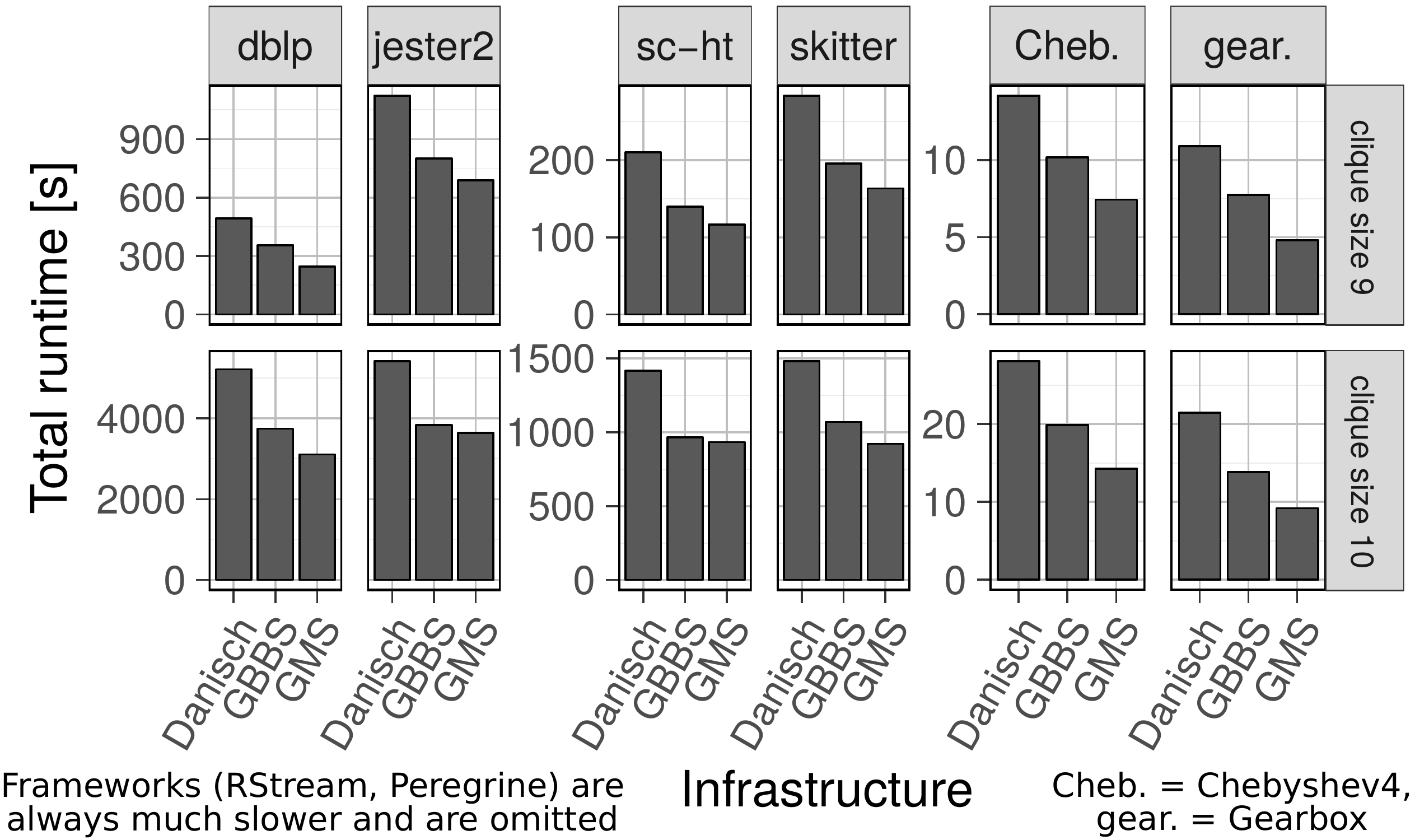}
\vspaceSQ{-1em}
\caption{\hl{\textmd{Comparison between GMS and the GBBS benchmark, for mining
\mbox{$k$}-cliques. We also include Danisch's algorithm (edge-centric)
for additional reference. System: Einstein (full parallelism).}}}
\label{fig:k-clique-more}
\vspaceSQ{-2em}
\end{figure}

\marginpar{\vspace{4em}\colorbox{yellow}{\textbf{R-2}}\\\colorbox{yellow}{\textbf{R-3}}}

\iftr
\subsection{{GMS and Pattern Matching Frameworks}}

\else
\textbf{\hl{GMS and Pattern Matching Frameworks}}
\fi
\hl{There is also little overlap between GMS and pattern matching frameworks,
cf.~Table~\mbox{\ref{tab:comparison_problems}}. While they support
mining patterns, they focus on patterns of fixed sizes (e.g.,
\mbox{$k$}-cliques).
We compare GMS to two very recent frameworks that, similarly to GMS, target
shared-memory parallelism, Peregrine~\mbox{\cite{jamshidi2020peregrine}} and
RStream~\mbox{\cite{wang2018rstream}}. Peregrine can only list
\mbox{$k$}-cliques. It does not offer a native scheme for maximal clique
listing and we implement it by iterating over \mbox{$k$}-cliques of different
sizes (we consult the authors of Peregrine to find the best scheme). RStream is
only able to find \mbox{$k$}-cliques.
Overall, GMS is much faster in all considered schemes (10-100\mbox{$\times$}
over Peregrine and more than 100\mbox{$\times$} over RStream). This is because
these systems focus on programming abstractions, which improves programmability
but comes with performance overheads. GMS enables maximizing performance of
parallel algorithms targeting specific problems.}

\marginpar{\vspace{-17em}\colorbox{yellow}{\textbf{R-2}}\\\colorbox{yellow}{\textbf{R-3}}}

\ifall

\subsection{Key Takeaways}

We conclude this section with a design and an algorithmic takeaway.
In the former, we observe that the GMS platform facilitates developing 
the considered algorithms. In the latter, GMS helped to discover
that the ADG ordering benefits clique listing graph mining problems,
both theoretically and empirically.

\fi

\iftr
\section{Related Work \& Discussion}

We already exhaustively analyzed a large body of works related to {graph
processing benchmarks}~\cite{DBLP:journals/ppl/LumsdaineGHB07, besta2017push,
besta2015accelerating} and {graph pattern matching frameworks}, see 
Section~\ref{sec:intro} and Tables~\ref{tab:comparison_problems}
and~\ref{tab:comparison_functionalities}.
General graph processing is summarized in a recent overview~\cite{sakr2020future}.
In general, GMS complements these works by delivering the first benchmarking
suite that specifically targets graph mining. 

While in the current GMS version we focus on the parallel shared-memory
setting, GMS could be extended into multiple directions as future work. This
includes moving into distributed processing~\cite{gonzalez2014graphx,
gonzalez2012powergraph} and incorporating high-performance
techniques~\cite{thebault2016scalable, firoz2018runtime, gregor2005parallel}
such as Remote Direct Memory Access~\cite{besta2015active, besta2014fault,
fompi-paper, gerstenberger2018enabling, schweizer2015evaluating,
schmid2016high} combined with using general high-performance networks that work
well with communication-intensive workloads~\cite{besta2014slim, di2019network,
besta2020highperformance, besta2020fatpaths}.
We are also working on variants of graph mining algorithms in GMS that harness
the capabilities of the underlying hardware, such as low-diameter
on-chip networks~\cite{besta2018slim, moscibroda2009case, grot2011kilo}, NUMA
and focus on data locality~\cite{schweizer2015evaluating, tate2014programming},
near- and in-memory processing~\cite{seshadri2017ambit, ahn2015scalable,
hassan2016chargecache, mutlu2015research, hsieh2016accelerating, seshadri2015fast, boroumand2016lazypim, 
pattnaik2016scheduling, mutlu2013memory, seshadri2013rowclone, lee2013tiered,
lee2010phase, ahn2015pim}, 
various architecture-related compression techniques~\cite{pekhimenko2012base, pekhimenko2013linearly},
and others~\cite{de2018transformations, kim2012case}.
One could incorporate various forms of recently proposed lossy {graph
compression} and {summarization}~\cite{besta2019slim, besta2018survey,
liu2018graph}, and graph neural networks~\cite{ben2019modular,
wu2020comprehensive, xu2018powerful}.
\fi

\ifall
As we already broadly discussed related work over Sections~\ref{sec:intro},
\ref{sec:overview}, \ref{sec:bench_spec}, and~\ref{sec:hpc_algs}, we now only
summarize key differences.

\macb{Benchmarks}
We describe related benchmarks in Section~\ref{sec:intro}.  GMG is the first
benchmark suite for graph mining. Moreover, it is the only one that comes with
a design platform that facilitates constructing high-performance graph mining.
In GMB, we use as a dependency several routines from the GAP Benchmark
Suite~\cite{beamer2015gap} for loading graphs from files and constructing the
CSR representation.

\macb{Graph Processing Frameworks}
GMB is not a processing framework. Instead, it is a benchmark specification and
thus is can serve as support when selecting the scope of a potential framework,
or appropriate comparison baselines.

\macb{Graph Mining Algorithms}
We broadly described graph mining algorithms in Sections~\ref{sec:bench_spec}
and~\ref{sec:hpc_algs}. The purpose of GMB's design platform is to facilitate
developing high-performance graph mining algorithms, by making it easier to
design, develop, and evaluate different algorithms, optimizations, data layouts
and representations, load balancing, and many others.

\macb{Complex Network Analysis}
GMB does \emph{not} aim to cover advanced statistical methods that -- for
example -- analyze power laws in input graphs.  For this, we recommend to use
specialized software, for example iGraph~\cite{csardi2006igraph}. 

\macb{Discussion: What Is Included \& Excluded?}
We fix GMB's scope to include problems and algorithms related to ``graph
mining'', often also referred to as ``graph analytics'', in the \emph{offline}
setting, with a \emph{single} input graph. Thus, we do \emph{not} focus on
streaming or dynamic graphs (as they usually come with vastly different design
and implementation challenges~\cite{besta2019practice}) and we do \emph{not}
consider problems that operate on multiple \emph{different} input graphs.  We
leave these two domains for future work.
Finally, we do \emph{not} consider low-complexity problems such as BFS or
Connected Components, as we find them of low interest for the domain of graph
mining.
However, one can easily add these problems to GMB. 
\fi

\ifall
\maciej{Check: koutra2017individual}
\fi

\section{CONCLUSION}

We introduce GraphMineSuite (GMS), the first benchmarking suite for
graph mining algorithms. GMS offers an extensive \emph{benchmark specification}
and taxonomy that distill more than 300 related works and can aid in selecting
appropriate comparison baselines. 
Moreover, GMS delivers a highly modular \emph{benchmarking platform}, with dozens of
parallel implementations of key graph mining algorithms and graph
representations.  Unlike \emph{frameworks for pattern matching} which focus on
abstractions and programming models for expressing mining specific patterns,
GMS simplifies designing high-performance \emph{algorithms for solving specific
graph mining problems} from a \emph{wide} graph mining area.
Extending GMS towards distributed-memory systems or dynamic workloads are
interesting future lines of work. 
Third, GMS' \emph{concurrency analysis} illustrates
theoretical tradeoffs between time, work, storage, and accuracy, of several
representative problems in graph mining; it can be used as a guide when rapidly
analyzing the scalability of a planned graph mining scheme\tr{, or to obtain
performance insights independent of implementation details}. 
Finally, we show GMS' potential by using it to \emph{enhance state-of-the-art
graph mining algorithms}, leading to theoretical and empirical advancements in
maximal clique listing (speedups by $>$9$\times$ and better work-depth bounds
over the \emph{fastest known Bron-Kerbosch baseline}), degeneracy reordering
and core decomposition (speedups by $>$2$\times$), $k$-clique listing (speedups
by up to 1.1$\times$ and better bounds), and subgraph isomorphism (speedups by
2.5$\times$).
\all{
GMS may propel research into fast and scalable graph mining algorithms on
today's and future massively parallel architectures. }


\iftr

\vspace{1em}

\footnotesize

%

\macb{Acknowledgements: }
We thank Hussein Harake, Colin McMurtrie, Mark Klein, Angelo Mangili, and the
whole CSCS team granting access to the Ault and Daint machines, and for their
excellent technical support.
We thank Timo Schneider for immense help with computing infrastructure at SPCL.
We thank Maximilien Danisch, Oana Balalau, Mauro Sozio, Apurba Das, Seyed-Vahid
Sanei-Mehri, and Srikanta Tirthapura for providing us with the implementations
of their algorithms for solving $k$-clique and maximal clique listing.
We thank Dimitrios Lekkas, Athina Sotiropoulou, Foteini Strati, Andreas
Triantafyllos, Kenza Amara, Chia-I Hu, Ajaykumar Unagar, Roger Baumgartner,
Severin Kistler, Emanuel Peter, and Alain Senn for helping with the
implementation in the early stages of the project.

\normalsize

\fi

\iftr
\appendix

\section*{Appendix}

We now provide extensions of several sections.

\section{Details of Problems and Algorithms in Graph Mining}
\label{sec:app-details}

We additionally provide more details of considered graph mining
problems and the associated algorithms.

\textbf{$\bullet$ {Maximal Cliques Listing }}
For finding maximal cliques, we use the established {Bron-Kerbosch (BK)
algorithm}~\cite{bron1973algorithm}, a recursive backtracking algorithm often
used in practice, with well-known {pivoting} and {degeneracy
optimizations}~\cite{manoussakis2018output, cazals2008note,
DBLP:conf/isaac/EppsteinLS10, DBLP:journals/tcs/TomitaTT06}.

\textbf{$\bullet$ {$k$-Clique Listing }}
GMS considers listing $k$-cliques. We select a state-of-the-art
{algorithm by Danisch et al.}~\cite{danisch2018listing}. The algorithm
is somewhat similar to Bron-Kerbosch in that it is also recursive backtracking.
The difference is that its work is polynomial.
\tr{We also separately consider {Triangle Counting} as it comes with a
plethora of specific studies~\cite{al2018triangle, shun2015multicore,
schank2007algorithmic}.}

\textbf{$\bullet$ {Dense Non-Clique Subgraph Discovery }}
We also incorporate a problem of discovering dense \emph{non-clique} subgraphs.
Relevant classes of subgraphs are {quasi-cliques}, $k$-cores, $k$-plexes,
$k$d-cliques, $k$-clubs, $k$-clans, dal$k$s, dam$k$s, d$k$s, $k$-clique-stars,
and others~\cite{lee2010survey, jabbour2018pushing}.
Here, we implemented a very recent {algorithm for listing
$k$-clique-stars}~\cite{jabbour2018pushing}; $k$-\emph{clique-stars} are dense
subgraphs that combine the characteristics of cliques \emph{and stars}
(relaxing the restrictive nature of $k$-cliques)~\cite{jabbour2018pushing}.
GMS also implements an {exact} and an {approximate algorithm for
$k$-core decomposition}~\cite{DBLP:conf/latin/Farach-ColtonT14}.

\textbf{$\bullet$ {Subgraph Isomorphism }}
Subgraph isomorphism (SI) is an important NP-Complete problem, where one finds
all embeddings of a certain \emph{query graph}~$H$ in another \emph{target
graph}~$G$.
SI can be \emph{non-induced} and \emph{induced}; we
consider {both}. Consider a case where an embedding of~$H$ is found
in~$G$, but there are some additional edges in~$G$ that connect some vertices
that belong to this embedding. In the non-induced variant, this situation is
permitted, unlike in the induced variant, where the found embedding cannot have such
additional edges.
In GMS, we consider recent algorithms: {VF2}~\cite{cordella2004sub} and
{adapted Glasgow}~\cite{mccreesh2015parallel} for induced SI, and
{VF3-Light}~\cite{carletti2018vf3} and
{TurboISO}~\cite{han2013turbo} for non-induced SI.
Our selection covers different approaches for solving SI: VF2 and VF3-Light
represent backtracking as they descent from the well-known ULLMAN
algorithm~\cite{ullmann1976algorithm}.  TurboISO uses a graph indexing as
opposed to backtracking while Glasgow incorporates implied constraints.

\textbf{$\bullet$ {Frequent Subgraph Mining }}
We separately consider the Frequent Subgraph Mining (FSM)
problem~\cite{jiang2013survey}, in which one finds all subgraphs that occur
more often than a specified threshold. An FSM algorithm consists of (1) a
strategy for exploring the tree of candidate subgraphs, and (2) a subroutine
where one checks if a candidate is included in the processed graph. (2)~usually
solves the {subgraph isomorphism} problem, covered above. (1)~uses
either a {BFS-based} or a {DFS-based} exploration strategy.

\textbf{$\bullet$ {Vertex Similarity }}
Vertex similarity measures can be used on their own, for example in graph
database queries~\cite{robinson2013graph}, or as a building block of more
complex algorithms such as clustering~\cite{jarvis1973clustering}. We consider
seven measures: {Jaccard}, {Overlap}, {Adamic Adar}, {Resource Allocation},
{Common Neighbors}, {Total Neighbors}, and {Preferential Attachment}
measures~\cite{leicht2006vertex, robinson2013graph}. All these measures
associate (in different ways) the degree of similarity between $v$ and $u$ with
the number of common neighbors of vertices $v$ and $u$.

\textbf{$\bullet$ {Link Prediction }}
Here, one is interested in developing schemes for predicting whether two
non-adjacent vertices can become connected in the future. There exist many
schemes for such prediction that are based on {variations of vertex
similarity}~\cite{liben2007link, lu2011link,
al2006link, taskar2004link}. We provide them in GMS, as well as a simple
algorithm for {assessing the accuracy} of a specific link prediction
scheme~\cite{wang2014robustness}, which assesses how well a given prediction
scheme works.

\textbf{$\bullet$ {Clustering and Community Detection }}
We consider graph clustering and community detection, a widely studied problem.
We pick {Jarvis-Patrick clustering}
(JP)~\cite{jarvis1973clustering}, a scheme that uses similarity of two vertices
to determine whether these two vertices are in the same cluster.
Moreover, we consider {Label Propagation}~\cite{raghavan2007near}
and the {Louvain method}~\cite{blondel2008fast}, two established
methods for detecting communities that, respectively, use the notions of
\emph{label dominance} and \emph{modularity} in assigning vertices to
communities. 

\textbf{$\bullet$ {Approximate Degeneracy Ordering }}
We also consider an easily parallelizable algorithm to compute an
\emph{approximate} degeneracy order (the algorithm has $O(\log n)$ iterations
for any constant $\epsilon>0$ and has an approximation ratio of
  $2+\epsilon$~\cite{besta2020high}). The algorithm is based on a streaming
  scheme for large graphs~\cite{DBLP:conf/latin/Farach-ColtonT14} and uses set
  cardinality and difference.
The derived degeneracy order can be directly used to compute the $k$-core of
$G$ (a maximal connected subgraph of $G$ whose all vertices have degree at
least $k$). This is done by iterating over vertices in the degeneracy order and
removing all vertices with out-degree less than $k$ (in the oriented graph).

\textbf{$\bullet$ {Optimization Problems }}
Third, we also consider some problems from a family of
\emph{{optimization problems}}, also deemed important in the
literature~\cite{aggarwal2010managing}. 
Here, we focus on {graph coloring} (GC), considering
several graph coloring algorithms that represent different approaches:
{Jones and Plassmann's}~\cite{jones1993parallel} and {Hasenplaugh
et al.'s}~\cite{hasenplaugh2014ordering} heuristics based on appropriate vertex
orderings and vertex prioritization,
{Johansson's}~\cite{johansson1999simple} and
{Barenboim's}~\cite{barenboim2016locality} randomized palette-based
heuristics that use conflict resolution, and {Elkin et
al.'s}~\cite{elkin20142delta} and {sparse-dense
decomposition}~\cite{harris2016distributed} that are examples of
state-of-the-art distributed algorithms.


\begin{figure*}[hbtp]
\centering
\includegraphics[width=1.0\textwidth]{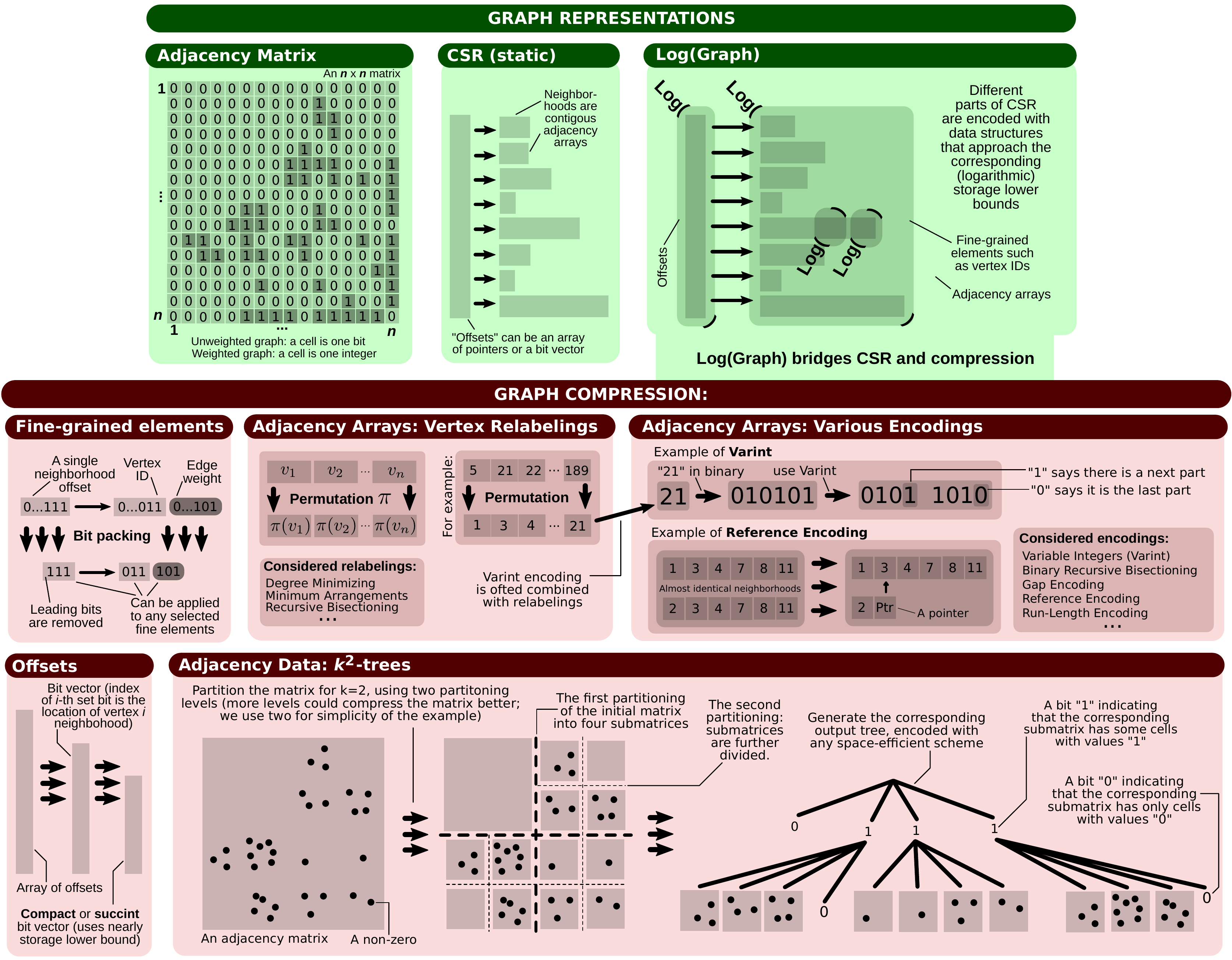}
%
\caption{\textmd{\textbf{An overview of models, representations, and
compression schemes.}} The GMS logo indicates the ones provided in
the current GMS platform. A recent survey provides full
details on all the representations~\cite{besta2018survey}}
%
\label{fig:reps-dets}
\end{figure*}

\section{Navigating the Maze of Graph Representations}
\label{sec:app-maze}

The right data layout is one of key enablers of high performance.
For this, we now overview the most relevant graph representations
and compression schemes.
%
%
We picture key designs in Figure~\ref{fig:reps-dets}.

\begin{table*}[t]
\centering
\setlength{\tabcolsep}{2pt}
\renewcommand{\arraystretch}{1.5}
\footnotesize
%
\begin{tabular}{@{}lllll@{}}
\toprule
\textbf{Algorithm} & \textbf{AL (sorted)} & \textbf{AM} & \textbf{EL (unsorted)} & \textbf{EL (sorted)} \\
\midrule
Node Iterator (TC) & $\bigO{n + m^{{3}/{2}} \log \Delta }^*$ & $\bigO{n + m^{3/2}}$ & $\bigO{n + m^{3/2}( \Delta  +\log m)}$ & $\bigO{n + m^{5/2}}$ \\
Rank Merge (TC) & $\bigO{n + n \Delta  + m^{3/2}}$ & $\bigO{n + n \Delta  + m^{3/2}}$ & $\bigO{n + n \Delta  + m^{3/2}}$ & $\bigO{n + n \Delta  + m^{3/2}}$ \\
BFS, top-down & $\bigTheta{n + m}$ & $\bigTheta{n + m}$ & $\bigO{n \log m + m}$ & $\bigO{nm + n + m}$ \\
PageRank, pushing & $\bigO{n + m^{3/2}\log  \Delta }^*$ & $\bigO{n + m^{3/2}}$ & $\bigO{n + m^{3/2}( \Delta  +\log m)}$ & $\bigO{n + m^{5/2}}$ \\
$D$--Stepping (SSSP) & $\bigO{n + m +  \frac{L}{D} + n_D + m_D}$ & $\bigO{n^2 +  \frac{L}{D} + nn_D + m_D}$ & $\bigO{nm +  \frac{L}{D} + n_D(\log m + \Delta) + m_D}$ & $\bigO{nm + m +  \frac{L}{D} + n_D m + m_D}$ \\
Bellman-Ford (SSSP) & $\bigO{n^2 + nm}$ & $\bigO{n^3}$ & $\bigO{n + nm}$ & $\bigO{n + nm}$ \\
Boruvka (MST) & $\bigO{m\log n}$ & $\bigO{n^2\log n}$ & $\bigO{nm\log n \log m}$ & $\bigO{n^2m}$ \\
Boman (Graph Coloring) & $\bigO{n + m}$ & $\bigO{n^2}$ & $\bigO{n^2}$  & $\bigO{n + nm}$ \\
Betweenness Centrality & $\bigO{nm}$ & $\bigO{n^3}$ & $\bigO{nm \log m}$ & $\bigO{nm^2}$ \\
\bottomrule
\end{tabular}
%
%
\caption{\textbf{Time complexity of graph algorithms for different graph
representations.}
``$^*$'' indicates that the $\log \Delta$ terms becomes $\Delta$ when the used AL representation is unsorted.
%
%
$D$ is a parameter of the Delta--Stepping algorithm that controls the
``bucket size'' and thus the amount of parallelism (for $D=1$ one obtains
Dijkstra's algorithm while for $D=\infty$ one obtains the Bellman-Ford algorithm).
$L$ is the maximum length of a shortest path between any two
vertices.
}
\label{tab:reps_algs}
\end{table*}

%

\subsection{Graph Representations}
\label{sec:app_reps_fundamental}

\ifall
We model a {simple undirected unweighted} graph $G$ as a tuple $(V,E)$;
$V$ is a set of vertices and $E \subseteq V \times V$ is a set of edges (or
arcs, if $G$ is {directed}); $|V|=n$, $|E|=m$. 
If $G$ is {weighted}, it is usually modeled with a tuple $(V,E,w)$,
where $w: E \to \mathbb{R}$ is a function that assigns weights to vertices.
\fi

We consider several graph representations. 
%
%

\subsubsection{Adjacency List and Adjacency Array}

AL is a very popular representation that uses $O(m \log n + n \log m)$ space,
with many implementations and variants. AM uses $O(n^2)$ space and is thus
rarely directly used. However, several interesting compression schemes are
based on AM, for example $k^2$-trees or some succinct
graphs~\cite{besta2018survey}.

%

\subsubsection{CSR aka Adjacency Array}

Compressed Sparse Row (CSR), also referred to as Adjacency Array (AA), usually
consists of $n$ arrays that contain neighborhoods of graph vertices. Each array
is usually sorted by vertex IDs. AA also contains a structure with offsets (or
pointers) to each neighborhood array.
%
%
%
AA is very popular in processing engines for static graphs~\cite{besta2018log}.
Due to its simplicity, it offers very low latency of accesses. Moreover, its
variants are used in graph streaming settings~\cite{besta2019practice}, where
edges and vertices may be inserted or deleted over time.

\subsubsection{Log(Graph)}

Log(Graph)~\cite{besta2018log} is a recently proposed variant of AA, in which
one separately compresses fine elements of the representation (vertex IDs, edge
weights, etc.) as well as coarse parts, such as a whole offset array. The main
compression method in Log(Graph) is encoding each considered graph element
using \emph{a data structure that approaches the corresponding logarithmic
storage lower bounds while simultaneously enabling fast accesses}.  An example
such structure used in Log(Graph) are succinct bit vectors.
One Log(Graph) advantage is \emph{low-overhead decompression}. Another benefit
is a \emph{tunable storage-performance tradeoff}: one can choose to compress
more aggressively at the cost of more costly decompression, and vice versa.
Third, Log(Graph) is modular: the user can select which parts of AA are
compressed.

\begin{table}[t]
\centering
\setlength{\tabcolsep}{2pt}
\sf
\footnotesize
%
\begin{tabular}{@{}lllll@{}}
\toprule
\textbf{Graph query} & \textbf{AL} & \textbf{AM} & \textbf{EL (unsorted)} & \textbf{EL (sorted)} \\
\midrule
Iterate over all vertices & $\bigTheta{n}$ & $\bigTheta{n}$ & $\bigTheta{n}$ & $\bigTheta{n}$ \\
Iterate over all edges & $\bigTheta{n+m}$ & $\bigTheta{n^2}$ & $\bigTheta{m}$ & $\bigTheta{m}$ \\
Iterate over a neighborhood & $\bigTheta{\Delta}$ & $\bigTheta{n}$ & $\bigTheta{m}$ & $\bigTheta{\log m + \Delta}^{\#}$ \\
Check vertex' degree & $\bigTheta{n}^*$ & $\bigTheta{n}^*$ & $\bigTheta{m}^*$ & $\bigTheta{\log m + \Delta}^{*\#}$ \\
Check edge's existence & $\bigO{\log \Delta}$ & $\bigO{1}$ & $\bigO{m}$ & $\bigO{\log m}$ \\
Check edge's weight & $\bigO{1}$ & $\bigO{n}$ & $\bigTheta{m}$ & $\bigTheta{\log m + \Delta}^{\#}$ \\
\bottomrule
\end{tabular}
%
%
\caption{\textbf{Time complexity of graph queries for different graph
representations.} ``$^*$'' indicates that a given complexity can be reduced to
$\bigO{1}$ with $\bigTheta{m + n}$ preprocessing and $n$ auxiliary storage.
``$^{\#}$'' indicates that a given complexity assumes that each edge $(u,v)$ is present
twice in the edge list (both in $u$'s and in $v$'s neighborhoods),
which requires double storage but does not increase the preprocessing
complexity.}
\label{tab:reps_accesses}
\end{table}

\ifall

\subsubsection{NUMA-Aware CSR}

A NUMA-aware representation maintains information on the physical location of
each vertex in the memory hierarchy of a compute node~\cite{zhang2015numa}. For
example, consider a compute node that has a CPU with two sockets; each socket
has an attached local DRAM bank while still being able to access a remote bank.
Here, a NUMA-aware graph representation could keep track of which vertex is
stored in which physical memory bank. Two established ways of partitioning a
graph across memory banks use {1D} or {2D partitioning} of the
associated adjacency matrix~\cite{ccatalyurek2001fine}.
Considering a NUMA-aware CSR variant~\cite{zhang2015numa} facilitates the
analysis of the impact of modern {hierarchical} multi and manycore
architectures on the performance of graph computations.

\fi

\subsection{Graph Compression Schemes}
\label{sec:back_compress}

In a survey on lossless graph compression and space-efficient graph
representations~\cite{besta2018survey}, we illustrate that the majority of
graph compression methods fall into two major families of
methods: relabelings (permutations) and transformations.

One type of the considered compression schemes for adjacency data are
\textbf{relabelings} that permute vertex IDs. Different permutations enable
more or less efficient compression of vertex IDs (e.g., when combining
permutations with a Varint compression and gap encoding~\cite{besta2018log}).
Established examples use shingle ordering~\cite{chierichetti2009compressing},
recursive bisection~\cite{Blandford:2003:CRS:644108.644219}, degree
minimizing~\cite{besta2018log}, and Layered Label
Propagation~\cite{boldi2011layered}.

Another type of compression schemes provided by GMS are
\textbf{transformations} that apply a certain function to the adjacency
data~\cite{besta2018log} after a relabeling is used. Here, in addition to the
above-mention Varint and gap encoding, GMS considers
$k^2$-trees~\cite{brisaboa2009k}, run-length and reference
encoding~\cite{besta2018log}, and implementation of certain schemes proposed in 
WebGraph~\cite{boldi2004webgraph}.

\subsection{Theoretical Analysis}

We offer a brief theoretical analysis on the impact of different
representations on the performance of graph queries and algorithms.  The
analysis for the former can be found in Table~\ref{tab:reps_accesses}.
For the latter, Table~\ref{tab:reps_algs} provides time complexities of
Triangle Counting (Node-Iterator and Rank-Merge
schemes~\cite{shun2015multicore}), Page Rank (pushing and
pulling~\cite{besta2017push}), BFS, Betweenness Centrality (Prountzos et al.'
algorithm~\cite{prountzos2013betweenness}), Single Source Shortest Paths
($\Delta$-Stepping~\cite{meyer2003delta} and
Bellman-Ford~\cite{bellman1958routing}), Minimum Graph Coloring (Boman et al.'
algorithm~\cite{boman2005scalable}), and Minimum Spanning Tree (Boruvka's
algorithm~\cite{boruuvka1926jistem}).

\section{Additional Results}

Figure~\ref{fig:more-data-app} shows additional results for the performance of various Bron-Kerbosch
variants, when measured in the mined cliques per time unit.

\begin{figure*}[hbtp!]
\vspace{-2em}
\centering
\includegraphics[width=1.0\textwidth]{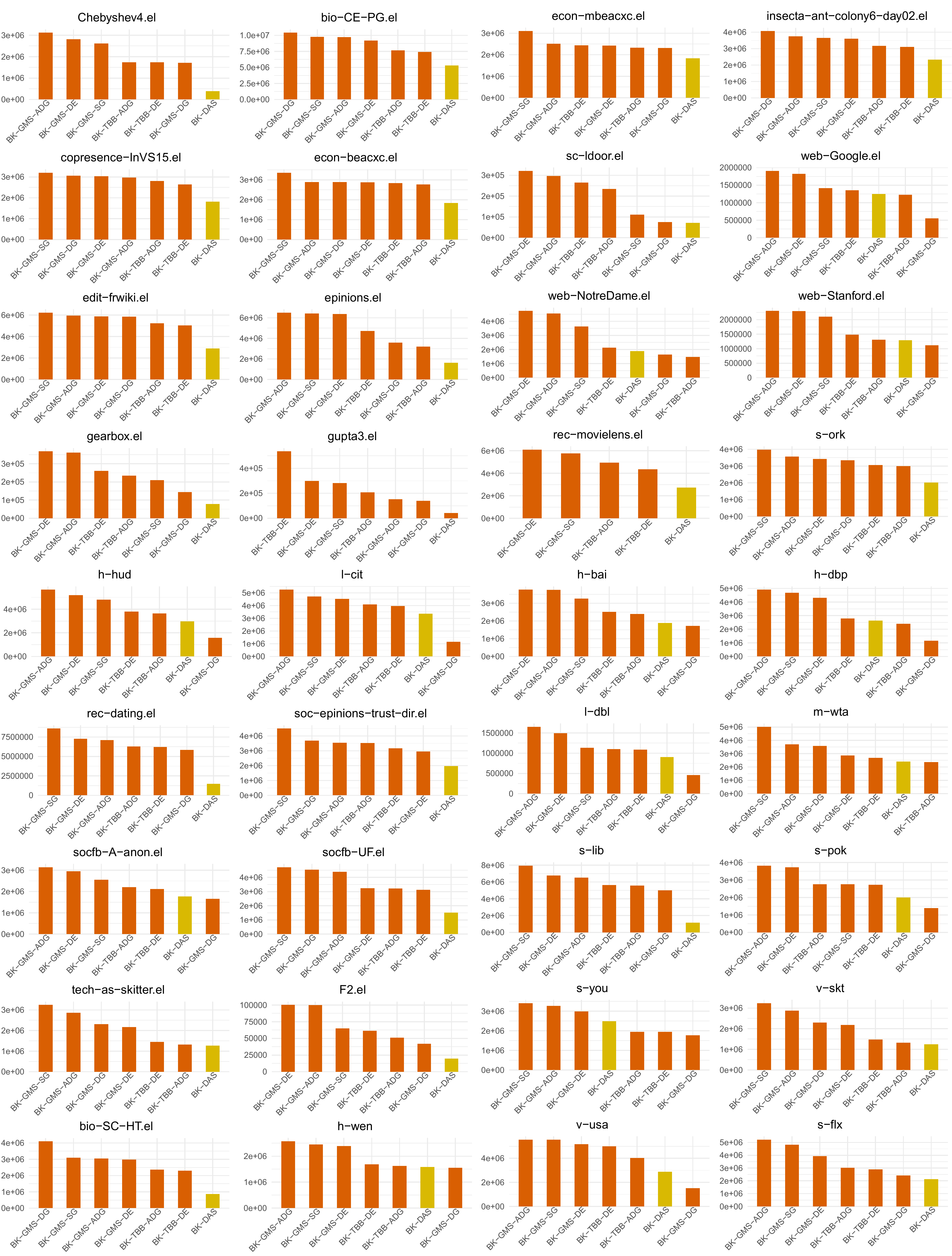}
\vspace{-1.75em}
\caption{\textmd{\textbf{Additional data from performance analysis related to
mining maximal cliques; \ul{the Y axis 
plots the number of maximal cliques mined per time unit.}}
The naming of the schemes is the same as in the main body of the document,
with the following exceptions: ``BK-GMS-SG'' indicates the subgraph
optimization (``BK-GMS-ADG-S''), ``BK-GMS-DG'' indicates the degeneracy
ordering (``BK-GMS-DGR''), and ``BK-GMS-DE'' as well as ``BK-TBB-DE''
are the degree orderings in GMS and TBB, respectively.}}
%
\label{fig:more-data-app}
\end{figure*}

\fi

\bibliographystyle{abbrv}
\bibliography{references}

\iftodos
\maciej{METRICS? others don't support?}

\maciej{!!!!!!!!!!!! add clustering coeff and power law! use Konect site}

\maciej{add balanced mincuts and use METIS?}

\maciej{See if the classes of problems are OK}

\maciej{TODO: go back in the past 10 (?) years of the important mining venues
(KDD, WWW, TKDE, CIKM, etc...) and see if we didn't miss anything?}

\maciej{Potential problems to add: Delaunay mesh / triangulation, balanced mincuts, maxflow, graph isomorphism}
\fi

\end{document}